\documentclass[11pt,a4paper]{article}
\usepackage{jheppub}

\usepackage{amsfonts}
\usepackage{float}
\usepackage{indentfirst}
\usepackage{wrapfig}
\usepackage{latexsym}
\usepackage{slashed}
\usepackage{bm}
\usepackage{cancel}
\usepackage{graphicx}
\usepackage{pdflscape}
\usepackage{caption}
\usepackage{subcaption}
\usepackage{feynmf}
\usepackage{physics}
\usepackage{tensor}
\usepackage{mathtools}
\usepackage{enumitem}
\usepackage{listings}
\usepackage[dvipsnames]{xcolor}
\usepackage{url}
\usepackage{multirow}
\usepackage{empheq}

\newcommand{\Op}{\mathcal{O}}

\newcommand{\DD}{\mathcal{D}}

\newcommand{\Lagr}{\mathcal{L}}

\newcommand{\Hilb}{\mathcal{H}}

\newcommand{\EE}{\mathcal{E}}
\newcommand{\p}{\partial}

\newcommand{\Dadj}[1]{\bar{#1}}

\newcommand{\RN}[1]{%
  \textup{\uppercase\expandafter{\romannumeral#1}}%
  }
\newcommand{\diag}{\mathop{\mathrm{diag}}}
\newcommand{\secref}[1]{Section \ref{#1}}

\title{Electric Field Decay Without Pair Production: Lattice, Bosonization and Novel Worldline Instantons}
\author{Xu-Yao Hu,}
\author{Matthew Kleban}
\author{and Cedric Yu}

\affiliation{Center for Cosmology and Particle Physics, New York University, \\
726 Broadway, New York, NY 10003, U.S.A}
\emailAdd{xuyao.hu@nyu.edu}
\emailAdd{kleban@nyu.edu}
\emailAdd{cedric.yu@nyu.edu}

\abstract{
Electric fields can spontaneously decay via the  Schwinger effect, the nucleation of a charged particle-anti particle pair separated by a critical distance $d$.  What happens if the available distance is smaller than $d$?
  Previous work on this question has produced   contradictory results.  Here, we study the quantum evolution of electric fields when the  field points in a compact direction with circumference $L < d$ using the massive Schwinger model, quantum electrodynamics in one space dimension with massive charged fermions.  We uncover a new and previously unknown set of instantons that result in novel physics that disagrees with all previous estimates.  In parameter regimes where the field value can be well-defined in the quantum theory, generic initial fields $E$ are in fact \emph{stable and do not decay}, while initial values that are quantized in half-integer units of the charge $E = (k/2) g$ with $k\in \mathbb Z$ \emph{oscillate in time} from $+(k/2) g$ to $-(k/2) g$, with exponentially small probability of ever taking any other value.  We verify our results with four distinct techniques: numerically by measuring the decay directly in Lorentzian time on the lattice, numerically using the spectrum of the Hamiltonian, numerically and semi-analytically using the bosonized description of the Schwinger model, and analytically via our  instanton estimate.
}

\makeatletter
\gdef\@fpheader{}
\makeatother

\begin{document}
\maketitle
\flushbottom

\section{Introduction}

\begin{figure}[H]
\centering
\begin{subfigure}{0.5\textwidth}
  \centering
  \hspace*{0.5cm}\includegraphics[width=.68\linewidth,keepaspectratio]{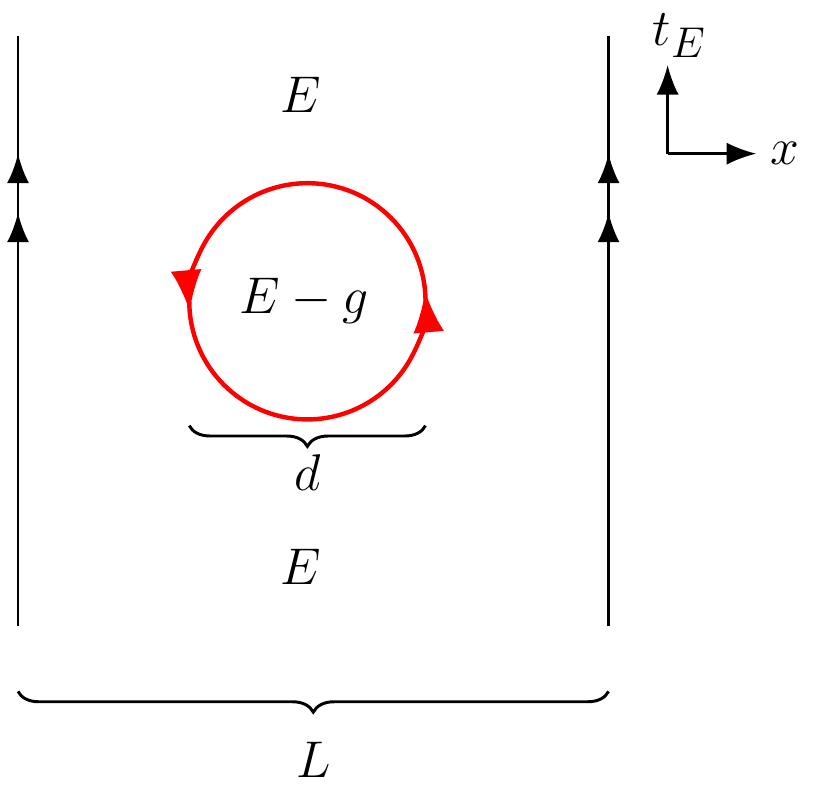}
\caption{Schwinger's worldline instanton for pair production in \cite{Schwinger1951,Affleck:1981bma}, on a non-compact or compact spatial dimension of length $L$ larger than the nucleation distance $d=2m/gE$, $L>d$.}\label{schwingerfig1}
\end{subfigure}%
\linebreak
\centering
\begin{subfigure}{0.5\textwidth}
  \centering
  \hspace*{0.5cm}\includegraphics[width=.68\linewidth,keepaspectratio]{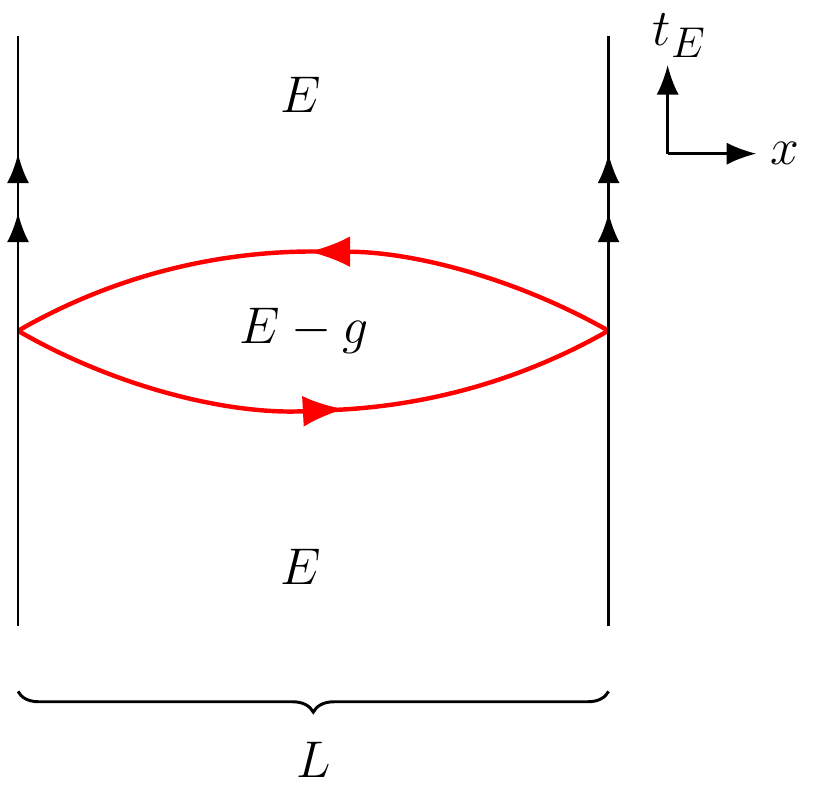}
\caption{WKB instanton on a compact spatial dimension in \cite{Brown2015}, which can be adapted to the finite temperature case.}\label{brownfig}
\end{subfigure}%
  \hspace*{0.5cm}
\begin{subfigure}{.5\textwidth}
  \centering
  \hspace*{0.5cm}\vspace*{0.4cm}\includegraphics[width=.68\linewidth,keepaspectratio]{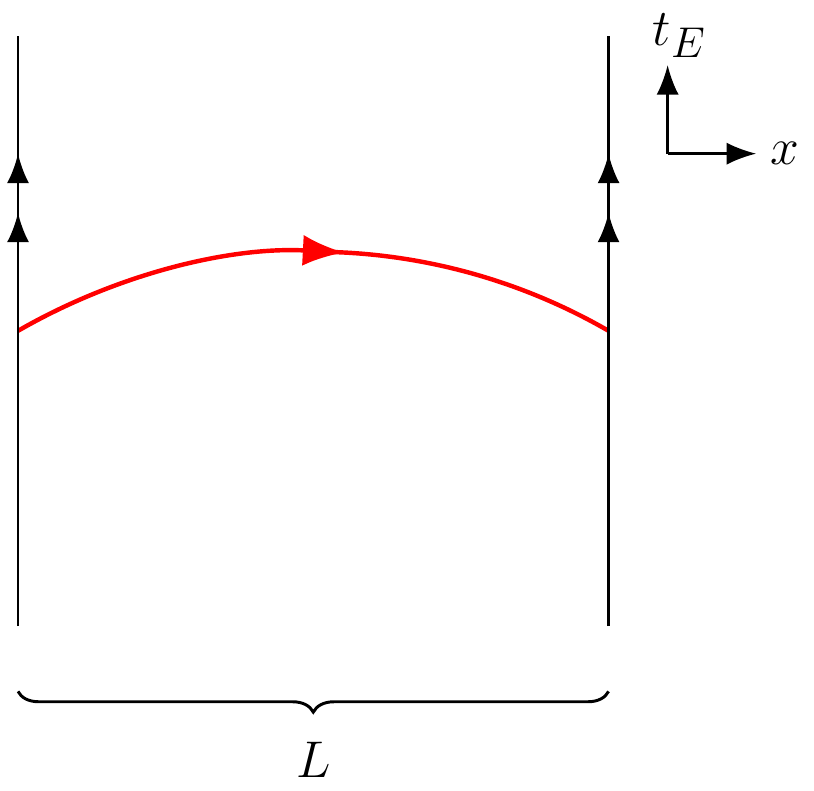}
  \caption{Worldline instanton on a compact spatial dimension in \cite{Draper2018}.}
  \label{draperfig}
\end{subfigure}
\linebreak
   \hspace*{-.5cm} \begin{subfigure}{.5\textwidth}
  \centering
  \includegraphics[width=.96\linewidth,keepaspectratio]{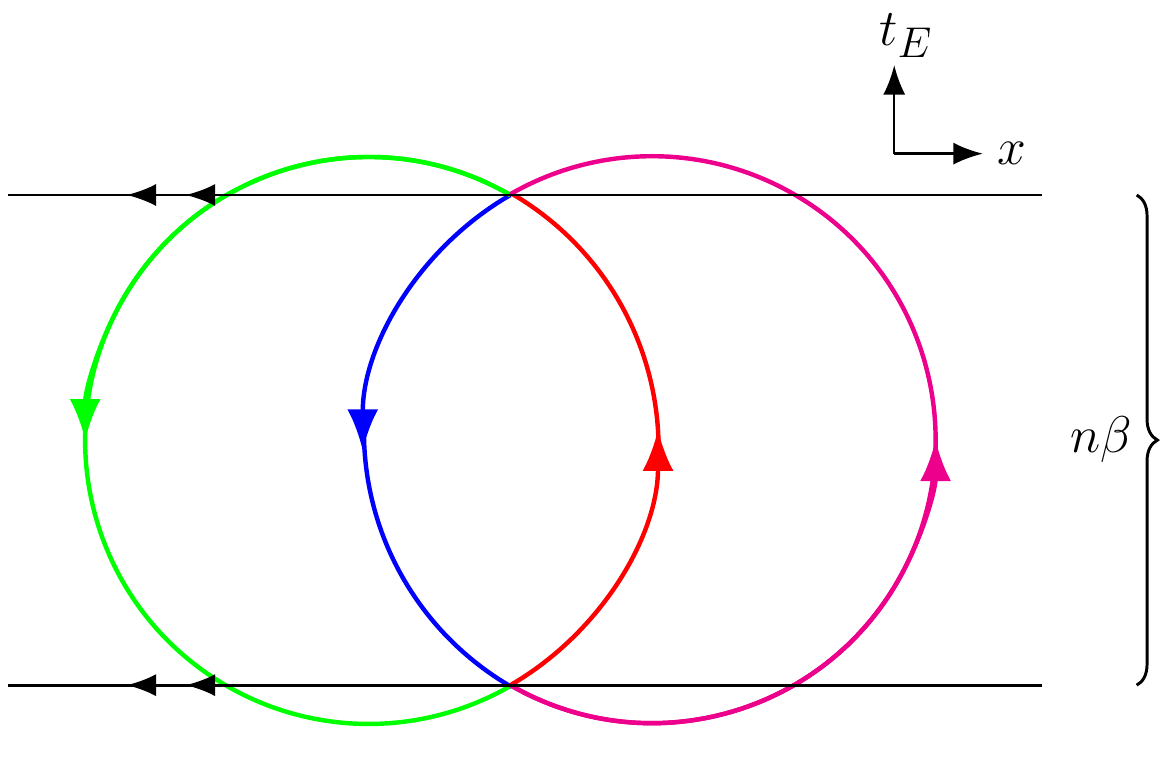}
 \caption{Worldline instantons at finite temperature $\beta^{-1}$ in \cite{MO2015}, $n\in\mathbb{Z}$.}
  \label{MOfig}
\end{subfigure}%
  \hspace*{0.5cm}
\begin{subfigure}{0.5\textwidth}
 \centering
  \hspace*{1.5cm}\includegraphics[width=.68\linewidth,keepaspectratio]{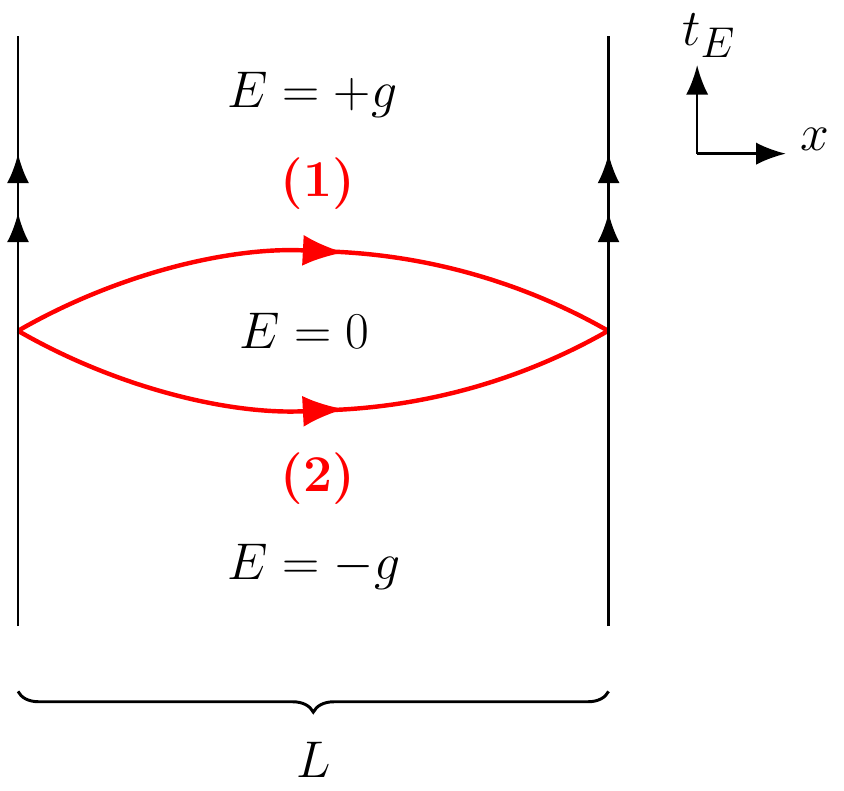}
\caption{One of our proposed worldline instantons.}\label{browntempfig}
\end{subfigure}%

\caption{Some of the instantons in higher dimensional space proposed in the literature \cite{Schwinger1951,Affleck:1981bma,Brown2015,Draper2018,MO2015}. \label{fig1}}
\label{previousworksfig}
\end{figure}

It is said that in our quantum world, the totalitarian principle applies:  everything not forbidden is compulsory.   In quantum electrodynamics, non-zero electric fields can decay via the quantum nucleation of charged particles.  This process has been very well-studied since the seminal work of Schwinger \cite{Schwinger1951}.  When the electric field is below the critical field $E_{c} \sim m^{2}/g$ (where $m, g$ are the mass and charge of the  particles) this process is non-perturbative and exponentially slow, with a rate that scales as $e^{- \pi m^2/g E}$.  The particle-anti-particle pairs responsible for this effect nucleate  a distance $d = 2 m/gE$ apart and are separated along the direction the field is pointing.  The instanton responsible for this process is simply an oriented circle with radius $d$ (Figure \ref{schwingerfig1}).

Many generalizations of \cite{Schwinger1951} have been considered over the years, incorporating inhomogeneities or finite extent in the electric field.
Without introducing inhomogeneities, another generalization is to consider the decay of an initial field when the direction along which it points is periodically identified.\footnote{Equilibrium at finite temperature corresponds to compactifying Euclidean time, and since electric fields always ``point'' in the time direction this description also applies finite temperature quantum electrodynamics.}  If the circumference $L$ of this compact dimension is larger than $d$ one expects the standard instanton in Figure  \ref{schwingerfig1} to govern the decay and the pair-production rate to be approximately unchanged.  However, when $L < d$ the instanton no longer ``fits'' in the compact space.

In the regime $L < d$ there is a stark disagreement in the recent literature over the decay rate and how to calculate it.  In \cite{Brown2015}, Brown argues that the  instanton responsible for the decay is similar to the standard circular instanton, but squeezed into a ``lemon'' so as to fit in the compact dimension (Figure \ref{brownfig}).  
 By contrast in \cite{MO2015}, Medina and Ogilvie argue for a  configuration that includes a different set of arcs of the circular instanton (Figure \ref{MOfig}).  The action for these configurations scales very differently from Brown's and is  discontinuous as a function of $L$.  The fact that these two estimates differ \emph{parametrically in the exponent} is remarkable for such a simple and classic problem.   
 
In this work we study this question in a simple case, namely the massive Schwinger model, quantum electrodynamics with massive charged fermions in one space and one time dimension.  Neither of the ``instantons'' proposed in \cite{Brown2015, MO2015} actually solve the equations of motion of the theory (due to the discontinuous first derivative at certain points in the trajectory) and so neither should necessarily be expected correctly describe the decay.  Here, we identify a novel set of instantons that do solve the equations of motion.  The resulting prediction for the rate differs (in the exponent) from both \cite{Brown2015, MO2015} (although it is closer to \cite{Brown2015}).  We check our prediction using multiple numerical techniques, including directly time evolving the initial state in a lattice version of the theory and identifying the physical properties of the eigenstates of the lattice Hamiltonian.

The massive Schwinger model has several features that distinguish it from QED in higher dimensions.  Most importantly, there is no magnetic field and the gauge field is non-propagating, so the dynamics of the theory is entirely due to the charged particles (that interact through the electric fields they produce).  Classically, this means that the electric field is almost entirely determined by the charge configuration (almost, because a constant background field is allowed).  In the quantum theory there is a constraint that, in a charge eigenstate, determines the field (see~\eqref{eq: E_field}).

In particular,  the possible field values are quantized in integer multiples of the charge $g$ plus the constant background.   One implication is that unlike for larger values, an initial field in the range $-g/2 < E < g/2$ cannot decay, because the energy density $E^{2}/2$ would be larger were the field to increase or decrease by $g$.  Indeed, one can think of a  field in this range as a parameter defining the specific theory under study.  This parameter is the famous $\theta$ angle, so-called because as the  field varies the spectrum of the Hamiltonian undergoes a spectral flow that is periodic under $E \rightarrow E + g$. 

A novel and surprising feature of our results is that when the theory is compactified on a small circle there is a very sharp distinction between generic values of the initial field and those equal to integer or half-integer multiples of the charge, $E = (k/2) g$ with $k \in \mathbb{Z}$.   In the regime of would-be slow decay and when $E$ is not equal to one of these quantized values, the field \emph{never} decays, in  contrast to the non-compact or large-field case.  This can be understood in our analysis from the fact that we do not find any instanton solution for such cases.  We perform multiple distinct numerical analyses that confirm that both the quantum expectation value of the field and its (small) variance remain very close to their initial values for arbitrarily long times.   By contrast, when $E = (k/2) g$ the field \emph{oscillates} coherently and sinusoidally in time between its initial value $(k/2)g$ and $-(k/2)g$, {with exponentially suppressed probability of taking any other value no matter how long one waits.}  

A  feature of the massive Schwinger model is the existence of a bosonized version of the  theory, the massive Sine-Gordon model.  This constitutes a type of strong-weak duality, because the fermionic field theory is weakly coupled when $g/m \ll 1$, while the bosonized description approaches a free field theory in the opposite limit $g/m \to \infty$.  The bosonized description gives an alternative view of electric field decay that can help account for the surprising features mentioned above.  Meta-stable values of the field correspond to states in which the boson is localized in one of the local minima of the cosine potential.  In the non-compact theory the field can decay via 1+1 dimensional bubble nucleation, where the walls of the bubble are a kink and an anti-kink (corresponding to a fermion and an anti-fermion).  However, if the theory is compactified on a sufficiently small circle there is not enough room for a critical bubble to form.  When the Kaluza-Klein modes can be ignored, the theory reduces to the quantum mechanics of the bosonic zero mode.  The distinction between half integer and non-half integer $E$ can then be understood as follows.  For half-integer $E$ the quantum mechanical potential has a $\mathbb{Z}_{2}$ reflection symmetry, so that for each local minimum at $E = (k/2) g$ there is a symmetric minimum at $E = -(k/2) g$.  In the semi-classical regime the energy eigenstates are  symmetric and anti-symmetric states localized in these two wells.  Hence (much like the classic symmetric double-well problem in quantum mechanics) the field oscillates sinusoidally with an exponentially long period between the values $E = \pm (k/2) g$.  By contrast, for non-half integer values of the field the potential is not symmetric and the energy eigenstates are localized in a single well, so the field remains constant and localized for all time.

In summary, we arrive at our conclusions with the following set of distinct and complementary techniques:
\begin{itemize}

\item We directly measure the time evolution of the electric field using a real-time lattice  code for the fermionic massive Schwinger model

\item We compute the Hamiltonian eigenvalues from the lattice code, identify the states relevant to the electric field evolution, and estimate the rate of oscillation (when it exists) via energy differences

\item We identify the bosonized counterpart of the massive Schwinger model and compute the time-dependence of the electric field numerically and semi-analytically in the bosonic theory (using the approximation where the circle is small enough to ignore Kaluza-Klein modes).

\item We estimate the rate using a novel set of field-theory instantons.\footnote{In the case $k=1$ the estimate based on our novel ``straight line'' instanton agrees very well with our numerical results.  For $k>1$ we found an infinite set of chains of instantons that contribute to a sum that does not converge.  The exponential dependence is consistent with our numerical results, and we believe this failure to converge is due to a  missing pre-factor we have  so far been unable to identify.}

\end{itemize}

\paragraph{Relation to previous work:}

As already mentioned above, Brown \cite{Brown2015} considered QED with a compact spatial or Euclidean time dimension and argued that the instanton illustrated in Figure \ref{brownfig} computes the rate of decay of the electric field.  In contrast, Medina and Ogilvie \cite{MO2015} proposed a different set of instantons with a different action and hence an exponentially different rate  (Figure \ref{MOfig}).  Korwar and Thalapillil \cite{Korwar_2018} considered the case of non-zero $E$ and $B$ fields; their results reduce to those of  \cite{MO2015}  when $B=0$.  Draper \cite{Draper2018} also computed the rate, relying on the instanton illustrated in Figure \ref{draperfig}.  None of these analyses coincide with our main results, although their focus is on higher-dimensional versions of QED.  Qiu and Sorbo in  \cite{QS2020} considered scalar QED in 1+1 dimensions in the linearized approximation around a background field.  Their results also do not coincide with ours, although this may be due to the fact that they considered scalar QED and focused on the short-time behavior, whereas we consider the fermionic model and are mostly concerned with the evolution at long times.  Electric field decay in the Schwinger model was studied numerically in e.g. \cite{Hebenstreit:2013baa, Buyens:2016hhu}.  Finally, \cite{Nagele:2018egu} investigated the phenomenon of ``flux unwinding'' in the compact Schwinger model where a pair of charges can discharge many units of flux by repeatedly traversing the circle.  Our present analysis differs from that of \cite{Nagele:2018egu}  because we focus on the regime where the circle is too small to accommodate pair production.

The structure of this work is as follows. In \secref{sec2}, we describe the massive Schwinger model in the continuum and on the lattice.  In \secref{sec3}, we outline the numerical techniques used on the lattice and present our  results. In \secref{sec4}, we study the bosonized description of the theory in the small circle regime and numerically compute the transition rates. In \secref{sec5} we use the worldline formalism of path integrals to identify a novel set of worldline instantons and compute the associated rates. 	
We conclude in \secref{sec6}. 	
In Appendix \ref{sec: lattice}, we review the lattice formulation of the massive Schwinger model. 
In Appendix \ref{sec: lmax and extrapolation}, we report technical details about the Hilbert space cutoff on lattice and the continuum extrapolation protocol for the energy spectrum in our numerics.
In Appendix \ref{sec: prefactor}, we discuss the continuum extrapolation of the prefactor $c$ appearing in the bosonized action in \secref{sec4}. Appendix \ref{appworldlinealt} contains  calculations that complement \secref{sec5}.

\section{Massive Schwinger model on a compact spatial dimension}\label{sec2}
The massive Schwinger model \cite{Schwinger1951,Coleman1976} is quantum electrodynamics with a massive Dirac fermion in one space and one time dimension. The Lagrangian is
\begin{align}
	\Lagr=
		-\frac{1}{4}F_{\mu\nu} F^{\mu\nu} +\Dadj{\psi}\left(i\gamma^{\mu} D_{\mu}-m\right)\psi \ ,
	\label{eq: lagrangian_Schwinger}
\end{align}
where $\psi$ is a two-component Dirac fermion, $m$ is the fermion mass, $D_{\mu}\equiv \p_{\mu}+i g A_{\mu}$ is the covariant derivative with
$g$ being the electric charge and $A_{\mu}$ being a $U(1)$ gauge field, and
$F_{\mu\nu}= \p_{\mu} A_{\nu}-\p_{\nu} A_{\mu}$ is the field strength.
Throughout this work, we work in natural units $\hbar = c =1$ with Minkowski metric $\eta_{\mu\nu} = \diag (+1,-1)$ where the indices $\mu$, $\nu$ runs from 0 to 1.
In the temporal gauge $A_0 = 0$ such that the electric field $E=F^{10} = -\dot A^1(t)$, the Hamiltonian reads
\begin{align}
	H=\int \dd x \left[- \Dadj{\psi} i\gamma^1\left(\p_1 + i g A_1\right)\psi + m \Dadj{\psi} \psi + \frac{E^2}{2}\right]\ .
	\label{eq: hamiltonian_Schwinger}
\end{align}

We are interested in this model on a compact spatial dimension of size $L$, i.e.~with identification $x \sim x+ L$. We impose periodic boundary condition on both the electric field and the fermion, $E(x+L)=E(x)$ and $\psi(x+L)=+\psi(x)$.  

Gauss' law $\p_1 E = g j^0$ implies that the
electric field is given by 
\begin{align}
	E(x)=F + \int^x dx'\ g j^0(x') \ ,
	\label{eq: E_field}
\end{align}
where $j^0 = \Dadj{\psi} \gamma^0 \psi$ and $F$ is a constant background field which is physically significant in $1+1$ dimensions \cite{Coleman1976}. Physical states satisfying Gauss' law have zero net charge on the circle.  When the space is non-compact, the constant $F$ is naturally fixed by the electric field value at infinity. On a circle of circumference $L$, we pick an arbitrary point to be the origin, and take it as the lower limit of the integral in \eqref{eq: E_field}; then $F=E(x=0)=E(x=L)$.

\subsection{Massive Schwinger model on a lattice}
 Unlike its massless counterpart, the massive Schwinger model is not exactly solvable (although as we will see, its bosonized form is weakly coupled when the dimensionless coupling $g/m$ is large).  For this reason it is useful to simulate the model on a lattice and compare numerical results against analytical predictions. In this work, we perform numerical studies on a lattice using the Kogut-Susskind formulation of lattice gauge theory \cite{KS1975,CKSS1975,Susskind1976}, which we review in Appendix \ref{sec: lattice}. Here we summarize the key elements of the construction. We put the model defined by the Hamiltonian (\ref{eq: hamiltonian_Schwinger}) on a spatial staggered lattice, on which electrons and positrons respectively occupy odd and even sites, while the gauge field is implemented by Wilson lines (also called \textit{link fields}) between adjacent sites. Further performing a Jordan-Wigner transformation which maps the fermions to spins \cite{BSK1976}, the lattice Hamiltonian takes the form 
\begin{align}
	H_{\text{lat}}=&\frac{1}{2a} \sum_n \left[\sigma^+(n) e^{i\theta(n)} \sigma^-(n+1) 
	+ \sigma^+(n+1) e^{-i \theta(n)} \sigma^-(n)\right] \notag \\
	&+\frac{m}{2}\sum_n (-1)^n  \left[1+\sigma_3(n)\right]
	+\frac{1}{2} a g^2 \sum_n \left[L(n)+ \alpha\right]^2 \ ,
	\label{eq: lattice Schwinger Hamiltonian}
\end{align}
where $a$ is the spatial lattice spacing, $n$ is the lattice site index, $\sigma_{1,2,3}$ are the Pauli matrices with $\sigma^{\pm}(n)\equiv \frac{1}{2}\left(\sigma_1(n)\pm i \sigma_2(n)\right)$, and $\alpha = F/g$ is the dimensionless background field. $L(n)\equiv (E(n)-F)/g$ is the dimensionless lattice electric field (with the background field subtracted) operator defined on the $n$-th link between $n$-th and $(n+1)$-th sites, and the link phase on the $n$-th link $\theta(n)$ is related to the gauge field by $\theta(n) \equiv a g A_1(n)$.\footnote{The link phase $\theta(n)$ is not to be confused with the background angle $\theta =2\pi F/g$, or the operator $L(n)$ and the circle size $L$. These distinctions should be clear from the context.} The total number of lattice sites $N$ has to be even in this formulation.

We introduce dimensionless lattice observables by defining the dimensionless parameters 
\begin{align}
x&\equiv \frac{1}{a^2g^2},\;\mu\equiv \frac{2m}{ag^2},
\end{align}
in terms of which the dimensionless Hamiltonian is $a H_{\text{lat}} \equiv W/2x$, where $W$ given by
\begin{align}
	W&\equiv \frac{2}{a g^2} H_{\text{lat}} = x \sum_n \left[ \sigma^+(n) e^{i\theta(n)} \sigma^-(n+1) 
	+  \sigma^+(n+1) e^{-i \theta(n)} \sigma^-(n)\right] \notag \\
	&+\frac{\mu}{2} \sum_n (-1)^n \left[1+\sigma_3(n)\right]
	+ \sum_n \left[L(n)+ \alpha\right]^2 .
	\label{eq: dimensionless hamiltonian}
\end{align}
The dimensionless lattice electric field $L(n)$ obeys the lattice Gauss law
\begin{align}
	L(n)-L(n-1)=\frac{1}{2} \left(\sigma_3(n)+(-1)^n\right) .
	\label{eq: lattice Gauss' law}
\end{align}
As in the continuum case, we impose periodic boundary conditions (PBC) on the lattice, so the integer $n$ in (\ref{eq: lattice Schwinger Hamiltonian}) and (\ref{eq: dimensionless hamiltonian}) runs from $0$ to $N-1$.

\paragraph{Hilbert space truncation and a basis\\}
Given the lattice Hamiltonian \eqref{eq: dimensionless hamiltonian}, it is  convenient to work with the spin and electric field eigenstates.  These take the form
\begin{align}
	\ket{\{l\},\{\sigma_3\}}&\equiv 
	\ket{\{l(0),\cdots, l(N-1)\}, \{\sigma_3(0),\cdots, \sigma_3(N-1)\}} ,
	\label{eq: eigen basis}
\end{align} 
where $l(n)$ is the eigenvalue of $L(n)$ on the $n$-th link, and $\sigma_3(n)$ is the spin eigenvalue on the $n$-th site. The lattice Gauss law (\ref{eq: lattice Gauss' law}) enforces zero net charge on the lattice, which for PBC translates to $\sum_n \sigma_3(n) = 0$, and moreover leaves only one quantum electric field degree of freedom $l \equiv l(N-1) \equiv l(-1) $ unfixed. Then, a physical eigenstate \eqref{eq: eigen basis} satisfying the lattice Gauss law (\ref{eq: lattice Gauss' law}) is uniquely labeled by 
\begin{align}
	\ket{\{l\},\{\sigma_3\}} \equiv \ket{l,\{\sigma_3(0),\cdots, \sigma_3(n),\cdots, \sigma_3(N-1)\}_{\sum_n \sigma_3(n) = 0}} \ .
	\label{eq: eigen basis_2}
\end{align}
Note that $l$ can be any integer, after we canonically quantize the gauge field sector--- this is explained in Appendix \ref{sec: lattice} after (\ref{eq: L_ladder}). As such, the lattice Hilbert space is actually infinite dimensional, even with a finite $N$.  We perform a truncation on the electric field, such that $|l(n)|\leq l_{\text{max}}$ on all links $n$. This results in a finite dimensional lattice Hilbert space $\Hilb_{\text{lat}}$. 
For the low energy physics of interest in this work, the error of lattice simulation is negligible as long as the cutoff $l_{\text{max}}$ is sufficiently large.
In practice, how large the cutoff $l_{\text{max}}$ needs to be depends on the coupling $m/g$ and the spatial size $mL$. We summarize in Appendix \ref{sec: lmax and extrapolation} the explicit values of $l_{\text{max}}$ used for various parameter choices in our lattice simulation.

The finite dimensional lattice Hilbert space $\Hilb_{\text{lat}}$ is spanned by the states in \eqref{eq: eigen basis_2}, $\ket{\{l\},\{\sigma_3\}}_j$, $j=1,\cdots,\dim \Hilb_{\text{lat}}$, with the truncation on $l$ so that $|l(n)|\leq l_{\text{max}}$ for all links $n$. For the $j$-th basis state $\ket{\{l\},\{\sigma_3\}}_j$, we denote the eigenvalues of the  electric field on the $n$-th link and the spin on the $n$-th site respectively by $l(n)_j$ and $\sigma_3(n)_j$. Then, we have matrix elements $\tensor[_i]{\mel{\{l\},\{\sigma_3\}}{L(n)}{\{l\},\{\sigma_3\}}}{_j}=l(n)_j \delta_{ji}$ and $\tensor[_i]{\mel{\{l\},\{\sigma_3\}}{\sigma_3(n)}{\{l\},\{\sigma_3\}}}{_j}=\sigma_3(n)_j \delta_{ji}$.

\paragraph{Observables and probability\\}
Given a state $\ket{\psi}$ expanded in the lattice basis,
\begin{align}
	\ket{\psi}=\sum_{j=1}^{\dim \Hilb_{\text{lat}}}
	v_j \ket{\{l\},\{\sigma_3\}}_j \ ,
	\label{eq: generic state_lat}
\end{align}
we compute the main lattice observables in this work.    
\begin{itemize}
	\item The expectation value of the spatially averaged electric field $\bar E$ is
		\begin{align}
			\frac{\ev{\bar E}}{g} = \frac{1}{N}\sum_{n=0}^{N-1} \mel{\psi}{L(n)}{\psi}+\frac{F}{g}
			= \frac{1}{N}\sum_{n=0}^{N-1}\sum_{j=1}^{\dim \Hilb_{\text{lat}}} \left|v_j\right|^2 l(n)_j + \alpha \ .
		\end{align}
	\item The standard deviation of the electric field is 
	\begin{align}
		\sigma_{E/g}\equiv \sqrt{\ev{\left(\frac{\bar E-\ev{\bar E}}{g}\right)^2}}
		=\sqrt{\sum_{j=1}^{\dim \Hilb_{\text{lat}}}|v_j|^2\left(\frac{1}{N}\sum_{n=0}^{N-1}l(n)_j+\alpha-\frac{\ev{\bar E}}{g}\right)^2} \ .
	\end{align}
	\item The probability of measuring the electric field value $E_m/g$ on the $n$-th link can be computed by first constructing the projection operator on the lattice
		\begin{align}
			\hat P(n,E_m)=\sideset{}{'}\sum_{j} \ket{\{l\},\{\sigma_3\}}_j \tensor[_j]{\bra{\{l\},\{\sigma_3\}}}{}
		\end{align}
		where the sum $\sideset{}{'}\sum_{j}$ is over all the basis states with $l(n)_j=\frac{E_m}{g}-\alpha$. The probability is then given by the expectation value of the projection,
		\begin{align}
			P\left(E(n) = E_m\right) = \mel{\psi}{\hat P(n,E_m)}{\psi} 
			= \sideset{}{'}\sum_j \sum_{i=1}^{\dim \Hilb_{\text{lat}}} |v_i|^2 \delta_{ji} 
			= \sideset{}{'}\sum_j |v_j|^2
			\ .
		\end{align}
The quantum state on all  odd (even) links is identical due to the periodic boundary conditions.  There is a slight difference between odd and even links due to the staggered nature of the lattice, but this brings no important modifications to our analysis.  Hence, without loss of generality we will focus  on the probability of measuring the electric field on the central link (i.e. $(\frac{N}{2}-1)$-th link) throughout this work.
	\item To calculate the number of pairs in a state, we recall that in the staggered fermion formalism, $(-1)^n \sigma_3(n) = +1$ when an electron (positron) appears at the $n$-th site where $n$ is odd (even), and $(-1)^n \sigma_3(n) = -1$ when the $n$-th site is empty.
	The basis states $\ket{\{l\},\{\sigma_3\}}_j$ are eigenstates of electron (positron) number operator $\hat N_e$ ($\hat N_p$):
	\begin{align}
		\hat N_{e}\ket{\{l\},\{\sigma_3\}}_j &= \sum_{n \text{ odd}} \delta_{(-1)^n\sigma_3(n)_j,1} \ket{\{l\},\{\sigma_3\}}_j \equiv N_{e,j} \ket{\{l\},\{\sigma_3\}}_j \ ,\\
		\hat N_{p}\ket{\{l\},\{\sigma_3\}}_j &= \sum_{n \text{ even}} \delta_{(-1)^n\sigma_3(n)_j,1} \ket{\{l\},\{\sigma_3\}}_j \equiv N_{p,j} \ket{\{l\},\{\sigma_3\}}_j \ .
	\end{align}
	The number of pairs operator is defined as $\hat N_{\text{pairs}} = \frac{1}{2}(\hat N_e + \hat N_p)$ with eigenvalues $N_{\text{pairs},j} = (N_{e,j}+N_{p,j})/2$. In the state $\ket{\psi}$ as given by the expansion (\ref{eq: generic state_lat}), the number of pairs is given by
	\begin{align}
		\mel{\psi}{\hat N_{\text{pairs}}}{\psi} 
		= \sum_{j=1}^{\dim \Hilb_{\text{lat}}} |v_j|^2 N_{\text{pairs},j} \ . 
	\end{align}
\end{itemize}

\paragraph{Exact Diagonalization and time evolution\\}
We study both static properties and dynamics of the compact massive Schwinger model on the lattice, using the simple technique of exact diagonalization:  after constructing the Hamiltonian matrix, we numerically diagonalize it to determine the energy levels and eigenstates.  Time evolution is done by decomposing the initial state into Hamiltonian eigenstates and  multiplying them by the phase factors $\exp{-i \EE_{i} t}$ with $\EE_i$ being the $i$-th Hamiltonian eigenvalue.

\paragraph{Bounds on lattice parameters\\}
In order to simulate the continuum theory on a lattice with  spacing $a$, we need to characterize the range of  lattice parameters that ensure the lattice results are physical in the continuum limit.
The two bounds we impose on the lattice simulation are as follows.
\begin{itemize}
 	\item Lattice artifacts are suppressed if:
 	\begin{align}
		a<\frac{1}{m} &=\frac{a\sqrt{x}}{m/g}\quad \implies \quad \sqrt{x} >\frac{m}{g} \ , \\
		a<\frac{1}{g} &=a\sqrt{x} \quad \implies \quad \sqrt{x}>1 \ .
		\label{eq: lattice bound_1}
	\end{align}
	\item Semiclassical instanton calculations that we will introduce in the later sections are valid if:
	\begin{align}
		mL=\frac{m}{g} \frac{N}{\sqrt{x}} \gtrsim 1 \quad \implies \quad \sqrt{x} \lesssim N \cdot \frac{m}{g}.
	\end{align}
 \end{itemize} 
Combining these constraints, valid lattice simulation results are obtained within the range for $x$:
\begin{align}
	\max\left\{\frac{m}{g},1\right\}<\sqrt{x} \lesssim N\cdot \frac{m}{g} \ .
	\label{eq: sensible x range}
\end{align}

In addition, an extrapolation procedure is required for the lattice results to approach the continuum limit.
Unlike most previous work \cite{SBH1999,BSBH2002} which explored the properties of the noncompact Schwinger model, we are interested in the finite size effects brought by the compact spatial dimension. The size of the spatial circle $L=Na$ is now a relevant quantity. Instead of using the two-step extrapolation, $N\to \infty$ followed by $x\to \infty$ as proposed in \cite{SBH1999,BSBH2002}, we keep
$\frac{N}{\sqrt{x}}$ fixed and take $N\to \infty$ to approach the continuum limit.
This extrapolation approach keeps fixed the size of the space with respect to the coupling $gL$, while taking the spacing $ga \to 0$. The continuum limit is thus approached with the finite size effects intact.
For more details about the continuum extrapolation, see Appendix \ref{sec: lmax and extrapolation}
and Appendix \ref{sec: prefactor}.

\section{Lattice dynamics and decay of an initial state electric field}\label{sec3}
In this section, we proceed to investigate the dynamics of the massive Schwinger model on a circle making use of the lattice simulation code. We focus on the time evolution of the electric field and measure the timescales for certain types of quantum quench. In addition, the energy splittings relevant to those typical quenches are identified. This provides a computationally cheaper way to study the dynamics of the electric field in the weak field regime.

\subsection{Initial state and time evolution}
To study the decay or time evolution of an initial state electric field we need to specify an initial quantum state that we will then evolve in time using the lattice Hamiltonian.  Throughout this paper we will use the ground state of the $\alpha = 0$ Hamiltonian as the initial state.  Being the ground state, this should be a ``minimum uncertainty state" in the sense that it minimizes some product of the variance in the field and its time derivative.  

Due to symmetry this $\alpha=0$ ground state has $\frac{\ev{\bar E}}{g} = 0$.  We then evolve this initial state using the (numerically diagonalized) Hamiltonian with $\alpha=k/2$ for some $k \in \mathbb R$ (this is a quantum quench).  The  spectrum of the Hamiltonian is periodic under $\alpha \to \alpha + 1$ via spectral flow, but the initial state defined this way has $\frac{\ev{\bar E}}{g}  = 0 + \alpha$ with respect to this Hamiltonian.  Hence the time evolution from this state reflects the behavior of an initial state electric field satisfying $\ev{ \bar E} = \alpha g$.\footnote{As we will see when we examine the bosonized version of the theory, this procedure is similar to translating an approximately Gaussian initial state so that it becomes a coherent state with non-zero initial position expectation value.}

The observables calculated on the lattice include the expectation value and the standard deviation of the spatially averaged electric field, the probability to measure a dimensionless electric field value on the central link (C-L) (i.e. $(\frac{N}{2}-1)$-th link) and the number of electron-positron pairs.

\begin{figure}[ht]
	\centering
		\includegraphics[width=0.33\textwidth]{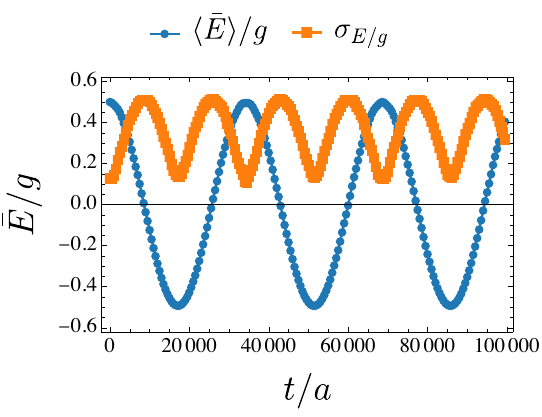}%
		\includegraphics[width=0.33\textwidth]{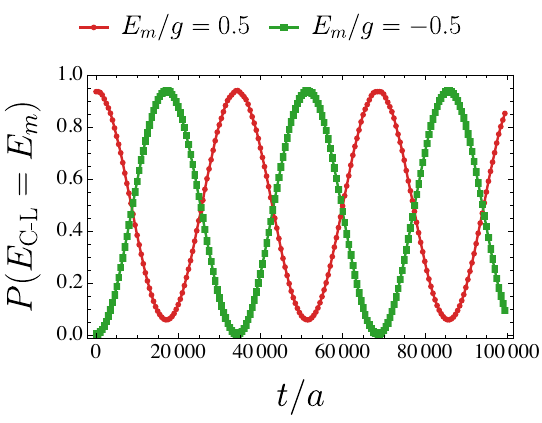}%
		\includegraphics[width=0.33\textwidth]{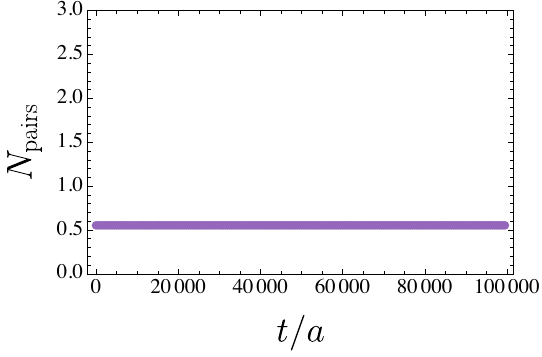}
		\includegraphics[width=0.33\textwidth]{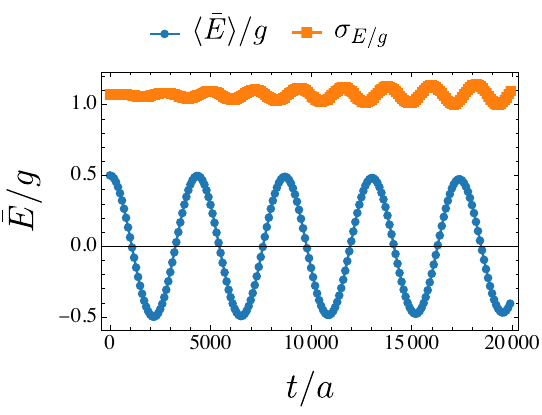}%
		\includegraphics[width=0.33\textwidth]{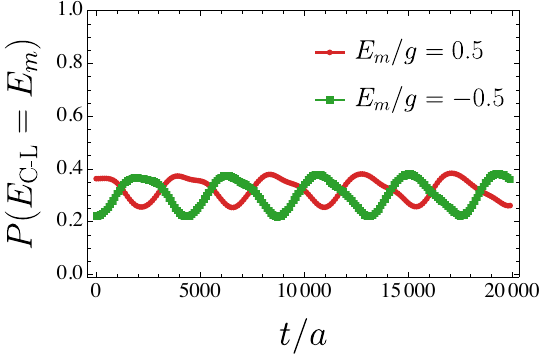}%
		\includegraphics[width=0.33\textwidth]{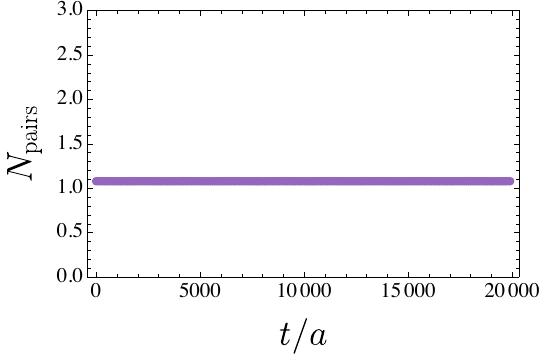}
		\includegraphics[width=0.33\textwidth]{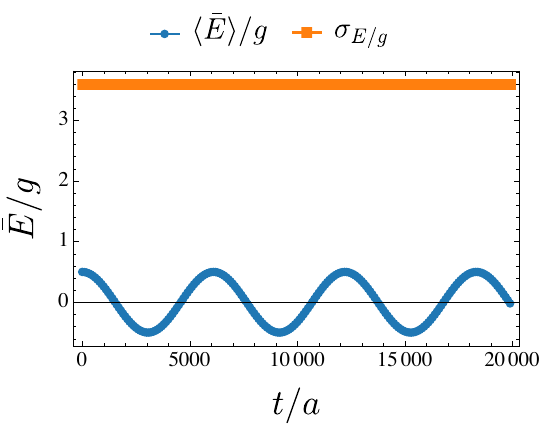}%
		\includegraphics[width=0.33\textwidth]{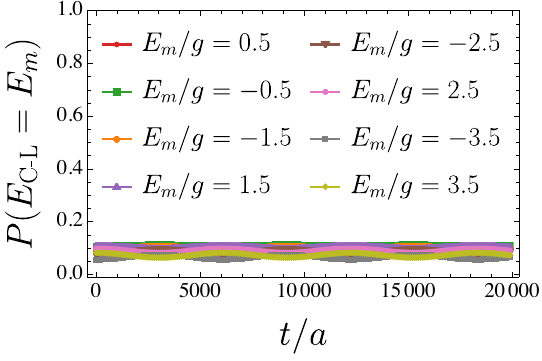}%
		\includegraphics[width=0.33\textwidth]{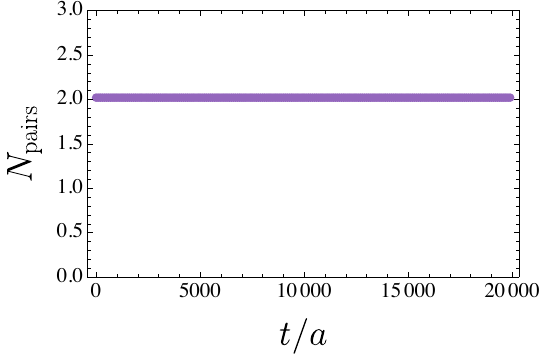}
	\caption{The expectation values of the spatially averaged electric field $\ev{\bar E}/g$, the standard deviation of the spatially averaged electric field $\sigma_{E/g}$, the probability to measure field $E_m$ on the central link $P\left(E_{\text{C-L}} = E_m\right)$, and the number of electron-positron pairs $N_{\text{pairs}}$ as functions of time in the $k=1$ quench for $\frac{m}{g}=50$, with three different spatial circle sizes $mL=8$ (top panel), $mL=4$ (middle panel) and $mL=1$ (bottom panel) on an $N=10$ lattice.
	On the lattice, we set $l_{\text{max}}=10$ for $mL=8,4$ simulations and $l_{\text{max}}=30$ for $mL=1$ simulation to achieve error $\lesssim \Op(10^{-10})$ for the lowest four energy levels $\EE/g$.
	}
\label{fig: nf=1 quench_1}
\end{figure}

\begin{figure}[ht]
	\centering
		\includegraphics[width=0.33\textwidth]{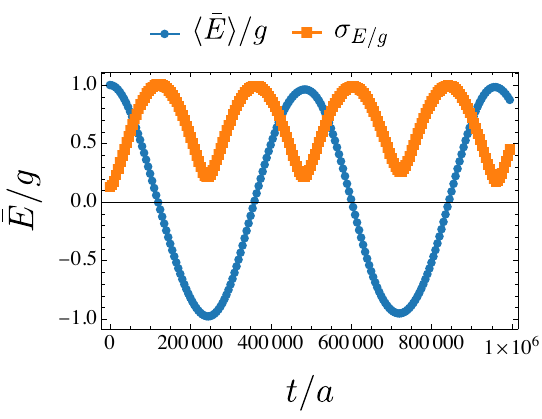}%
		\includegraphics[width=0.33\textwidth]{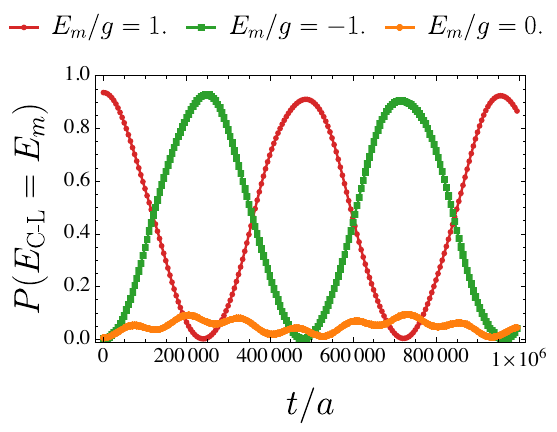}%
		\includegraphics[width=0.33\textwidth]{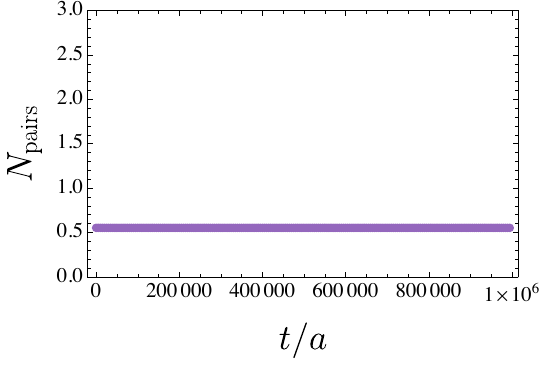}
		\includegraphics[width=0.33\textwidth]{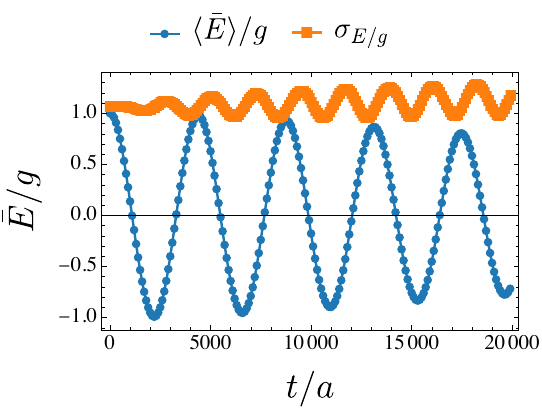}%
		\includegraphics[width=0.33\textwidth]{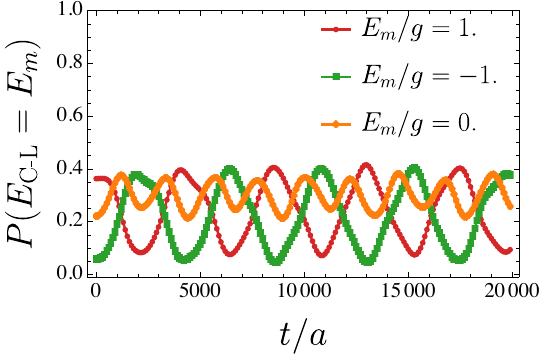}%
		\includegraphics[width=0.33\textwidth]{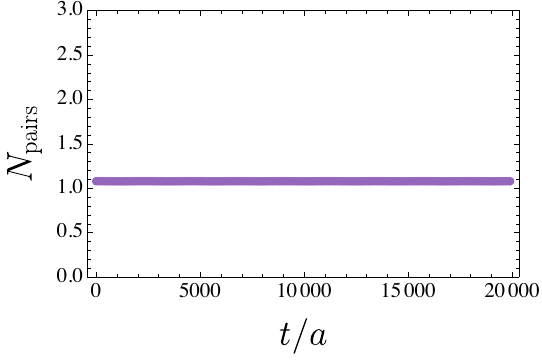}
		\includegraphics[width=0.33\textwidth]{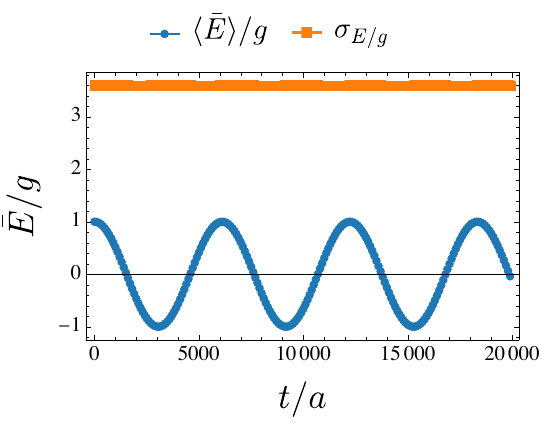}%
		\includegraphics[width=0.33\textwidth]{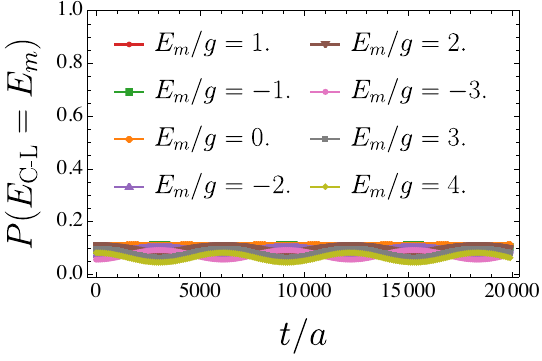}%
		\includegraphics[width=0.33\textwidth]{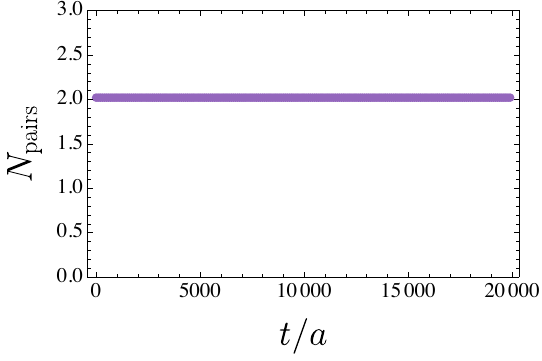}
	\caption{The expectation values of the spatially averaged electric field $\ev{\bar E}/g$, the standard deviation of the spatially averaged electric field $\sigma_{E/g}$, the probability to measure field $E_m$ on the central link $P\left(E_{\text{C-L}} = E_m\right)$, and the number of electron-positron pairs $N_{\text{pairs}}$ as functions of time in the $k=2$ quench for $\frac{m}{g}=50$, with three different spatial circle sizes $mL=8$ (top panel), $mL=4$ (middle panel) and $mL=1$ (bottom panel) on an $N=10$ lattice.
	For the same reason as in Figure \ref{fig: nf=1 quench_1}, we set $l_{\text{max}}=10$ for $mL=8,4$ simulations and $l_{\text{max}}=30$ for $mL=1$ simulation.}
\label{fig: nf=2 quench_1}
\end{figure}

\begin{figure}[ht]
	\centering
		\includegraphics[width=0.33\textwidth]{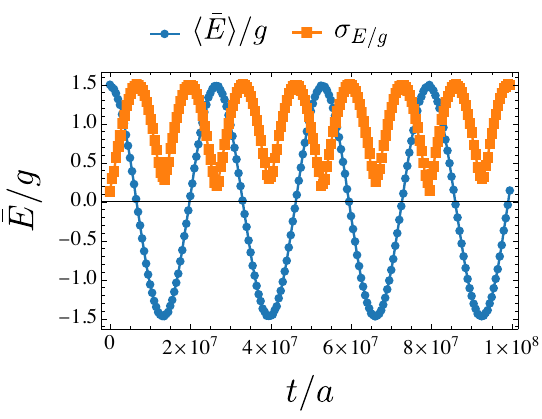}%
		\includegraphics[width=0.33\textwidth]{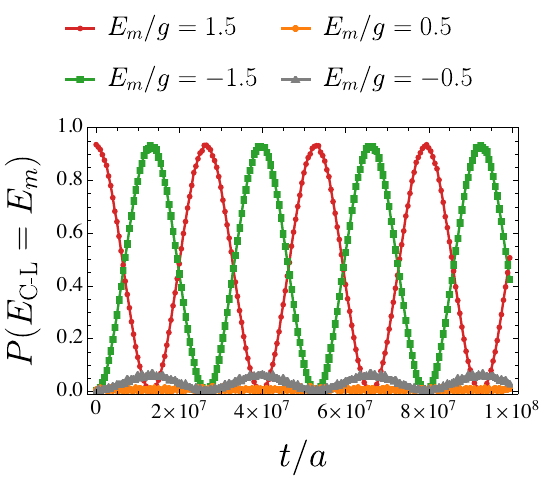}%
		\includegraphics[width=0.33\textwidth]{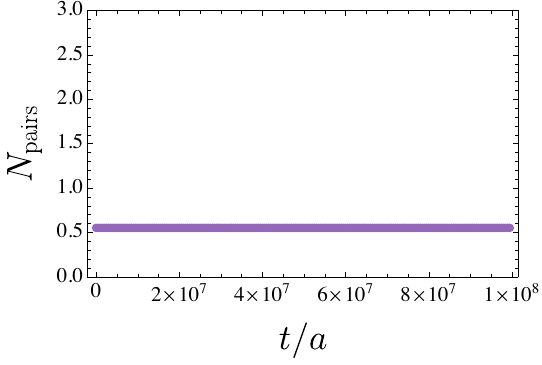}
		\includegraphics[width=0.33\textwidth]{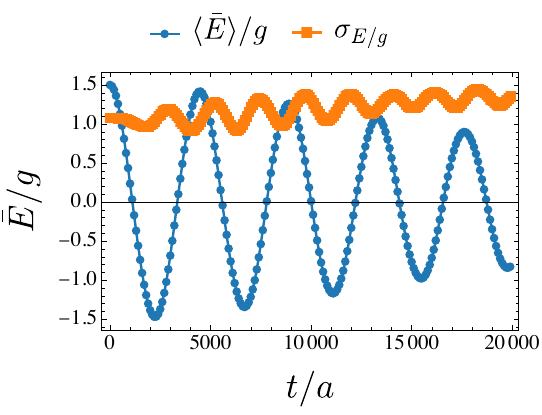}%
		\includegraphics[width=0.33\textwidth]{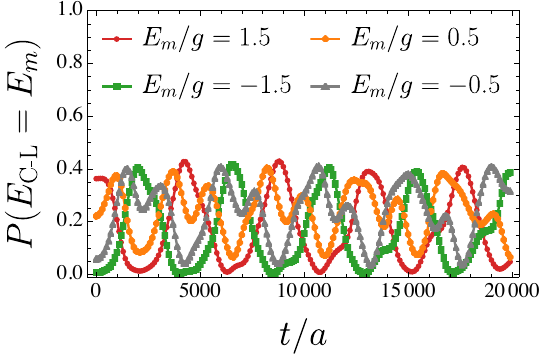}%
		\includegraphics[width=0.33\textwidth]{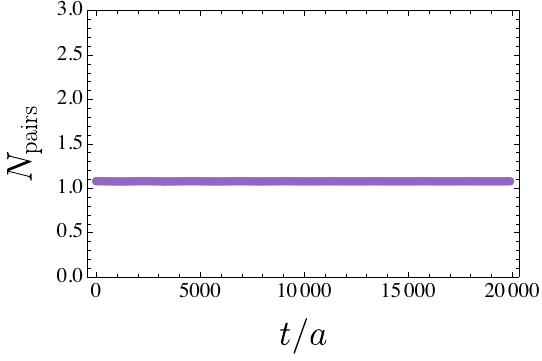}
		\includegraphics[width=0.33\textwidth]{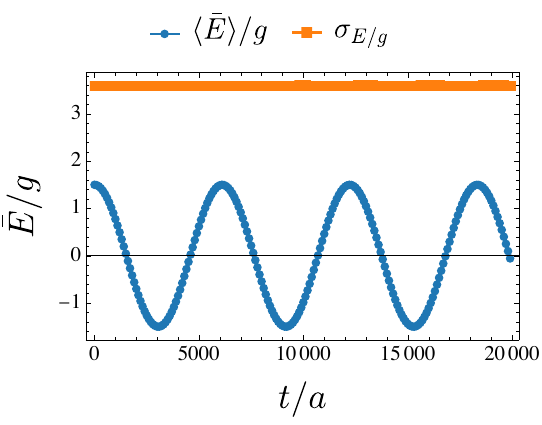}%
		\includegraphics[width=0.33\textwidth]{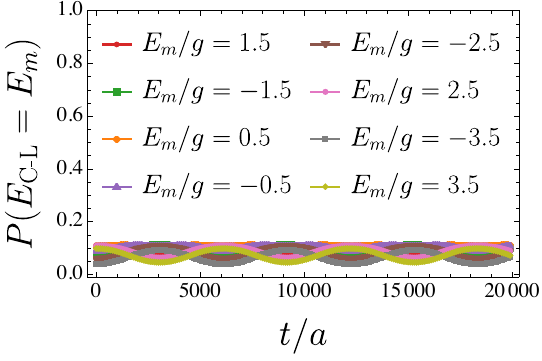}%
		\includegraphics[width=0.33\textwidth]{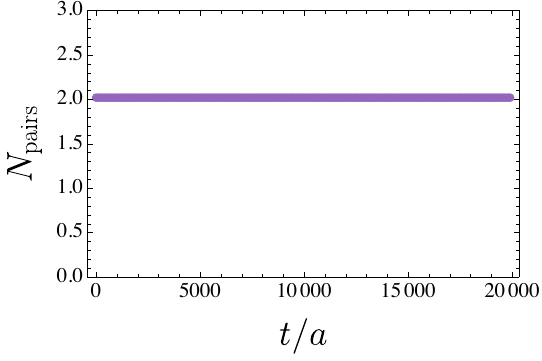}
	\caption{The expectation values of the spatially averaged electric field $\ev{\bar E}/g$, the standard deviation of the spatially averaged electric field $\sigma_{E/g}$, the probability to measure field $E_m$ on the central link $P\left(E_{\text{C-L}} = E_m\right)$, and the number of electron-positron pairs $N_{\text{pairs}}$ as functions of time in the $k=3$ quench for $\frac{m}{g}=50$, with three different spatial circle sizes $mL=8$ (top panel), $mL=4$ (middle panel) and $mL=1$ (bottom panel) on an $N=10$ lattice.
	For the same reason as in Figure \ref{fig: nf=1 quench_1}, we set $l_{\text{max}}=10$ for $mL=8,4$ simulations and $l_{\text{max}}=30$ for $mL=1$ simulation.}
\label{fig: nf=3 quench_1}
\end{figure}

\paragraph{$k \in \mathbb Z$ quenches\\}
The dynamics is particularly interesting when the background fields are equal to half-integer multiples of the coupling, $\alpha = k/2$ with $k \in \mathbb Z$. 
We consider the following three types of quenches: (1) $|k| = 1$ quench, (2) $|k| = 2$ quench, and (3) $|k| = 3$ quench. Without loss of generality, we will discuss quenches with $k>0$ only; a quench with $-k$ can be obtained immediately from the one with $k$ by reversing the electric field direction.
Figures \ref{fig: nf=1 quench_1} -- \ref{fig: nf=3 quench_1} respectively illustrate $k=1,2,3$ quenches,
the evolution behaviors of which vary with the spatial circle size $mL$.

By fitting the evolution data to the Ansatz
\begin{align}
	&\frac{\ev{\bar E}}{g} = L_0 \cos(\frac{t/a}{\tau/a}) + L_b \ ,
	\label{eq: Eog_cosine model}  \\
	&P\left(\frac{E_{\text{C-L}}}{g}=-\frac{k}{2}\right) =
	P_0 + c_P \sin^2\left(\frac{1}{2}\frac{t/a}{\tau/a}\right) \ ,
	\label{eq: Prob_cosine model}
\end{align}
we extract the timescales $\tau/a$, which are related to the periods $T$ of the oscillations by $T/a= 2\pi \tau /a$. As shown in Table \ref{table: tunneling timescale and energy splittings}, the timescales extracted respectively from the expectation values and measuring probabilities agree excellently.

It is striking that ${\ev{\bar E}}/{g}$ oscillates roughly sinusoidally for all values of $m L$ (left column, blue line of Figures \ref{fig: nf=1 quench_1}, \ref{fig: nf=2 quench_1} and \ref{fig: nf=3 quench_1}).  However, there are significant qualitative and quantitative differences as $m L$ varies which reveal the different origins of these oscillations.  For $m L = 8$, the initial field is well-localized in the sense that its standard deviation satisfies $\sigma_{E/g} \ll 1$ and the period of oscillation is very long.  From the probabilities for specific $E$-field values (middle column) and the time dependence of $\sigma_{E/g}$ (left column, orange line) one can see that the field is oscillating from $\frac{\ev{\bar E}}{g} = k/2$ to  $\frac{\ev{\bar E}}{g} = -k/2$, with only a very small probability of taking any other value.  However for $m L = 1$ the oscillation is much faster and $\sigma_{E/g} \gg 1$ with very little time-dependence.  The probabilities for specific values of $E/g$ are all small (in keeping with the large standard deviation) and oscillate with time.  The case $m L = 4$ is intermediate between these two extremes.

This behavior can be understood  in terms of the novel instantons we discover, which have an action proportional to $L$ (Section \ref{sec5}).  When $m L \gg 1$ the instanton action is large and the initial field is well-localized to a specific value, with exponentially suppressed probability to take any other value.  As we will see, these instantons lack a negative mode.  Rather than mediating a traditional decay, they predict an oscillation from $E$ to $-E$, precisely as is seen in the top row of these figures.  When $m L$ is smaller the instanton action is not large and the field is not localized to a specific initial value (in units of $g$).  In this case the timescale is perturbative (the oscillation frequency for $m L = 1$ is within $10 \%$ of $g/\sqrt{\pi}$).  One might wonder whether a different initial state with $\sigma_{E/g} \ll 1$ might be more pertinent to the question of electric field decay in this regime.    This is not the case -- any such state would include higher energy eigenstates and would very quickly evolve into a state with $\sigma_{E/g} \gg 1$.  Indeed, this regime is in some ways similar to the massless or light regime of the Schwinger model in non-compact space, where the instanton action is small and the field decays by perturbative nucleation of light pairs of particles (except that in this small-circle regime the decay does not occur by pair production, as can be seen in the right column from the fact that the number of pairs is constant in time).

These behaviors can also be precisely understood in the bosonized theory (Section \ref{section: bosonized QM}).   The bosonized potential is symmetric when $k$ is integer and has multiple local minima with a quadratic envelope.  The behavior in the top row is similar to that of a symmetric double well potential when the barriers are high.  The behavior in the lower row is that of a quadratic potential with small ``wiggles" superimposed.  The frequency and amplitude of these oscillations found on the lattice agree quantitatively with those predicted by the bosonized version of the theory.

\begin{figure}[ht]
	\centering
		\includegraphics[width=0.33\textwidth]{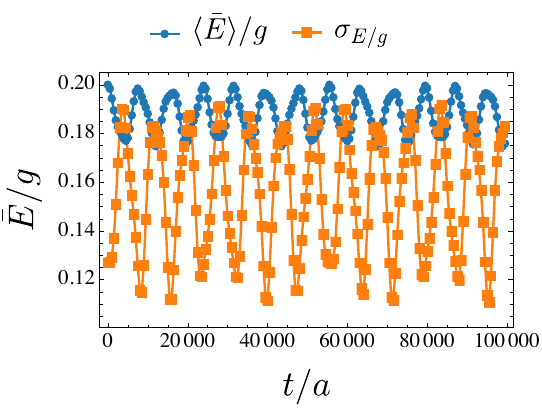}%
		\includegraphics[width=0.33\textwidth]{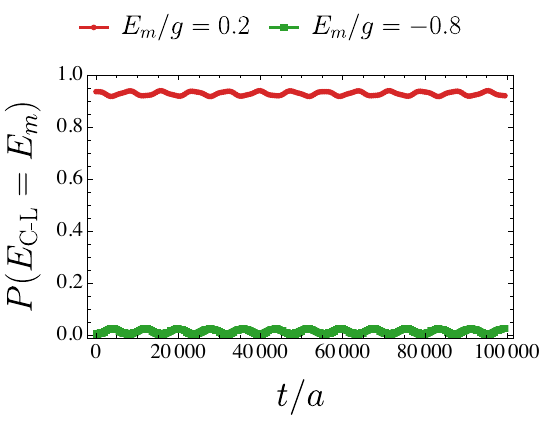}%
		\includegraphics[width=0.33\textwidth]{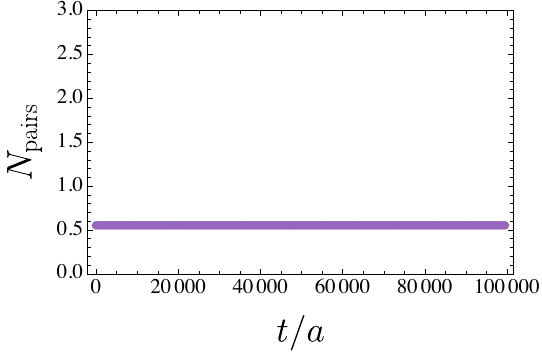}
		\includegraphics[width=0.33\textwidth]{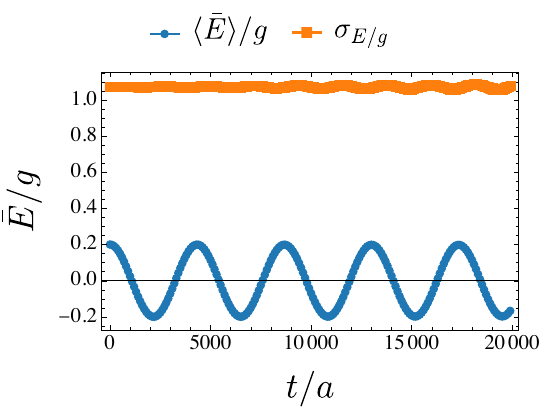}%
		\includegraphics[width=0.33\textwidth]{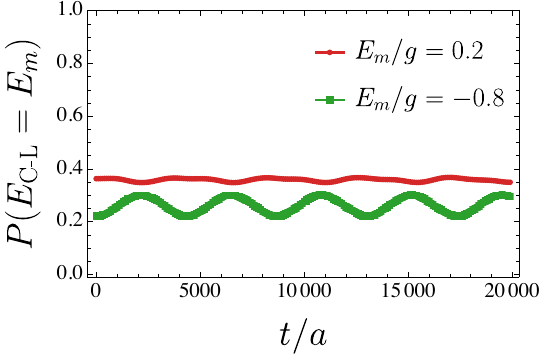}%
		\includegraphics[width=0.33\textwidth]{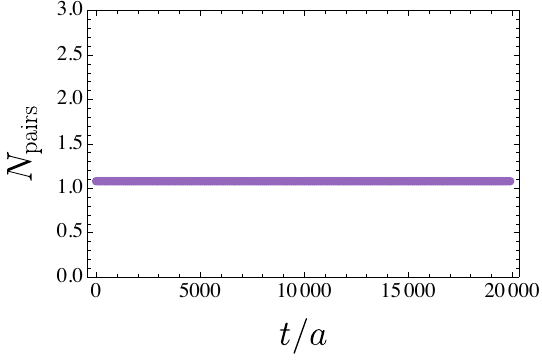}
		\includegraphics[width=0.33\textwidth]{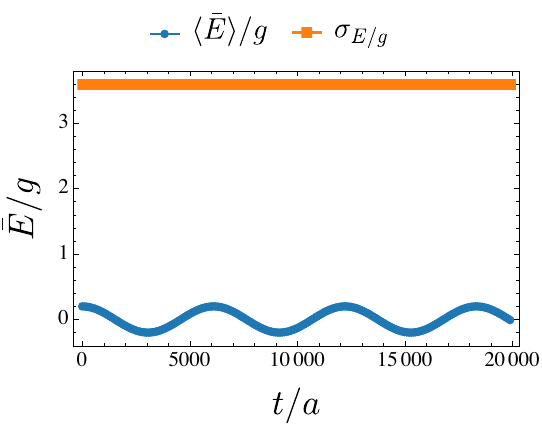}%
		\includegraphics[width=0.33\textwidth]{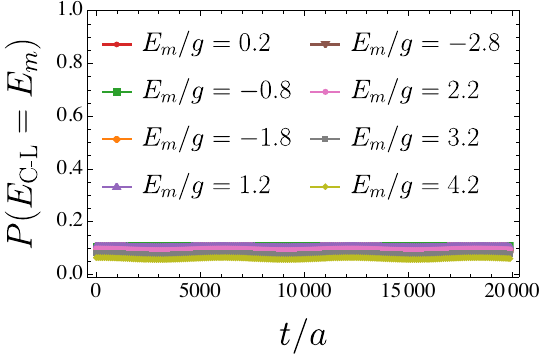}%
		\includegraphics[width=0.33\textwidth]{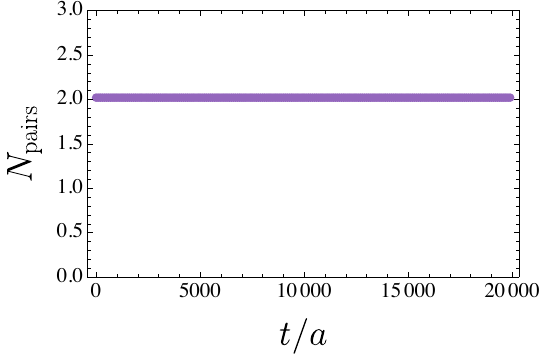}
	\caption{
	The expectation values of the spatially averaged electric field $\ev{\bar E}/g$, the standard deviation of the spatially averaged electric field $\sigma_{E/g}$, the probability to measure field $E_m$ on the central link $P\left(E_{\text{C-L}} = E_m\right)$, and the number of electron-positron pairs $N_{\text{pairs}}$ as functions of time in the $k=0.4$ quench for $\frac{m}{g}=50$, with three different spatial circle sizes $mL=8$ (top panel), $mL=4$ (middle panel) and $mL=1$ (bottom panel) on an $N=10$ lattice.
	For the same reason as in Figure \ref{fig: nf=1 quench_1}, we set $l_{\text{max}}=10$ for $mL=8,4$ simulations and $l_{\text{max}}=30$ for $mL=1$ simulation.}
\label{fig: nf=0.4 quench}
\end{figure}

\begin{figure}[ht]
  \centering
    \includegraphics[width=0.33\textwidth]{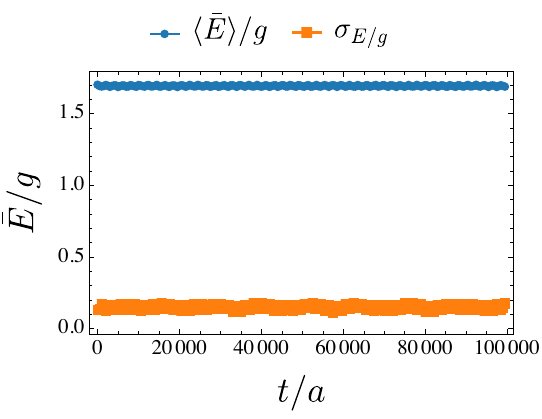}%
    \includegraphics[width=0.33\textwidth]{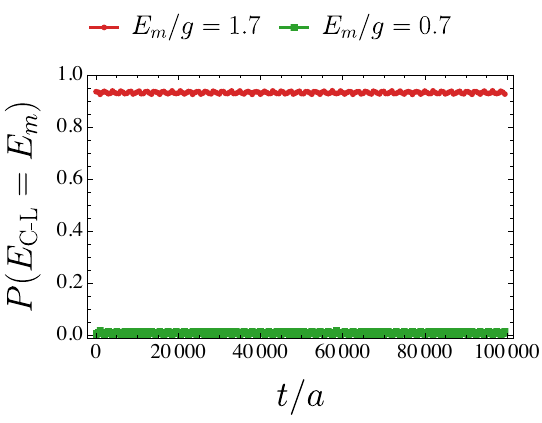}%
    \includegraphics[width=0.33\textwidth]{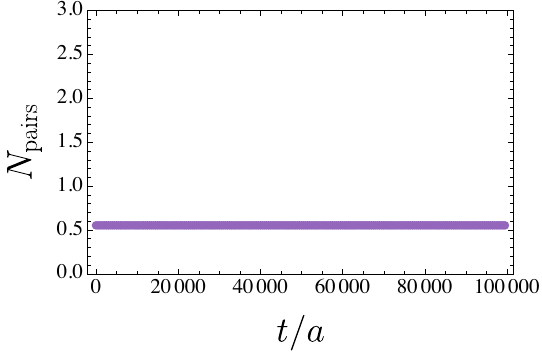}
    \includegraphics[width=0.33\textwidth]{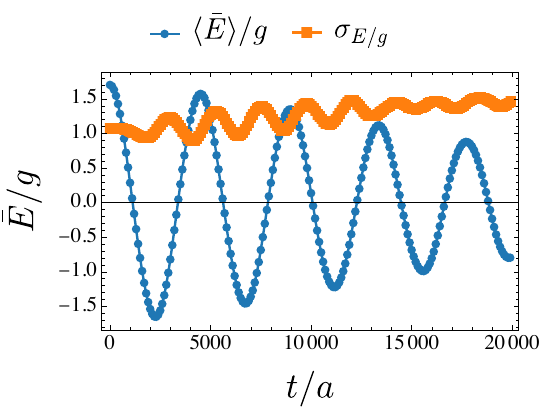}%
    \includegraphics[width=0.33\textwidth]{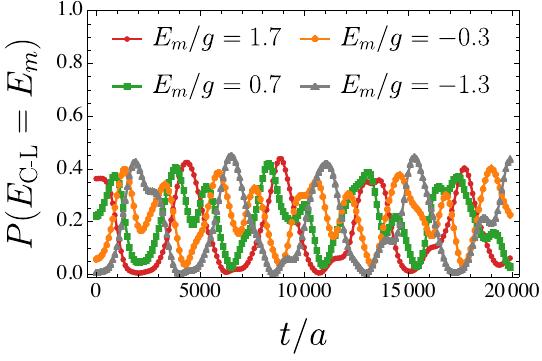}%
    \includegraphics[width=0.33\textwidth]{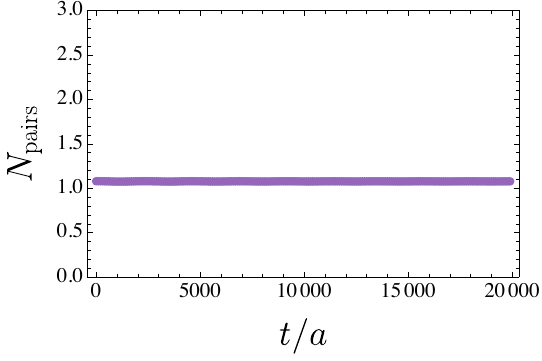}
    \includegraphics[width=0.33\textwidth]{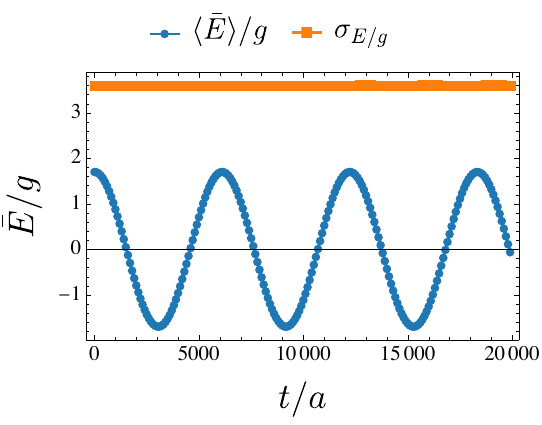}%
    \includegraphics[width=0.33\textwidth]{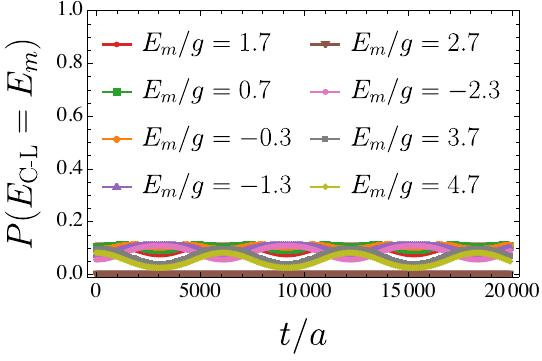}%
    \includegraphics[width=0.33\textwidth]{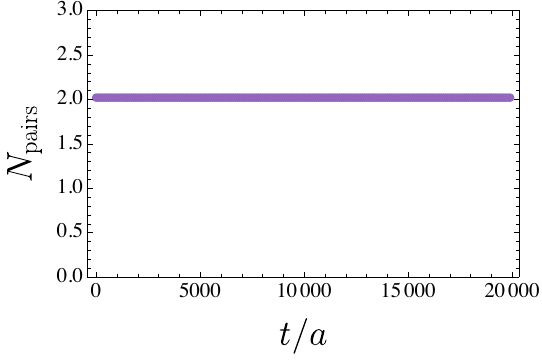}
  \caption{
  The expectation values of the spatially averaged electric field $\ev{\bar E}/g$, the standard deviation of the spatially averaged electric field $\sigma_{E/g}$, the probability to measure field $E_m$ on the central link $P\left(E_{\text{C-L}} = E_m\right)$, and the number of electron-positron pairs $N_{\text{pairs}}$ as functions of time in the $k=3.4$ quench for $\frac{m}{g}=50$, with three different spatial circle sizes $mL=8$ (top panel), $mL=4$ (middle panel) and $mL=1$ (bottom panel) on an $N=10$ lattice.
  For the same reason as in Figure \ref{fig: nf=1 quench_1}, we set $l_{\text{max}}=10$ for $mL=8,4$ simulations and $l_{\text{max}}=30$ for $mL=1$ simulation.}
\label{fig: nf=3.4 quench}
\end{figure}

\paragraph{$k \notin \mathbb Z$ quenches\\}
Let us turn to consider $k \notin \mathbb Z$ quench. As before the initial state is evolved using the final Hamiltonian with $\alpha = k/2$, but now for $k \notin \mathbb Z$. In Figures \ref{fig: nf=0.4 quench} and \ref{fig: nf=3.4 quench} we show the lattice simulation results of $k \notin \mathbb Z$ quenches in both large and small $mL$ regimes.

In the large $m L$ regime (cf. the top row of Figures \ref{fig: nf=0.4 quench} and \ref{fig: nf=3.4 quench}), the electric field $\ev{\bar E}/g$ undergoes small amplitude and relatively rapid oscillations  with probability $\gtrsim 90\%$ to coincide with its initial value.  This behavior is in sharp contrast to the case $k \in \mathbb Z$.  It is also instructive to compare to the behavior of the massive ($m/g \gg 1$) Schwinger model in the non-compact case.  There, initial fields with $-1 < k < 1$ ($-0.5 < \bar E/g < 0.5$) are  stable because the energy density in the field between a pair of nucleated particles would exceed that in the background field.  By contrast, $k=3.4$ would decay by Schwinger pair production.  Instead, our numerical results show that in the small-circle regime this field is stable (as we will see, there is indeed no instanton that can mediate its decay).	

For small $m L$ (cf. the middle and bottom rows of Figures \ref{fig: nf=0.4 quench} and \ref{fig: nf=3.4 quench}) the behaviors are qualitatively similar to $k$ integer. 
Again, these behaviors are simple to explain both in terms of our novel instantons and the bosonized description.  When $m L \gg 1$ the would-be instanton action is large and the field is well-localized.  However, for non-integer $k$ the instanton ceases to exist and there  is no decay channel for the field.  In the bosonized description this corresponds to an asymmetric potential with local minima; an initial state localized in one minimum with high barriers has no symmetric state to oscillate to, or any other decay channel.  The non-compact massless Schwinger model would behave in a qualitatively similar fashion.

\begin{figure}[ht]
	\centering
		\includegraphics[width=0.7\textwidth]{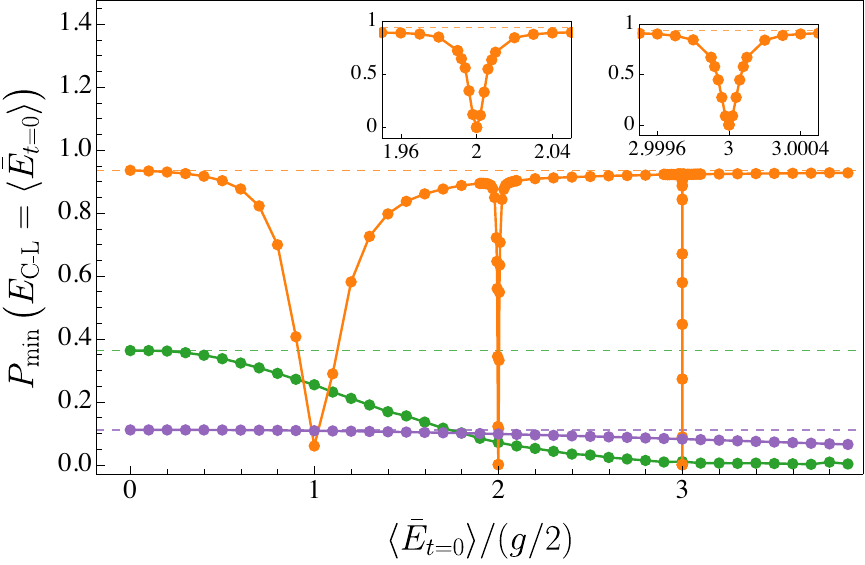}%
	\caption{The minimum probability of measuring the dimensionless electric field on the central link to be $k/2$ as a function of the background field at $t=0$ for $m/g=50$ with $mL=8$ (orange), $mL=4$ (green), $mL=1$ (purple) on an $N=10$ lattice.
	Here we take the total evolution time $(t/a)_{\text{max}}=10^8$ and 200 sampling points each simulation for searching the minimum probability.
	The dashed lines indicate the invariant probabilities for the trivial ($k=0$) quenches: $0.93537$ for $mL=8$, $0.36266$ for $mL=4$, and $0.11055$ for $mL=1$.
	The dips around integer $k$ in orange demand much longer $(t/a)_{\text{max}}$ than other points to extract. And all other points (i.e. orange points away from the dips, and all green and purple points) are insensitive to the choice of a long enough $(t/a)_{\text{max}}$. For our examples, these points remain essentially the same for $(t/a)_{\text{max}}=3\times 10^4, 10^5, 10^6, 10^8$.
	}
\label{fig: Probmin_vs_nfinal_Np=10}
\end{figure}

In Figure \ref{fig: Probmin_vs_nfinal_Np=10} we plot the minimum probability over a long time period of measuring the central link electric field to be $k/2$ as a function of the background field at $t=0$, i.e. $F/(g/2)=\ev{\bar E_{t=0}}/(g/2)=k$, which is again equal to the initial spatially averaged electric field.  If this probability remains close to 1 it means the initial field is stable, while if it is close to zero it means the field decays or oscillates from its initial value to some other values.

In the large $mL$ regime, e.g. $mL=8$ for $m/g=50$
as shown by the orange points in Figure \ref{fig: Probmin_vs_nfinal_Np=10}, when $k$ is not (close to) an integer value the field is very likely to remain at its initial value for an extremely long time (e.g. $(t/a)_{\text{max}}=10^8$ in our lattice simulation). 
The sharp dips in $P_{\text{min}}\left(E_{\text{C-L}}/g=k/2\right)$ appear at integer $k$, corresponding to the long-timescale quenched tunnelings shown in the top panels of Figures \ref{fig: nf=1 quench_1}, \ref{fig: nf=2 quench_1} and \ref{fig: nf=3 quench_1}. For $k \notin \mathbb Z$, the timescale associated with the dominating oscillation mode is much shorter than the one for $k \in \mathbb Z$.  One can also see (insets) that the width of this resonance decreases exponentially with increasing $k$.  This is expected because the instanton action scales linearly with $k$, and (as we will see) in the bosonized description the number of barriers separating $E$ and $-E$ also scales linearly with $k$.

In the small $mL$ regime, e.g. $mL=4$ for $m/g=50$
as denoted in Figure \ref{fig: Probmin_vs_nfinal_Np=10} by the green points, the minimum probability for the field on the central link to be equal to the initial average electric field is noticeably away from unity even for the trivial quench with $k=0$.
As $k$ increases, the central link field gradually loses the memory of the initial value of $\ev{\bar E}/g$ in the sense that the probability $P\left(E_{\text{C-L}}=\ev{\bar E_{t=0}}\right)$ almost vanishes at some later times. In contrast to the large $mL$ case, there are no dips in $P_{\text{min}}\left(E_{\text{C-L}}/g=k/2\right)$ at integer $k$.

\subsection{Relation to energy splittings}
\label{sec: energy splittings}
For large $mL$, the long-period sinusoidal oscillations in time suggest that only two Hamiltonian eigenstates with small energy splitting are involved, as is the case for  tunneling between two symmetric wells in quantum mechanics (this is indeed the case in the bosonized description, as we will see in Section \ref{section: bosonized QM}). 
Consider two states $\ket{\bar E/g=\pm k/2} \equiv \ket{\bar E_{\pm}}$ in which the
electric field  is localized around $\bar E/g = \pm k/2$.  Define $|S\rangle$ and $|A\rangle$ respectively to be the symmetric and antisymmetric superpositions of $\ket{\bar E_{\pm}}$. 
If $\ket{S}$ and $\ket{A}$ are Hamiltonian eigenstates, the  transition amplitudes between the wells are given by
\begin{align}
	G(\bar E_+,\bar E_{\pm},t) &=\mel{\bar E_{\pm}}{e^{-iHt}}{\bar E_+} \notag \\
	&\approx \mel{\bar E_{\pm}}{\left(\ket{S}e^{-i\EE_S t}\bra{S}
	+\ket{A}e^{-i\EE_A t}\bra{A}
	\right)}{\bar E_+} \notag \\
	&= \frac{C}{2}\left(e^{-i(\bar \EE-\frac{\Delta \EE}{2})t} \pm e^{-i(\bar \EE + \frac{\Delta \EE}{2})t}\right) \notag \\
	&= C' e^{-i \bar \EE t} \times \begin{cases}
		\cos(\Delta \EE t/2) \\
		\sin(\Delta \EE t/2)
	\end{cases}
  \label{eq: amplitude_DW_QM}
\end{align}
where $\EE_S$ and $\EE_A$ are Hamiltonian eigenvalues for  $\ket{S}$ and $\ket{A}$ respectively, with $\bar\EE\equiv (\EE_S+\EE_A)/2$ and $\Delta \EE\equiv \EE_A - \EE_S$.
It follows that the transition probability from one well to the other at time $t$ is
\begin{align}
	P(\bar E_+\to \bar E_-,t) = \left|G(\bar E_+,\bar E_{-},t)\right|^2 = |C'|^2 \sin^2 \left(\frac{\Delta \EE t}{2}\right) \ .
	\label{eq: transition prob_DW}
\end{align}
In Table \ref{table: tunneling timescale and energy splittings} we compare the oscillation timescale $\tau$  from a cosine fit to the lattice real-time data to the 
inverse of the energy difference $\Delta \EE$ between the $k$-th eigenstate and the $(k-1)$-th eigenstate.  We find very precise agreement.
(This numerical agreement
is robust for lattices with varying numbers of sites $N$, in the large $mL$ regime with $1\lesssim m/g <50$)

\begin{table}[th]
	\hspace{-24pt}
	\begin{tabular}{ccccccccccc}
	\hline
		$N$ & $\frac{m}{g}$ & $mL$ & $\left(\frac{\tau}{a}\right)_{\bar E/g}$ & $\left(\frac{\tau}{a}\right)_P$ & which $\Delta \EE$? & $a \Delta \EE$ 
		& $\left(a \Delta \EE\right)^{-1}$
		& $k$
		\\ 
	\hline
		10 & 50 & 8 & 5460.96 & 5460.97 & $\Delta \EE_{1,0}$ & $1.83118\times 10^{-4}$ & 5460.97 & 1
	\\
		10 & 50 & 8 & 76592.8 & 76597.5 & $\Delta \EE_{2,1}$ & $1.30556\times 10^{-5}$  & 76595.5 & 2
	\\
		10 & 50 & 8 & $4.20673\times10^6$ & $4.20673\times10^6$ & $\Delta \EE_{3,2}$ & $2.37714\times 10^{-7}$ & $4.20674\times10^6$ & 3
	\\
	\hline
	\end{tabular}
	\caption{The timescale $\tau/a$ and the relevant energy splittings $\Delta \EE$ for different types of quenched tunneling in the large $mL$ regime as shown in Figures \ref{fig: nf=1 quench_1}, \ref{fig: nf=2 quench_1}, \ref{fig: nf=3 quench_1}. The timescales $\left(\frac{\tau}{a}\right)_{\bar E/g}$ and $\left(\frac{\tau}{a}\right)_P$ are extracted from lattice data of the electric field $\ev{\bar E}/g$ and the quantum mechanical probability using fits (\ref{eq: Eog_cosine model}) and (\ref{eq: Prob_cosine model}) respectively.
 $\Delta\EE_{1,0}$, $\Delta\EE_{2,1}$ and $\Delta\EE_{3,2}$ are respectively the energy differences between the first excited and the ground states, between the second and the first excited states, and between the third and the second excited states of the Hamiltonian with $\alpha=k/2$.
	The column of $a\Delta \EE$ lists the lattice data of the relevant energy differences. The type of a quench is indicated by $k$.}
	\label{table: tunneling timescale and energy splittings}
\end{table}

The massive Schwinger model spectrum is periodic in $\alpha$ with period one, so one only needs to look at $\alpha =0$ or $\alpha =1/2$ spectrum for $k\in \mathbb Z$ quenches.
The energy splittings $\Delta \EE/g$ converge as we perform the fixed $N/\sqrt{x}$ continuum ($N\to \infty$) extrapolation. This implies that the energy differences and the quenched tunneling effects are physical rather than lattice artifacts.  We also verified numerically that the symmetric and anti-symmetric combinations of these Hamiltonian eigenstates $\ket{\bar E_{\pm}}$ indeed correspond to states  with expectation value for the spatially averaged field $\ev{\bar E}/g= \pm k/2$ and standard deviation $\sigma_{{E/g}} \ll 1$.

For small $mL$, the oscillations in time deviate from pure sinusoidal functions. It is instructive to perform a discrete Fourier analysis of the evolution data to extract the timescale of the quench and determine which energy differences dominantly govern the dynamics. Consider a sequence of time evolution data $\{q_{n_t}\}$ where $n_t=1, \cdots N_t$ with $N_t$ being the total number of time steps. In our case, $q_{n_t}$ represents either $\ev{\bar E}/g$ or the relevant transition probability at time $t_{n_t}/a$. The discrete Fourier transform from $\{q_{n_t}\}$ to $\{Q_{k_t}\}$ is given by $Q_{k_t}=\frac{1}{N_t}\sum_{n_t=1}^{N_t} q_{n_t} \exp[-\frac{2\pi i (n_t-1)(k_t-1)}{N_t}]$, where $k_t=1,\cdots N_t$.
The power spectrum for the quantity $q$ is then defined by $S_q(k) = \left|Q_k\right|^2$.
After a standard rescaling, we obtain the power spectrum for observable $q$ as a function of dimensionless frequency, $S_q(f/a^{-1})$. The significant peaks in the power spectra of the electric field expectation value $S_{\ev{\bar E}/g}(f/a^{-1})$ and the relevant transition probability $S_{P}(f/a^{-1})$ correspond to the characteristic frequencies of the quench. Each peak $(f/a^{-1})_{\text{peak}}$ is related to a characteristic energy difference $\Delta \EE_{c}/g$ by $\Delta \EE_{c}/g = \sqrt{x}\left(\tau/a\right)^{-1}_{\text{peak}} = 2\pi \sqrt{x} (f/a^{-1})_{\text{peak}}$. 
Comparing energy differences $\Delta \EE_{c}/g$ with those in the Hamiltonian spectrum, we find that the $\Delta \EE_{c}/g$ derived from the most predominant peak in the power spectra is $\Delta \EE_{1,0}/g$ for all three types of quenches, in the small $mL$ regime (e.g. the middle panels of Figures \ref{fig: nf=1 quench_1}, \ref{fig: nf=2 quench_1} and \ref{fig: nf=3 quench_1}). 
The energy differences of the higher excited states such as $\Delta \EE_{2,1}$ and $\Delta \EE_{3,2}$ play a minor role in these quenches, serving to modulate the amplitude of oscillations.

\section{The bosonized description}\label{sec4}
\label{section: bosonized QM}

The massive Schwinger model defined by the Lagrangian (\ref{eq: lagrangian_Schwinger}) admits an equivalent bosonized description \cite{Coleman1974,CJS1975,Coleman1976,Mandelstam1975,Naon1984} --- a theory of a massive real scalar field with a cosine interaction potential. The Euclidean action of the bosonized theory takes the form
\begin{align}
	S_E = \int \dd^2 x_E \left[\frac{1}{2}\left(\p^E_{\mu}{\phi}\right)^2 + \frac{g^2}{2\pi} \phi^2 
	- c m g \cos(2\sqrt{\pi}\phi-\theta)
	\right] \ ,
	\label{eq: bosonized action}
\end{align}
where $\phi$ is a scalar, $\left(\p^E_{\mu} \phi\right)^2\equiv \left(\p_{t_E}\phi\right)^2+\left(\p_x \phi\right)^2$ and $t_E=i t$ is  Euclidean time. Here $g$ and $m$ are the gauge field coupling and the mass of the Dirac fermion in the original theory (\ref{eq: lagrangian_Schwinger}), and $c$ is a dimensionless prefactor of the cosine potential. The background angle $\theta=2\pi F/g$ \cite{Coleman1976}.

On a compact spatial dimension with identification $x\sim x+L$, the scalar field is periodic, $\phi(x+L) = \phi(x)$, and admits a mode expansion
\begin{align}
	\phi(t_E,x)=\sum_{p=-\infty}^{\infty} \phi_{(p)}(t_E) \exp(\frac{2\pi i p x}{L}) = \phi_{(0)}(t_E)+\sum_{p\neq 0}\phi_{(p)}(t_E) \exp(\frac{2\pi i p x}{L}) \ .
\end{align}
If the size of the spatial circle is small enough and we are interested in the low energy physics, we can truncate the massive Kaluza-Klein (KK) modes with $p\neq 0$. The $(1+1)$-dimensional field theory then reduces to a $(0+1)$-dimensional quantum mechanics of the zero mode. As we will see, this quantum mechanics generates accurate predictions for the energy differences relevant to the quenches discussed in Section \ref{sec3}. The dimensional reduction yields
\begin{align}
	S_E&=\int \dd t_E \left[\frac{1}{2}\dot{\tilde \phi}^2 + \frac{1}{2} \frac{g^2}{\pi} \tilde \phi^2
	-c m g L \cos(2\sqrt{\frac{\pi}{L}}\tilde \phi-2\pi\alpha) \right] +\cdots \ ,
	\label{eq: QM action_schwinger}
\end{align}
where $\tilde \phi \equiv \sqrt{L}\phi_{(0)}$ and `$\cdots$' stands for terms involving massive KK modes. 
The effective potential in the quantum mechanical theory (\ref{eq: QM action_schwinger}) is read off from (\ref{eq: QM action_schwinger}) as $V(\tilde \phi) \equiv \frac{1}{2} \frac{g^2}{\pi} \tilde \phi^2
	-c m g L \cos(2\sqrt{\frac{\pi}{L}}\tilde \phi-2\pi\alpha)$.
For future convenience, we introduce the dimensionless variable $\varphi \equiv \tilde \phi/\sqrt{\pi L}$ and the 
dimensionless potential $U \equiv \left[V(\tilde \phi)-V_{\text{min}}\right]/(g^2 L)$ with $V_{\text{min}}$ being the global minima of $V(\tilde \phi)$. 
The dimensionless potential in terms of $\varphi$ is explicitly given by
\begin{align}
	U(\varphi) = \frac{1}{2} \varphi^2 - c \frac{m}{g} \cos(2\pi \varphi - 2\pi \alpha) + U_0 \ ,
	\label{eq: dimensionless potential_QM}
\end{align}
where $U_0$ is chosen such that the global minima of $U(\varphi)$ vanishes, i.e. $U_{\text{min}}=0$.

\subsection{Numerical bosonized quantum mechanics}
\label{sec: nbQM}
We first study the bosonized quantum mechanics by numerically solving the time-independent Schr\"odinger equation
\begin{align}
	\frac{1}{2}u''(\tilde \phi) + \left[\EE-V(\tilde \phi )\right] u(\tilde \phi) = 0 \ .
	\label{eq: QM_SCH}
\end{align}
In the above $u(\tilde \phi) \equiv \ip{\tilde \phi}{\psi}$ is the wavefunction of the state $\ket{\psi}$ in the 
$\tilde \phi$ representation, and $\EE$ is a Hamiltonian eigenvalue. 
We further write the wavefunction as $u(\tilde \phi)=C_{\tilde \phi, \varphi} u (\varphi)$ with $u(\varphi) \equiv \ip{\varphi}{\psi}$ and 
$C_{\tilde \phi, \varphi}$ a dimensionful constant determined by the normalization condition.\footnote{The constant $C_{\tilde \phi, \varphi}$ is 
determined such that
$\int \dd \tilde \phi \left|u(\tilde \phi)\right|^2 = \int \dd \varphi \left|u(\varphi)\right|^2 =1$.}
To make connections to the lattice simulation data, we rewrite (\ref{eq: QM_SCH}) in terms of lattice parameters and the dimensionless wavefunction $u(\varphi)$ as
\begin{align}
	\frac{1}{2\pi N} u''(\varphi) + \left(a \EE - \frac{N}{x} U(\varphi)\right) u(\varphi) = 0 \ .
	\label{eq: dimensionless schroedinger}
\end{align}
By solving (\ref{eq: dimensionless schroedinger}), one obtains the dimensionless Hamiltonian eigenvalues and dimensionless eigenstates. From these we immediately know the energy difference $\Delta \EE_{i,j}/g=(a \Delta \EE_{i,j})\sqrt{x}$ between the $i$-th and the $j$-th eigenstates.\footnote{The ground state is referred to as $0$-th state.} Additionally, under bosonization the spatially-averaged electric field operator is proportional to the $\varphi$ operator,
\begin{align}
	\frac{\hat {\bar E}}{g} = \frac{\hat \phi}{\sqrt{\pi}}
	=\frac{\hat {\tilde \phi}}{\sqrt{\pi L}} = \hat \varphi \ .
	\label{eq: L=varphi}
\end{align}
It follows that the expectation value and the standard deviation of the electric field for the $i$-th Hamiltonian eigenstates $\ket{\psi_i}$ are 
\begin{align}
	\ev{\hat {\bar E}}/g =& \mel{\psi_i}{\hat {\bar E}}{\psi_i}/g=\int \dd \varphi\ \varphi u^*_i(\varphi) u_i(\varphi) \\
	\sigma_{E/g,i} =&\sqrt{\frac{1}{g^2}\left[\mel{\psi_i}{\hat {\bar E}^2}{\psi_i}-\mel{\psi_i}{\hat {\bar E}}{\psi_i}^2\right]}  \notag \\
	=&\left[\int \dd \varphi\ \varphi^2 u_i^*(\varphi)u_i(\varphi)-\left(\int \dd \varphi\ \varphi  u_i^*(\varphi)u_i(\varphi)\right)^2\right]^{\frac{1}{2}} \ .
	\label{eq: sigma_L_n}
\end{align}
With the Schr\"odinger equation and the observables to compute at hand, we can use the standard  shooting method to solve the quantum mechanics numerically. To do so we need one more input, the value of the prefactor $c$.
This factor hasn't been computed analytically for arbitrary coupling $g$. 
Instead, we determine the prefactor $c$ for given $m/g$ and $mL$ numerically by comparing the predictions of a single benchmark quantity from the numerical quantum mechanics and the lattice simulation data. 
We choose this benchmark quantity to be the energy difference between the first excited and ground states $\Delta \EE_{1,0}/g$ for $\alpha=0.5$. 
By  numerically solving the quantum mechanics for a set of trial values of $c$, we find the optimal value of $c$ up to $\Op(0.01)$ that generates the closest value of the benchmark quantity to the one obtained from lattice data for specified $m/g$, $mL$ and $N$.
Let us denote the prefactor obtained in this way by $c(m/g,mL;N)$. 
Tables \ref{table: c_nQM_vs_LAT_1_mog=50} and \ref{table: c_nQM_vs_LAT_2_mog=50} summarize the optimal $c(m/g,mL;N)$ obtained from numerical quantum mechanics (nQM), the lowest few energy differences and $\sigma_{E/g,0}$ and $\sigma_{E/g,1}$ computed using the same optimal $c$, together with the same quantities calculated using lattice simulation (LAT) for the examples of $m/g=50$, $mL=8,4,1$ on $N=10$ lattice.

Once the best-fit value $c(m/g,mL;N)$ is determined, we compute other observables with this $c$ and find that the bosonized theory indeed corresponds to the original fermionic theory.
For the examples presented in the tables, the energy differences between higher excited states and $\sigma_{E/g,0}$, $\sigma_{E/g,1}$ are  close to their lattice counterparts even for $\alpha\neq0.5$. 
We confirm that the consistency holds true throughout the ranges of $m/g$ and $mL$ we explored.
In this sense, the prefactors $c(m/g,mL;N)$ are quite reliable. The physical prefactor, which is independent of lattice parameters, is attained by a continuum extrapolation of $c(m/g,mL;N)$. More details about the extrapolation procedure and the prefactor's dependence on $m/g$ and $mL$ are summarized in Appendix \ref{sec: prefactor}.

\begin{landscape}
\begin{table}[ht]
	\hspace{-48pt}
	\begin{tabular}{cc|ccc|cc|cc|cccccccccc}
	\hline
		\multirow{2}{*}{$mL$} & \multirow{2}{*}{$\alpha$} & optimal $c$ 
		& optimal $c$
		& optimal $c$
		& \multicolumn{2}{c|}{$\Delta \EE_{1,0}/g$}
		& \multicolumn{2}{c|}{$\Delta \EE_{2,1}/g$}
		& \multicolumn{2}{c}{$\Delta \EE_{3,2}/g$}
		\\
		& & (nQM) & (Garg (\ref{eq: ground splitting over g_Garg})) & (WCL (\ref{eq: ground splitting over g_weak coupling limit_Garg})) & (nQM) & (LAT) & (nQM) & (LAT) & (nQM) & (LAT) 
		
		 \\ 
	\hline
		8 & 0.5 & 17.51 & 17.86 & 17.86 & 0.0114371 & 0.0114448 & 0.154576 & 0.154575 & $1.42260\times 10^{-5}$ & $1.48571\times 10^{-5}$
	\\
		4 & 0.5 & 29.75 & 42.61 ($\cross$) & 42.61 ($\cross$) & 0.181174 & 0.181212 & 0.17036 & 0.170388 & 0.15772 & 0.157739
	\\
		1 & 0.5 & 98.61 & 751.65 ($\cross$) & 751.65 ($\cross$) & 0.514044 & 0.514045 & 0.513532  & 0.513483 & 0.513010 & 0.512911
	\\
	\hline
		8 & 0 & -- & -- & -- & 0.0806698 & 0.0806740 & $8.15740\times 10^{-4}$ & $8.15975\times 10^{-4}$ & 0.239369 & 0.239372
	\\
		4 & 0 & -- & -- & -- & 0.181175 & 0.181213 & 0.170339 & 0.170367 & 0.158087 & 0.158106
	\\
		1 & 0 & -- & -- & -- & 0.514044 & 0.514045 & 0.513532 & 0.513483 & 0.513010 & 0.512911
	\\
	\hline
		8 & 0.2 & -- & -- & -- & 0.0494704 & 0.0494736 & 0.0636598 & 0.0636608 & 0.143890 & 0.143893
	\\
		4 & 0.2 & -- & -- & -- & 0.181468 & 0.181213 & 0.171687 & 0.170374 & 0.161436 & 0.157979
	\\
		1 & 0.2 & -- & -- & -- & 0.514045 & 0.514045 & 0.513532 & 0.513483 & 0.513010 & 0.512911
	\\
	\hline 
	\end{tabular}
	\caption{{ The lowest three energy gaps and the prefactor $c$ in the bosonized action (\ref{eq: bosonized action}) for $\frac{m}{g}=50$ with various $mL$ and diffrent background field $\alpha$ on $N=10$ lattice. The values in the `(nQM)' columns are calculated using numerical bosonized quantum mechanics while those in the `(LAT)' columns are obtained from lattice field simulation. The `optimal $c$ (nQM)' is the prefactor $c(m/g,mL;N)$ elaborated in Subsection \ref{sec: nbQM}. The `optimal $c$ (Garg (\ref{eq: ground splitting over g_Garg}))' and the `optimal $c$ (WCL (\ref{eq: ground splitting over g_weak coupling limit_Garg}))' are the prefactors $c$ that produce the lattice simulated value of $\Delta \EE_{1,0}/g$ according to (\ref{eq: ground splitting over g_Garg}) and (\ref{eq: ground splitting over g_weak coupling limit_Garg}), respectively, at $\alpha=0.5$. The `$\cross$' sign indicates that the formula is invalid for certain $mL$ as expected, and thus the value of $c$ there is listed for comparison purpose only.
  For the same $mL$ but different $\alpha$, we use the same optimal $c$ to carry out the numerical quantum mechanics calculation. }}
	\label{table: c_nQM_vs_LAT_1_mog=50}
\end{table}
\end{landscape}

\begin{table}[ht]
	\centering
	\begin{tabular}{cc|cc|ccccccccccccccc}
	\hline
		\multirow{2}{*}{$mL$} & \multirow{2}{*}{$\alpha$} 
		& \multicolumn{2}{c|}{$\sigma_{E/g,0}$}
		& \multicolumn{2}{c}{$\sigma_{E/g,1}$}
		\\
		& 
		& (nQM) & (LAT) & (nQM) & (LAT) 
		
		 \\ 
	\hline
		8 & 0.5 & 0.512736 & 0.508357 & 0.51319 & 0.508808     
	\\
		4 & 0.5 & 1.07152 & 1.07018 & 1.81127 & 1.81058 
	\\
		1 & 0.5 & 3.58543 & 3.58543 & 6.20741 & 6.20722
	\\
	\hline
		8 & 0 & 0.143362 & 0.126801 & 1.00609 & 1.00388  
	\\
		4 & 0 & 1.07151 & 1.07018 & 1.81131 & 1.81062
	\\
		1 & 0 & 3.58543 & 3.58543 & 6.20741 & 6.20722
	\\
	\hline
		8 & 0.2 & 0.162915 & 0.148566 & 0.158326 & 0.143513 
	\\
		4 & 0.2 & 1.06404 & 1.07018 & 1.77152 & 1.81060
	\\
		1 & 0.2 & 3.58547 & 3.58543 & 6.20741 & 6.20722
	\\
	\hline 
	\end{tabular}
	\caption{{The standard deviations $\sigma_{E/g}$ for $\frac{m}{g}=50$ with various $mL$ and diffrent background field $\alpha$ on $N=10$ lattice. The values in the `(nQM)' columns are calculated using numerical bosonized quantum mechanics while those in the `(LAT)' columns are obtained from lattice field simulation.}}
	\label{table: c_nQM_vs_LAT_2_mog=50}
\end{table}

\begin{figure}[ht]
	\centering
		\includegraphics[width=0.5\textwidth]{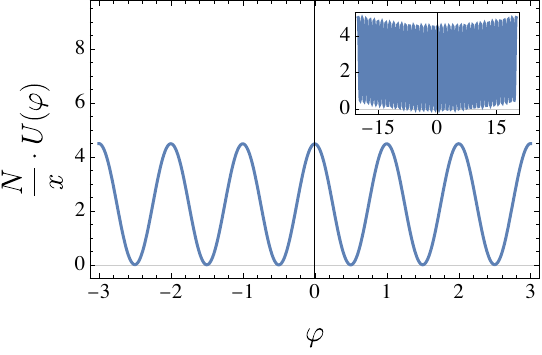}%
		\includegraphics[width=0.5\textwidth]{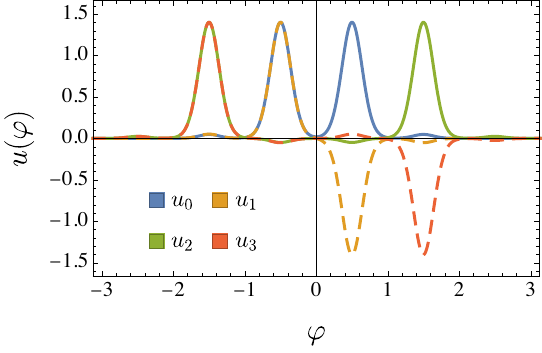}
		\includegraphics[width=0.5\textwidth]{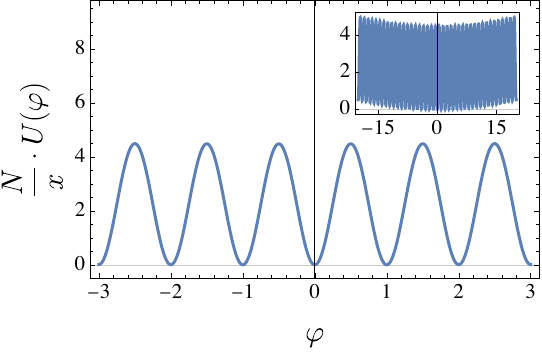}%
		\includegraphics[width=0.5\textwidth]{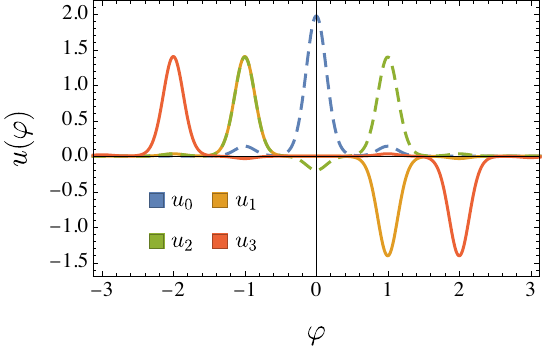}
		\includegraphics[width=0.5\textwidth]{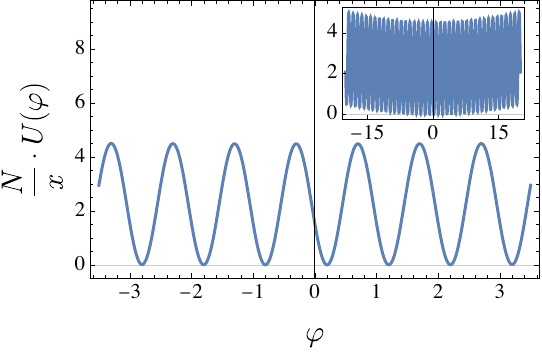}%
		\includegraphics[width=0.5\textwidth]{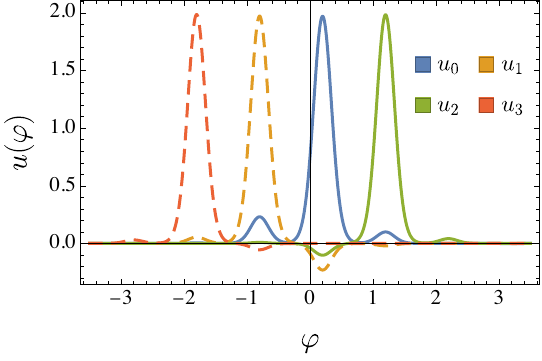}
	\caption{The (dimensionless) effective quantum mechanical potential in (\ref{eq: dimensionless schroedinger}) and the (dimensionless) wavefunctions of the lowest four Hamiltonian eigenstates for $m/g=50$, $mL=8$, a double-well regime example, on $N=10$ lattice.
	Upper panel: $\alpha=0.5$, middle panel: $\alpha=0$, lower panel: $\alpha=0.2$.
	The wavefunctions $u_0$, $u_1$, $u_2$, $u_3$ are respectively the ones in $\varphi$-representation for the ground, the first-, the second-, and the third-excited states.}
\label{fig: potential_and_wavefunction_Np=10_DW}
\end{figure}

Figure \ref{fig: potential_and_wavefunction_Np=10_DW} shows the effective quantum mechanical potential and the wavefunctions $u(\varphi)$ of relevant states for the quenches with $k=1,2,3$ in large $mL$ regime (cf. the top panels of Figures \ref{fig: nf=1 quench_1}, \ref{fig: nf=2 quench_1} and \ref{fig: nf=3 quench_1}). 
For the $k\in \mathbb Z$ quench, the relevant states are $\ket{\psi_{k}}$ and $\ket{\psi_{k-1}}$. 
Due to sufficiently high barriers between the wells,
the relevant wavefunctions are locally supported in the two wells of the potential, i.e. around $\varphi= \pm k/2$. 
These states exactly fit in the role of the symmetric and the antisymmetric states proposed in Section \ref{sec: energy splittings}, confirming the double-well  tunneling scenario mentioned there. The corresponding parameter regime of $m/g$ and $mL$ is referred to as the \emph{double-well regime}.

\begin{figure}[ht]
	\centering
		\includegraphics[width=0.5\textwidth]{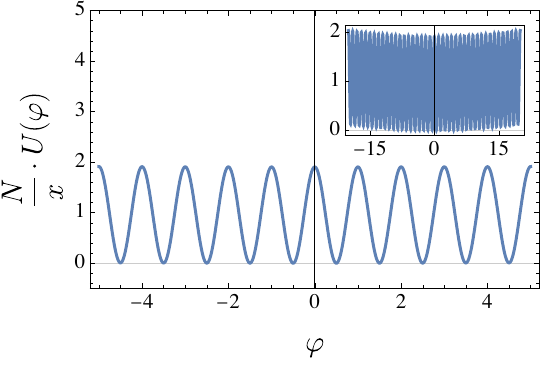}%
		\includegraphics[width=0.5\textwidth]{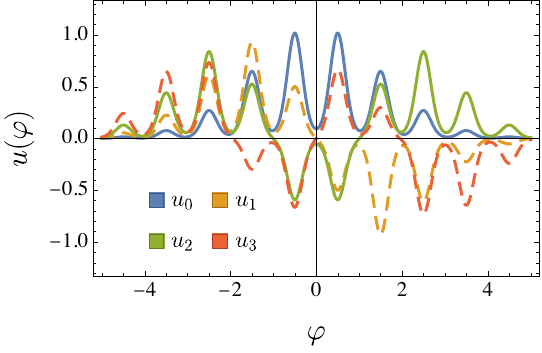}
		\includegraphics[width=0.5\textwidth]{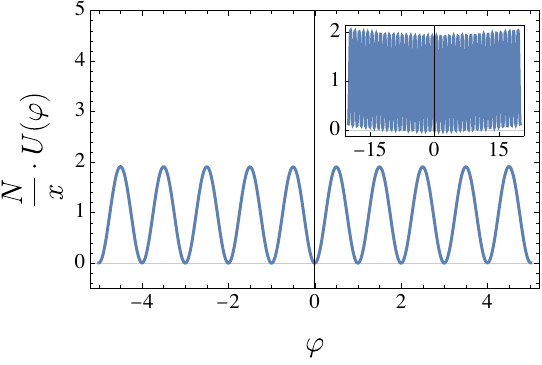}%
		\includegraphics[width=0.5\textwidth]{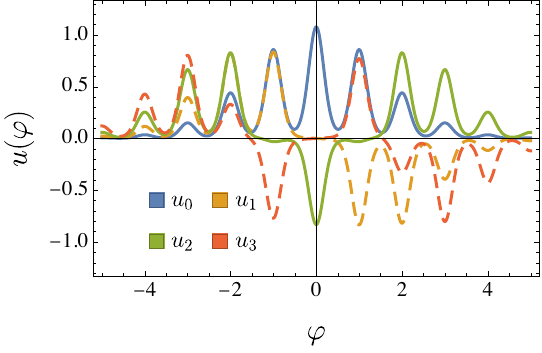}
		\includegraphics[width=0.5\textwidth]{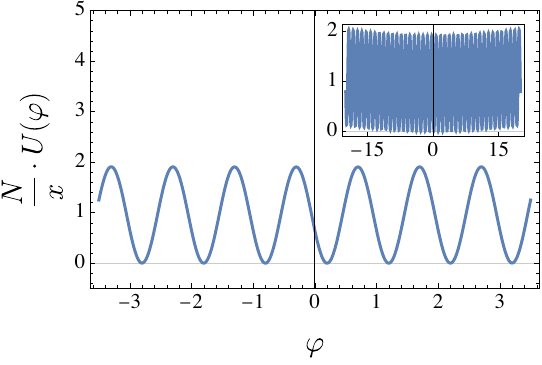}%
		\includegraphics[width=0.5\textwidth]{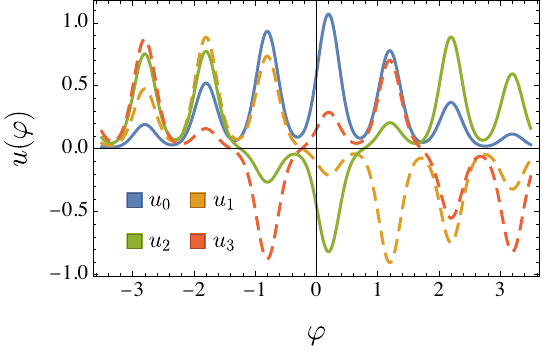}
	\caption{The (dimensionless) effective quantum mechanical potential in (\ref{eq: dimensionless schroedinger}) and the (dimensionless) wavefunctions of the lowest two Hamiltonian eigenstates for $m/g=50$, $mL=4$, a multiwell regime example, on $N=10$ lattice.
	Upper panel: $\alpha=0.5$, middle panel: $\alpha=0$, lower panel: $\alpha=0.2$.
	The wavefunctions $u_0$, $u_1$, $u_2$, $u_3$ are respectively the ones in $\varphi$-representation for the ground, the first-, the second-, and the third-excited states.}
\label{fig: potential_and_wavefunction_Np=10_MW}
\end{figure}

Figure \ref{fig: potential_and_wavefunction_Np=10_MW} illustrates the potential, and the lowest four wavefunctions $u(\varphi)$ for the $k=1,2,3$ quenches in small $m L$ regime (cf. the middle panels of Figures \ref{fig: nf=1 quench_1}, \ref{fig: nf=2 quench_1} and \ref{fig: nf=3 quench_1}).
In this intermediate regime the potential  barriers are still significant but are no longer high enough to confine the probability to a single well or symmetric pair of wells, and the wavefunctions spread across multiple wells. We  refer to this parameter regime as the \emph{multiwell regime}.

\begin{figure}[ht]
	\centering
		\includegraphics[width=0.5\textwidth]{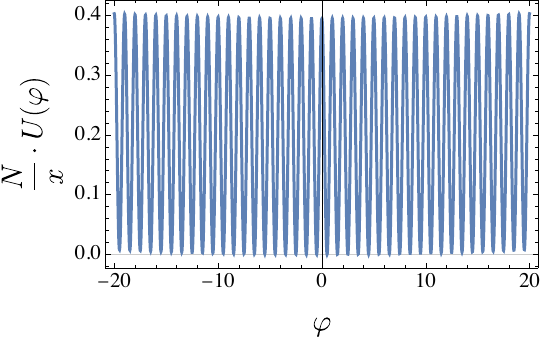}%
		\includegraphics[width=0.5\textwidth]{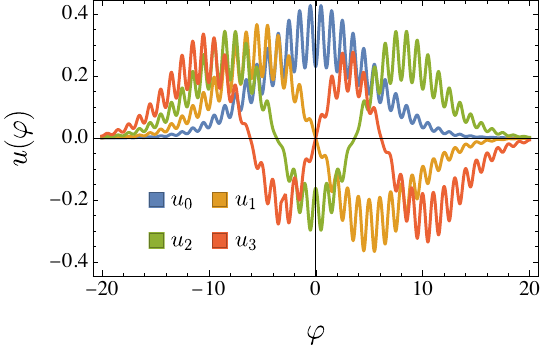}
		\includegraphics[width=0.5\textwidth]{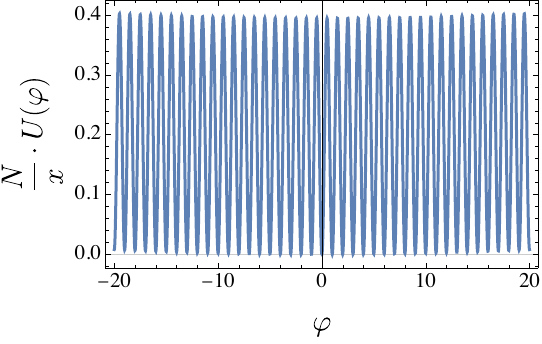}%
		\includegraphics[width=0.5\textwidth]{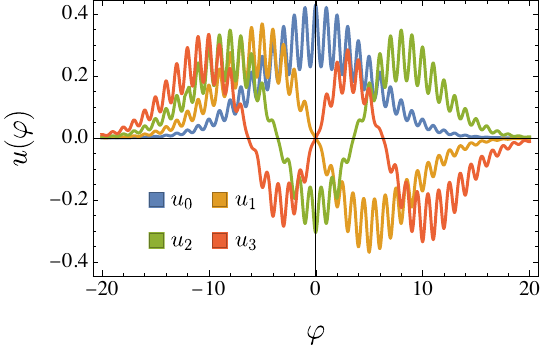}
		\includegraphics[width=0.5\textwidth]{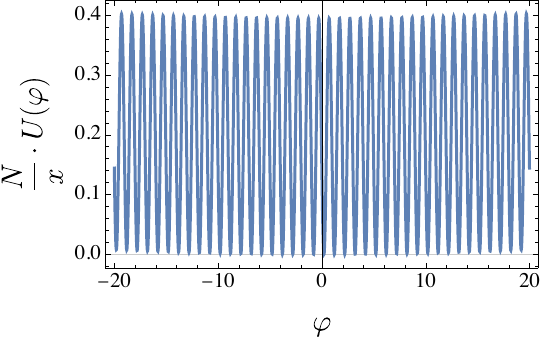}%
		\includegraphics[width=0.5\textwidth]{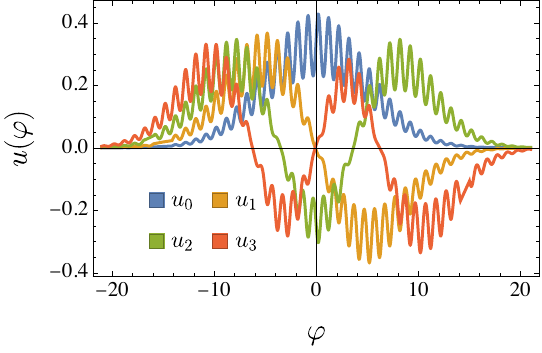}
	\caption{The (dimensionless) effective quantum mechanical potential in (\ref{eq: dimensionless schroedinger}) and the (dimensionless) wavefunctions of the lowest two Hamiltonian eigenstates for $m/g=50$, $mL=1$, a nearly SHO regime example, on $N=10$ lattice.
	Upper panel: $\alpha=0.5$, middle panel: $\alpha=0$, lower panel: $\alpha=0.2$.
	The wavefunctions $u_0$, $u_1$, $u_2$ and $u_3$ are respectively the ones in $\varphi$-representation for the ground, the first-, the second- and the third-excited states.}
\label{fig: potential_and_wavefunction_Np=10_SHO}
\end{figure}

When $m L$ is even smaller, the cosine term in the potential $V(\tilde \phi)$ becomes negligible in comparison to the quadratic term and potential approximately reduces to that of a simple harmonic oscillator. This scenario is illustrated in Figure \ref{fig: potential_and_wavefunction_Np=10_SHO}, and we refer to it as the \emph{nearly SHO regime}.

\subsection{Semi-analytical bosonized quantum mechanics}
\label{sec: semi-analytical bQM}

In this subsection, we present semi-analytical calculations of the ground splitting $\Delta \EE_{1,0}/g$ in the double-well regime, and of the standard deviations $\sigma_{E/g,0}$ and $\sigma_{E/g,1}$ in all three regimes for a $k=1$ quench.

For a $k=1$ quench in the double-well regime, the tunneling is occurring between the lowest two neighboring wells of $V(\tilde \phi)$. The states relevant to the tunneling are the ground and the first excited states. When it comes to these two states, $V(\tilde \phi)$ approximately reduces to a symmetric double-well potential. For the symmetric double well potential problem, one can solve it using either WKB \cite{LANDAU197750} or instanton method \cite{Coleman:1978ae}. Here we take the instanton approach due to \cite{doi:10.1119/1.19458}.

Let us denote the locations of the two global minima of $V(\tilde \phi)$ by $\tilde \phi_{\pm}$.
The action of the single instanton involving the quantum tunneling between $\tilde \phi_-$ and $\tilde \phi_+$ is
\begin{align}
	S_{\text{inst}} &= \int_{\tilde \phi_-}^{\tilde \phi_+} \dd \tilde \phi \sqrt{2\left[V(\tilde \phi)-V(\tilde \phi_{\pm})
	\right]}
	=\sqrt{2\pi} gL \int_{\varphi_-}^{\varphi_+} \dd \varphi\ \sqrt{U(\varphi)}  \ ,
\end{align}
where $\varphi_{\pm}=\tilde \phi_{\pm}/\sqrt{\pi L}$.
Expanding $U(\varphi)$ [i.e. eq.(\ref{eq: dimensionless potential_QM})] with $\alpha=1/2$ around $\varphi_+$ gives the frequency of the small oscillation around this well:
\begin{align}
	\omega^2_{\varphi} \equiv \left.\pdv[2]{U}{\varphi}\right|_{\varphi=\varphi_+} 
	= 1- 4 c \pi^2 \frac{m}{g} \cos(2\pi \varphi_+) \ .
\end{align}
The ground splitting is then given by \cite{doi:10.1119/1.19458}
\begin{align}
	\Delta \EE_{1,0} = 2\omega_{\tilde \phi} \left(\frac{ \omega_{\tilde \phi} \tilde \phi_+^2}{\pi}\right)^{\frac{1}{2}} 
	e^A e^{-S_\text{inst}} \ ,
\end{align}
where $\omega_{\tilde \phi} = \frac{g}{\sqrt{\pi}} \omega_{\varphi}$ and
\begin{align}
	A&=\int_0^{\tilde \phi_+} \dd \tilde \phi \left[\frac{\omega_{\tilde \phi}}{\sqrt{2\left[V(\tilde \phi)-V(\tilde \phi_{\pm})
	\right]}}-\frac{1}{\tilde \phi_+ - \tilde \phi}\right] \notag \\
	&= \int_0^{\varphi_+} \dd \varphi \left[\frac{\omega_{\varphi}}{\sqrt{2U(\varphi)}}
	-\frac{1}{\varphi_+-\varphi}\right] \ .
\end{align}
The ground energy splitting in the unit of coupling is then 
\begin{align}
	\frac{\Delta \EE_{1,0}}{g}= 2 \pi^{-\frac{3}{4}}\left(gL\right)^{\frac{1}{2}}\omega_{\varphi}^{\frac{3}{2}}\varphi_+ e^A e^{-S_{\text{inst}}} 
	=2 \pi^{-\frac{3}{4}}N^{\frac{1}{2}} x^{-\frac{1}{4}}\omega_{\varphi}^{\frac{3}{2}}\varphi_+ e^A e^{-S_{\text{inst}}} \ .
	\label{eq: ground splitting over g_Garg}
\end{align}
In the weak coupling regime $m/g \gg 1$, $\varphi_+ \approx 1/2$ and then
\begin{align}
	\omega_{\varphi}&\approx \sqrt{1+ 4 c \pi^2 \frac{m}{g}} \approx 2\pi \sqrt{c \frac{m}{g}}\ , \\
	S_{\text{inst}} &\approx \sqrt{2\pi} gL \int_{-\frac{1}{2}}^{\frac{1}{2}} \dd \varphi\ \sqrt{\frac{1}{2}\left(\varphi^2-\frac{1}{4}\right)+c\frac{m}{g}\left(1+\cos(2\pi \varphi)\right)} \notag \\
	&\approx 4 \pi^{-\frac{1}{2}} \sqrt{c m g L^2} \ .
	\label{eq: S_inst_weak coupling}
\end{align}
where in the second line of eq.(\ref{eq: S_inst_weak coupling}) the $\frac{1}{2}\left(\varphi^2-\frac{1}{4}\right)$ term in the integrand is neglected.
The integral $A$ is then simplified to
\begin{align}
	A 
	\approx \int_{0}^{\frac{1}{2}} \dd \varphi\ \frac{\sqrt{2}\pi}{\sqrt{1+\cos(2\pi \varphi)}} - \int_{0}^{\frac{1}{2}} \dd \varphi\ \frac{1}{1/2-\varphi} 
	=\ln(\frac{4}{\pi}) \ .
\end{align}
The ground splitting in the weak coupling limit (WCL) reduces to
\begin{align}
	\frac{\Delta \EE_{1,0}}{g} &= 8\sqrt{2} \pi^{-\frac{1}{4}} 
	c^{\frac{3}{4}} \left(gL\right)^{\frac{1}{2}} \left(\frac{m}{g}\right)^{\frac{3}{4}} \exp[-4\pi^{-\frac{1}{2}} c^{\frac{1}{2}} \left(mL\right)^{\frac{1}{2}} \left(gL\right)^{\frac{1}{2}}]  \\
	&= 8\sqrt{2} \pi^{-\frac{1}{4}} 
	c^{\frac{3}{4}} \left(\frac{m}{g}\right)^{\frac{3}{4}} \left(\frac{N}{\sqrt{x}}\right)^{\frac{1}{2}} \exp[-4\pi^{-\frac{1}{2}} c^{\frac{1}{2}} \left(\frac{m}{g}\right)^{\frac{1}{2}} \frac{N}{\sqrt{x}}] \ .
	\label{eq: ground splitting over g_weak coupling limit_Garg}
\end{align}
We determine the prefactor $c$ by comparing (\ref{eq: ground splitting over g_Garg}) and (\ref{eq: ground splitting over g_weak coupling limit_Garg}) with lattice data. The values of $c$ are summarized in Table \ref{table: c_nQM_vs_LAT_1_mog=50} for the examples of $m/g=50$, $mL=8,4$ on $N=10$ lattice. They are consistent with those derived from numerical bosonized quantum mechanics in the double-well regime.
For $m/g=50$ and $mL=4$ in the multiwell regime, the prefactor $c$ derived from (\ref{eq: ground splitting over g_Garg}) and (\ref{eq: ground splitting over g_weak coupling limit_Garg}), as distinct from the numerical quantum mechanical calculation, is not valid anymore.

Another set of useful observables are the standard deviations in electric field for the lowest eigenstates. As we will see in Subsection \ref{sec: PT}, they play the role of order parameters characterizing double-well-to-multiwell and multiwell-to-SHO phase transitions. Here we compute the standard deviation for the ground state $\sigma_{E/g,0}$ and for the first excited state $\sigma_{E/g,1}$.

In the double well regime for $k=1$, the wavefunctions of the ground state $u_0(\varphi)=\ip{\varphi}{\psi_0}$ and the first excited states $u_1(\varphi)=\ip{\varphi}{\psi_1}$ are locally supported around $\varphi =\pm\frac{1}{2}$, symmetric and antisymmetric about $\varphi=0$ respectively. Therefore for both states we have $\ev{\hat {\bar E}}/g \approx 0$ and $\ev{\hat {\bar E}^2}/g^2\approx 1/4$. It then follows from (\ref{eq: sigma_L_n}) that $\sigma_{E/g,0}\approx 1/2$ and $\sigma_{E/g,1}\approx 1/2$. Because each of the wells is not perfectly symmetric about the local minimum, and the wavefunctions $u(\varphi)$ have small but non-vanishing supports in the wells beyond the central double wells, these values are not exact.

In the multiwell regime, as the wavefunctions of the ground and first excited states are nonvanishing in higher wells $|\varphi|>\frac{1}{2}$, $\sigma_{E/g,0}$ and $\sigma_{E/g,1}$ exceeds $1/2$ significantly.

In the nearly SHO regime, the quadratic term dominates the effective potential $V(\tilde \phi)$. In the crudest approximation, one may neglect the cosine term and treat the potential as a pure quadratic one. 
The dimensionless Schr\"odinger equation (\ref{eq: dimensionless schroedinger}) reduces to
\begin{align}
	-\frac{1}{2}u''(\varphi) + \frac{1}{2}\frac{\pi N^2}{x} \varphi^2 u(\varphi) = \pi N a \EE u(\varphi) \ .
\end{align}
This is nothing but the Schr\"odinger equation for a simple harmonic oscillator of unit mass with angular frequency $\Omega_{\varphi} = \sqrt{\pi}\frac{N}{\sqrt{x}} =\sqrt{\pi} gL $ and energy eigenvalue $\EE_{\varphi} = \pi N a \EE =\pi \frac{N}{\sqrt{x}}\left(\frac{\EE}{g}\right) = \pi gL \left(\frac{\EE}{g}\right)$. 
With facts about a quantum SHO and the operator relation (\ref{eq: L=varphi}), the standard deviations $\sigma_{E/g,0}$ and $\sigma_{E/g,1}$ are readily obtained as
\begin{align}
	\sigma_{E/g,0} &=\sigma_{\varphi,0} = \sqrt{\mel{0}{\hat \varphi^2}{0}-\mel{0}{\hat \varphi}{0}^2} 
	= \sqrt{\frac{1}{2\Omega_{\varphi}}} = \sqrt{\frac{{\sqrt{x}}}{2 N\sqrt{\pi}}} = \sqrt{\frac{m/g}{2\sqrt{\pi}mL}}\ , 
	\label{eq: sigma_0_SHO}
	\\
	\sigma_{E/g,1} &=\sigma_{\varphi,1} = \sqrt{\mel{1}{\hat \varphi^2}{1}-\mel{1}{\hat \varphi}{1}^2} 
	= \sqrt{\frac{3}{2\Omega_{\varphi}}} = \sqrt{\frac{{3\sqrt{x}}}{2 N\sqrt{\pi}}} = \sqrt{\frac{3m/g}{2\sqrt{\pi}mL}} \ .
	\label{eq: sigma_1_SHO}
\end{align}
In the above $\ket{0}$ and $\ket{1}$ are the ground and the first excited eigenstates associated with the lowest two $\EE_{\varphi}$ eigenvalues, respectively. They correspond to the physical Hamiltonian eigenstates $\ket{\psi_0}$ and $\ket{\psi_1}$ for a fixed $gL$. In addition, from the SHO's equally spaced energy levels $\Delta \EE_{\varphi}=\Omega_{\varphi}$, one deduces that $\Delta \EE_{i+1,i}/g =1/\sqrt{\pi}\approx0.56419$. This is exactly the energy gap $\Delta \EE_{1,0}/g$ in the massless ($m\to0$) Schwinger model where the bosonized potential $V(\tilde \phi)$ is exactly quadratic. As we argued, in the massive Schwinger model on a circle of sufficiently small size $mL$ (i.e. in the nearly SHO regime), the lowest energy difference $\Delta \EE_{1,0}/g$ is approximately $1/\sqrt{\pi}$.

\subsection{Three qualitative regimes}
\label{sec: PT}

In the preceding subsections, we recognized three regimes that we termed the double-well (DW), multiwell (MW) and nearly SHO regimes, according to the behavior of electric field evolution and to the configurations of potential and lowest-lying wavefunctions.  Qualitatively the distinction is that in the DW regime the barriers are high enough that the electric field can be localized to a specific value and will remain at this value for a long time with high probability, whereas in the MW and nearly SHO regimes it will rapidly evolve to other values.

For a fixed $m/g$, the three regimes can be reached by tuning the size of the spatial circle $mL$. The right columns of Figures \ref{fig: potential_and_wavefunction_Np=10_DW}, \ref{fig: potential_and_wavefunction_Np=10_MW} and \ref{fig: potential_and_wavefunction_Np=10_SHO}  illustrate the transitions from DW to MW and from MW to nearly SHO in terms of wavefunctions of the ground and the first excited states.
We now proceed to determine the boundaries separating the regimes in terms of $mL$ by using the electric field deviation $\sigma_{E/g}$ as an indicator.

\begin{figure}[ht]
	\centering
		\includegraphics[width=1\textwidth]{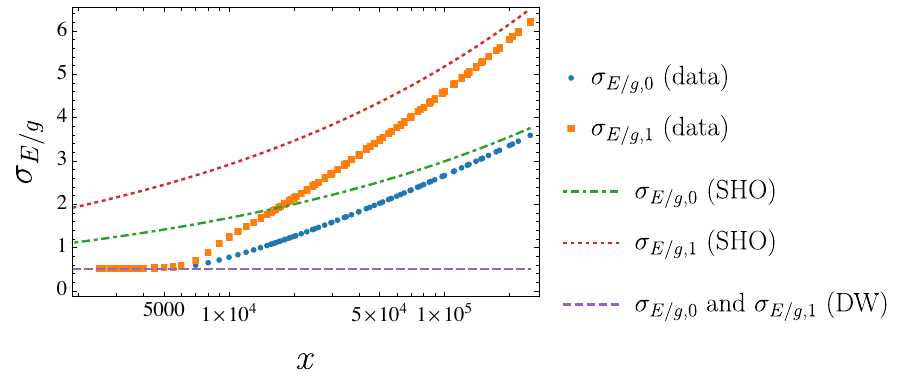}
	\caption{The standard deviations in the spatially averaged electric field for the ground state $\sigma_{E/g,0}$ and for the first excited state $\sigma_{E/g,1}$ as functions of $x\equiv 1/(ga)^2$ on $N=10$ lattice with $m/g=50$ and $\alpha=0.5$. A smaller spatial circle size corresponds to a larger $x$ according to $mL =\frac{m}{g}\frac{N}{\sqrt{x}}$. The blue and orange points are respectively the lattice data of $\sigma_{E/g,0}$ and $\sigma_{E/g,1}$.
	The green dot-dashed curve denotes $\sigma_{E/g,0}$ for a pure quantum SHO given by (\ref{eq: sigma_0_SHO}) whereas the red dotted curve is $\sigma_{E/g,1}$ for the SHO given by (\ref{eq: sigma_1_SHO}). The purple dashed line denotes the constant $\sigma_{E/g,0}=\sigma_{E/g,1} = 1/2$, which corresponds to the perfect DW case.
	}
\label{fig: sigmaEog-x_N=10_mog=50}
\end{figure}

\begin{figure}[ht]
	\centering
		\includegraphics[width=0.7\textwidth]{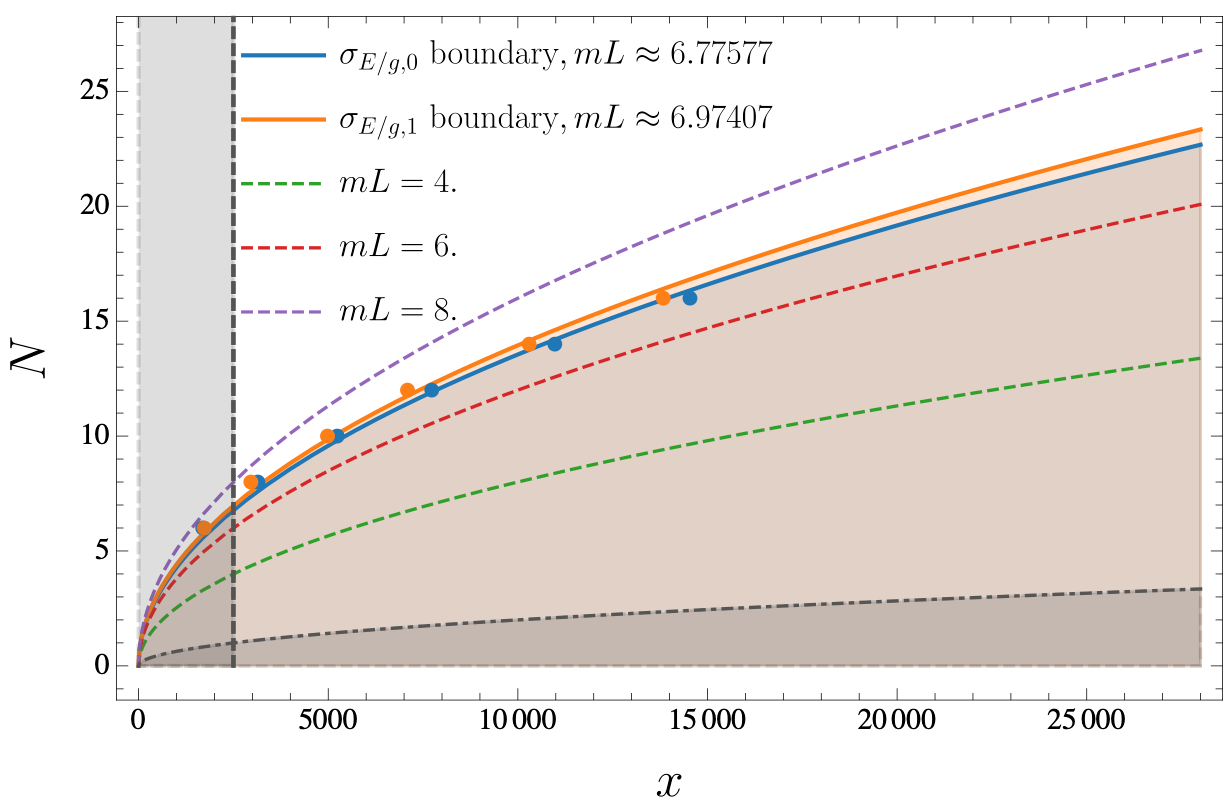}
	\caption{The boundary separating the double-well 
	regime and the multiwell regime for
	 $\frac{m}{g}=50$. The boundaries $x_{\text{bdy}}$ derived from $\sigma_{E/g,0}$ and $\sigma_{E/g,1}$ on the lattices of $N=6,8,10,12,14,16$ sites are shown as blue and orange dots respectively.
	 Blue and orange solid curves are the fitting curves (\ref{eq: DW-MW_boundary_continuum extrapolation}) of $\sigma_{E/g,0}$ and $\sigma_{E/g,1}$ respectively.
	 The boundary in terms of the physical space size, $mL_{\text{bdy}}$, is derived from the fitting parameter $\mathcal A = \frac{mL_{\text{bdy}}}{m/g}$.
	 The white region above the fitting curves is the double well regime while the orange/brown region below the fitting curves is the multiwell regime.
	 The gray shaded rectangular region ($x<\left(m/g\right)^2$) is ruled out due to the violation
	 of eq.(\ref{eq: sensible x range}).
	 The gray shaded region in the bottom ($x>\left(N\cdot m/g\right)^2$, i.e. $mL<1$), which is beyond the semiclassical regime, requires much larger $l_{\text{max}}$ to explore and is not presented in this work.
	 }
\label{fig: DW-MW boundary}
\end{figure}

As discussed in subsection \ref{sec: semi-analytical bQM}, $\sigma_{E/g} \approx 1/2$ for both the ground and the first excited state in DW regime while it significantly exceeds $1/2$ in MW regime for $\alpha=1/2$ (i.e. $k=1$). 
For a given $m/g$ with $\alpha=1/2$, we define the DW-MW boundary $x_{\text{bdy}}$ on the $N$-site lattice as the value of $x$ such that 
\begin{align}
	\frac{\sigma_{E/g,i}(x;N)-\sigma^{\text{DW}}_{E/g,i}}{\sigma^{\text{DW}}_{E/g,i}} \equiv 
	\frac{\sigma_{E/g,i}(x;N)-0.5}{0.5} = 5\% \ ,
\end{align}
where $i=0,1$ corresponds to the ground and the first excited states respectively. Figure \ref{fig: sigmaEog-x_N=10_mog=50} plots $\sigma_{E/g,0}$ and $\sigma_{E/g,1}$ as functions of $x$ for $m/g=50$ on $N=10$ lattice.
Figure \ref{fig: DW-MW boundary} shows the DW-MW boundaries $x_{\text{bdy}}$ based on both $\sigma_{E/g,0}$ and $\sigma_{E/g,1}$ for $m/g=50$ and $N=6,8,10,12,14,16$. 
The data points $(x_{\text{bdy}},N)$ are fit pretty accurately to
\begin{align}
	N=\mathcal A x_{\text{bdy}}^{\frac{1}{2}} \ ,
	\label{eq: DW-MW_boundary_continuum extrapolation}
\end{align}
with $\mathcal A$ a fitting constant.
Together with $\frac{N}{\sqrt{x}} = \frac{mL}{m/g}$, which is fixed in our continuum extrapolation, we identify the factor $\mathcal A = \frac{mL_{\text{bdy}}}{m/g}$. In this way, we obtained the physical DW-MW boundary $mL_{\text{bdy}}$. The fitting curves and the corresponding $mL_{\text{bdy}}$ are plotted and compared against various $mL$ in Figure \ref{fig: DW-MW boundary}.

The transition between MW and the nearly SHO regimes can be seen in Figure \ref{fig: sigmaEog-x_N=10_mog=50} as well. As $x$ gets very large and thus the spatial circle size $L$ shrinks to about a single fermion Compton wavelength, both $\sigma_{E/g,0}$ and $\sigma_{E/g,1}$ approaches the values for a SHO (cf. (\ref{eq: sigma_0_SHO}) and (\ref{eq: sigma_1_SHO})).

\section{Worldline instantons in the massive Schwinger model}\label{sec5}
In this section, we use the worldline formalism of path integrals to study tunneling processes in the massive Schwinger model on a circle. We identify the worldline instantons responsible for $k \in \mathbb Z$ quenched tunneling without pair production when $\frac{2m^2}{gE}>mL\gg 1$, i.e. the DW regime (e.g. top rows of Figures \ref{fig: nf=1 quench_1}, \ref{fig: nf=2 quench_1} and \ref{fig: nf=3 quench_1}), and compute the corresponding one-loop transition amplitudes.

We begin with a review on the worldline formalism applied to quantum electrodynamics, following \cite{Dunne:2005sx,Dunne:2006st,Affleck:1981bma}; readers who are familiar with it may skip directly to the next subsection. Consider quantum electrodynamics in $d$-dimensional \textit{Euclidean} spacetime with the gauge field $A$. The Euclidean action is obtained from the Lorentzian action by a Wick rotation, in which the Euclidean time $x_d$ is defined by $t=-ix_d$ with $t$ being the Lorentzian time. Accordingly the Euclidean gauge field $A_{\mu}$ is related to its Lorentzian counterpart $A^{L}_{\mu}$ as $A^L_0= iA_d$, $A^L_j= A_j,\ (j=1,\cdots,d-1)$. Let us first consider scalar electrodynamics, with $A$ coupled to a complex scalar $\phi$, and look at the normalized vacuum survival probability amplitude expressed as a path integral:
\begin{align}
  \langle 0_{\text{out}}| 0_{\text{in}}\rangle&=\int_{\text{B.C.}}\DD A \DD\phi \DD\phi^*\ \exp(-S_A[A]-S[\phi,\phi^*,A])\left/\int_{\text{B.C.}}\DD A\ \exp(-S_A[A])\right.,\label{vacamp}
\end{align}
where
\begin{align}
S[\phi,\phi^*,A]&=\int d^dx\ \phi^* (\Delta_A^2+m^2)\phi,\;\Delta_A\equiv -(\partial_\mu+igA_\mu)^2,\\
S_A[A]&=+\frac{1}{4}\int d^dx\  F^{\mu\nu}F_{\mu\nu}-\int d^dx\ \partial_\mu (F^{\mu\nu}A_\nu),
\end{align}
where $\mu,\ \nu = 1,\cdots,d$ and $F_{\mu\nu}=\p_{\mu} A_{\nu}-\p_{\nu} A_{\mu}$ is the Euclidean field strength.
Here an appropriate gauge-fixing is to be performed, and we have included the boundary term in $S_A[A]$ in order to have a well-defined variational problem with $F_{\mu\nu}$ fixed at infinity as boundary condition \cite{Brown:1988kg,Garriga:1993fh}.
Integrating out $\phi$, we obtain, at fixed $A$, the Euclidean effective action $\Gamma_{E,\text{scalar}}[A]$, given by
\begin{align}
\langle 0_{\text{out}}| 0_{\text{in}}\rangle [A]&\equiv \exp(-\Gamma_{E,\text{scalar}}[A]),
\end{align}
\begin{align}
-\Gamma_{E,\text{scalar}}[A]&=\log{\left(\frac{\det(\Delta_0+m^2)}{\det(\Delta_A+m^2)} \right)}=-\tr \log{\left(\frac{\Delta_A+m^2}{\Delta_0+m^2} \right)}\label{trlog}\\
&=+\tr\int_0^\infty \frac{ds}{s}  \left(e^{-(\Delta_A+m^2)s}-e^{-(\Delta_0+m^2)s} \right)\\
&=\int_0^\infty \frac{ds}{s} \int d^dx \langle x | e^{-(\Delta_A+m^2)s} |x\rangle +\text{constant}\\
&=\int_0^\infty \frac{ds}{s} e^{-m^2s}\int_{x(s)=x(0)} \DD x(u)\ \exp\left(-\int_0^s du \left( \frac{\dot{x}^2}{4}+igA\cdot \dot{x}\right)\right) \label{eucSeff}\\
&\equiv \int_0^\infty \frac{ds}{s}e^{-m^2s}Z_{\text{scalar}}(s,A) , \label{Zs}
\end{align}
where $\dot x^{\mu} \equiv dx^{\mu}/du$,
This expresses the effective action $\Gamma_{E,\text{scalar}}[A]$, at fixed $A$, as a path integral of a closed worldline $x^\mu(u)$ with periodic boundary condition, integrated over the proper time $s$. 

For spinor quantum electrodynamics, we introduce Feynman's spin factor $\Phi[x(u),A]$:
\begin{align}
\Phi[x(u),A]&\equiv \frac{1}{2} \Tr_{\gamma}\mathcal{P}\exp\left( \frac{ig}{4}[\gamma^\mu,\gamma^\nu]\int^s_0 du\ F_{\mu\nu}(x(u)) \right)\label{spinPhi}\\
&=\int_{\psi(s)=-\psi(0)}\DD\psi(u)\ \exp\left( -\int_0^s du\ \left( \frac{1}{2}\psi^\mu \dot{\psi}_\mu-ig \psi^\mu\psi^\nu F_{\mu\nu}(x(u)) \right) \right),
\end{align}
where $\Tr_{\gamma}$ is the trace over the (Euclidean) Gamma matrices, $\mathcal{P}$ denotes path-ordering, $\dot{\psi}_{\mu} \equiv d\psi_{\mu}/du$, and in the second line, we have expressed it as a fermionic path integral using the coherent state formalism \cite{SchubertReview2001,DG1995a,DG1995b}, which is necessary for a consistent semi-classical analysis \cite{Witten:1988hf,Beasley:2009mb}.
The effective action is obtained by augmenting \eqref{eucSeff} with $\Phi[x(u),A]$,
\begin{align}
\Gamma_{E,\text{spinor}}[A]&=\int_0^\infty \frac{ds}{s} e^{-m^2s}\int_{x(s)=x(0)} \DD x(u) e^{-\int_0^s du \left( \frac{\dot{x}^2}{4}+igA\cdot \dot{x}\right)}\Phi[x(u),A]\label{eucSeffspinor}\\
&\equiv \int_0^\infty \frac{ds}{s} e^{-m^2s}Z_{\text{spinor}}(s,A),\nonumber\\
Z_{\text{spinor}}(s,A)&\equiv  \int_{x(s)=x(0)} \DD x(u)\ \int_{\psi(s)=-\psi(0)}\DD\psi(u)\ e^{-S[x,\psi,A](s)},\label{Zsspinor}\\
S[x,\psi,A](s)&\equiv \int_0^s du\ \left( \frac{\dot{x}^2}{4}+igA\cdot \dot{x} +\frac{1}{2}\psi^\mu \dot{\psi}_\mu-ig \psi^\mu\psi^\nu F_{\mu\nu}(x(u)) \right).\label{Sspinor}
\end{align}

Including the dynamics of the gauge field, the vacuum survival amplitude \eqref{vacamp} (for both scalar and spinor electrodynamics) becomes 
\begin{align}
\langle 0_{\text{out}}| 0_{\text{in}}\rangle&=\int_{\text{B.C.}} \DD A\ \exp\left(-S_A[A] -\Gamma_E[A]\right)\left/\int_{\text{B.C.}} \DD A\ \exp(-S_A[A])\right.\\
&\equiv \left\langle \exp\left(-\Gamma_E[A]\right) \right\rangle\longrightarrow \exp\left\langle \left(-\Gamma_E[A]\right) \right\rangle.\label{dilutegas0}
\end{align}
The last line requires some explanation. In general, the exponential and the expectation value in the second equality do \textit{not} commute. And since, from \eqref{eucSeff}, $\Gamma_E[A]$ has a natural interpretation as a sum contributed by worldline single-instanton solutions, the non-commutativity is the result of correlations between single-instantons at different locations in the Euclidean spacetime, due to the dynamical gauge field \cite{Affleck:1981bma}. In the weak-field limit, we can make the dilute instanton gas approximation, wherein these correlations are ignored and one arrives at \eqref{dilutegas0}. Explicitly, in this approximation, \eqref{dilutegas0} becomes \cite{Gould:2017fve}
\begin{align}
\langle 0_{\text{out}}| 0_{\text{in}}\rangle=1+\left[\sum_{n=1}^\infty\frac{1}{n!}\int_{\text{B.C.}} \DD A\ e^{-S_A[A]}\prod_{j=1}^n \int_0^\infty \frac{ds_j}{s_j}e^{- m^2s_j}Z(s_j,A)\left/\int_{\text{B.C.}}\DD A\ e^{-S_A[A]}\right.\right] .\label{gould1}
\end{align}
At each $n$, there are $n$ uncorrelated single-instantons at different locations in the Euclidean spacetime, with the factor $1/n!$ accounting for the exchange of identical instantons. The integral over $A$ is evaluated semi-classically by expanding around a background classical solution $A_{\text{cl},n}$, and integrating over the fluctuation $\delta A$.

From now on, we focus on the spinor case. In this case, it is more convenient to evaluate the effective action in \eqref{gould1} by first performing the path integral $Z(s)$ in \eqref{Zsspinor} for each worldline, following \cite{Dunne:2006st}.  \footnote{ In Appendix \ref{appworldlinealt}, we provide alternative calculations for the worldline instantons studied in this section, but for \textit{scalar} quantum electrodynamics; there, we first integrate over the proper time, and then perform the path integral, as in \cite{Affleck:1981bma}.}  Factoring out $x(0)=x(s)=\tilde{x}$ from the path integral, $Z_{\text{spinor}}(s)$ reads
\begin{align}
Z(s)&= \int d^d\tilde{x} \int_{x(s)=x(0)=\tilde{x}} \DD x(u)\ \int_{\psi(s)=-\psi(0)}\DD\psi(u)\ e^{-S[x,\psi,A](s)}\equiv \int d^d\tilde{x}\ \tilde{Z}_{\text{spinor}}(s,\tilde{x}).\label{Zs0}
\end{align}
The classical equations of motion satisfied by the worldline instantons $(x^\mu(u),\psi^\mu(u))$ are
\begin{align}
\begin{cases}
    \ddot{x}_{\text{cl} \mu}&=2ig F_{\mu\nu}(x_{\text{cl}}(u)) \dot{x}_{\text{cl}}^\nu-2ig \psi_{\text{cl}}^\alpha \psi_{\text{cl}}^\beta \p_{\mu} F_{\alpha\beta}(x_{\text{cl}}(u))\\
\dot{\psi}_{\text{cl}\mu}&=2ig F_{\mu\nu}(x_{\text{cl}}(u))\psi^\nu_{\text{cl}}
\end{cases},\label{spinoreom}
\end{align}
which are to be supplemented by that of the gauge field. One obvious  class of solutions is with $\psi_{\text{cl}}=0$, so that at the classical level, the problem reduces to that of scalar electrodynamics. We assume that there are no non-trivial classical fermion solutions contributing to the path integral. In \eqref{Zs0}, we expand the worldline $x(u)$ as a sum of the classical path $x_{\text{cl}}(u)$ and fluctuations $\delta x(u)$,
\begin{align}
x^\mu(u)&=x^\mu_{\text{cl}}(u)+\delta x^\mu(u),\; \delta x^\mu (0)=\delta x^\mu (s)=0.\label{DBC}
\end{align}
Here the fluctuation $\delta x^\mu (u)$ satisfies the Dirichlet boundary condition, since we have factored out the endpoint $x(0)=x(s)=\tilde{x}$ from the path integral in \eqref{Zs0}. This fluctuation gives the following contribution to the path integral:
\begin{align}
\frac{1}{(4\pi s)^{\frac{d}{2}}} \sqrt{\frac{\det\nolimits_D \left(-\frac{1}{4}\delta_{\mu\nu}\frac{d^2}{du^2} \right) }{\det\nolimits_D \left(-\frac{1}{4}\delta_{\mu\nu}\frac{d^2}{du^2}+ \frac{1}{2} ig F_{\mu\nu}(x_{\text{cl}}(u)) \frac{d}{du} \right) } }\equiv \frac{1}{(4\pi s)^{\frac{d}{2}}} \sqrt{\frac{\det\nolimits_D\Lambda_{\mu\nu}^{\text{free}}}{\det\nolimits_D\Lambda_{\mu\nu}}},\label{scalardetnorm}
\end{align}
The subscript $D$ in the determinants refers to Dirichlet boundary condition. The factor $1/(4\pi s)^{\frac{d}{2}}$ is a normalization factor; when $gF_{\mu\nu}(x_{\text{cl}}(u))=0$, the path integral reduces to a free particle path integral with mass $m=1/2$. The fermionic path integral around the trivial solution gives the following contribution \cite{schubert2012lectures}
\begin{align}
\sqrt{\frac{\det\nolimits_A{\left(\frac{d}{du}-2ig F_{\mu\nu}(x_{\text{cl}}(u))\right)}}{\det\nolimits_A{\left(\frac{d}{du}\right)}}},
\end{align}
where the subscript $A$ refers to anti-periodic boundary condition. We have set the normalization of this ratio of determinants to be unity to be consistent with the fact that, when $F=0$ on the worldline, the spin factor defined in \eqref{spinPhi} evaluates to $\Phi=1$.  Including the on-shell contribution $\exp(-S[x_{\text{cl}}](s))$, $\tilde{Z}(s,\tilde{x})$ in \eqref{Zs0} becomes
\begin{align}
\tilde{Z}(s,\tilde{x})&=\frac{e^{-S[x_{\text{cl}}](s)}}{(4\pi s)^{\frac{d}{2}}} \sqrt{\frac{\det\nolimits_D\Lambda_{\mu\nu}^{\text{free}}}{\det\nolimits_D\Lambda_{\mu\nu}}}\sqrt{\frac{\det\nolimits_A{\left(\frac{d}{du}-2ig F_{\mu\nu}(x_{\text{cl}}(u))\right)}}{\det\nolimits_A{\left(\frac{d}{du}\right)}}}\label{Zs2}.
\end{align}

\subsection{Tunneling on a circle}
On a small circle of circumference $L<2m/gE$, pair production is exponentially suppressed, and a transition of the electric field without pair production dominates. We study the problem with the dynamics of the gauge field turned on, taking into account the backreaction of the worldline instantons to the gauge field. We are interested in finding the worldline instanton solutions that mediate the transition amplitude from one value of the electric field to another. 

As in \eqref{gould1}, we express an (un-normalized) transition amplitude as a sum of multi-instantons, which involves a path integral over the dynamical gauge field,
\begin{align}
\langle E_{f,\text{out}} |E_{i,\text{in}}\rangle &= \int_{\substack{E(t_E\rightarrow -\infty)=E_i\\E(t_E\rightarrow +\infty)=E_f}} \DD A\ \exp\left(-S_A[A]\right)\times\left( \sum_n (-\Gamma_E^{(n)}[A])\right)\\
&=\sum_n \int_{\substack{E(t_E\rightarrow -\infty)=E_i\\E(t_E\rightarrow +\infty)=E_f}} \DD A\ \exp\left(-S_A[A]\right)\times\left( -\Gamma_E^{(n)}[A]\right),\label{Eif}\\
\text{where }S_A[A]&= +\frac{1}{4}\int d^2x\ F_{\mu\nu}F^{\mu\nu}-\int d^2x\ \partial_\mu (F^{\mu\nu}A_\nu),
\end{align}
and $E$ is the Lorentzian electric field which is related to the Euclidean field strength $F_{\mu\nu}$ as $E=iF_{12}=-iF_{21}$.
In \eqref{Eif}, we used the dilute instanton gas approximation to commute the gauge field path integral with the sum over multi-instantons. Each $\Gamma_E^{(n)}[A]$ is contributed by $n$ worldline instantons, satisfying the boundary conditions $E(x_2\equiv t_E\rightarrow + \infty)=E_{f}$, $E(x_2\equiv t_E\rightarrow - \infty)=E_{i}$.

For each $n$, we have $n$ charged point particles $x_i(u_i)$ coupled to the gauge field by the Wilson line action along their worldlines $C_i$. The terms in the action involving $A$ read 
\begin{align}
S_{A,n}&\equiv +\frac{1}{4}\int d^2x\  F_{\mu\nu}F^{\mu\nu}-\int d^2x\ \partial_\mu (F^{\mu\nu}A_\nu)\notag \\
&\qquad\qquad+ig \sum_{i=1}^n\int_{C_i} du_i \left[A_\mu(x_i(u_i))\frac{dx_i^\mu(u_i)}{du_i}-\psi^\mu (u_i)\psi^\nu (u_i)F_{\mu\nu}(x_i(u_i))\right].\label{Sm2d}
\end{align}
Varying the action wrt.~$A$ gives Gauss' law; focusing on the trivial fermion solution $\psi_{\text{cl}}=0$,
\begin{align}
\partial_{\mu}(F_{\text{cl},n}^{\mu\nu}(x))&=+ig \sum_{i=1}^n\int_{C_i} du_i\ \delta^{(2)}(x-x_i(u_i))\frac{dx_i^\nu(u_i)}{du_i},\label{gauss}
\end{align}
which simply says that the on-shell electric field $F_{\text{cl}}$ changes by one unit of $g$ as it crosses a charged worldline. Substituting \eqref{gauss} back to the action \eqref{Sm2d}, the on-shell action is
\begin{align}
S_{A,n}[A_{{\text{cl}},n}]&=-\frac{1}{4}\int d^2x\ F^{\text{cl},n}_{\mu\nu}F^{\mu\nu}_{\text{cl},n}=+\frac{1}{2}\int d^2x\  E^2_{\text{cl},n}.\label{Sm2d2}
\end{align}
After that, one integrates over the fluctuation $\delta A= \bar{\delta}A+\tilde{\delta}A$ in \eqref{Eif}. We have factored out the variation $\bar{\delta}A$ due to the variations of the worldlines and the fermions; the corresponding contributions are accounted for in $\Gamma^{(n)}_E[A]$. As it turns out, the quadratic action of the remaining fluctuation $\tilde{\delta}A$ is that of a free field, decoupled from the worldlines and their fluctuations, as long as the worldlines do not intersect. Therefore, we can absorb its contributions in the normalization of the amplitude.

\subsubsection{\texorpdfstring{$|k|=1$}{|k|=1}: Straight line instantons}
By calculating the transition amplitude $\ip{E_f=-g/2}{E_i=+g/2}$, we will argue that it is the straight line instantons that account for $|k|=1$ quenched tunneling without pair production when $mL\gg 1$ (cf. the top row of Figure \ref{fig: nf=1 quench_1}).

We first consider the survival probability amplitude at $E=+g/2$, $\langle +g/2 | +g/2 \rangle$. 
The gauge field satisfies the boundary condition $E(t_E\rightarrow \pm \infty)=+g/2$. We wish to express the amplitude as a sum of multi-instantons,
\begin{align}
\langle +g/2 |+g/2\rangle &= \sum_n \exp\left(-S_{A,n}[A_{{\text{cl}},n}]\right)(-\Gamma_E^{(n)}),\label{go2}
\end{align}
where $\Gamma_E^{(n)}$ is contributed by $n$ worldline instantons.

\begin{figure}[H]
\centering
\begin{subfigure}{0.5\textwidth}
  \centering
  \hspace*{0.5cm}\includegraphics[width=1.\linewidth,keepaspectratio]{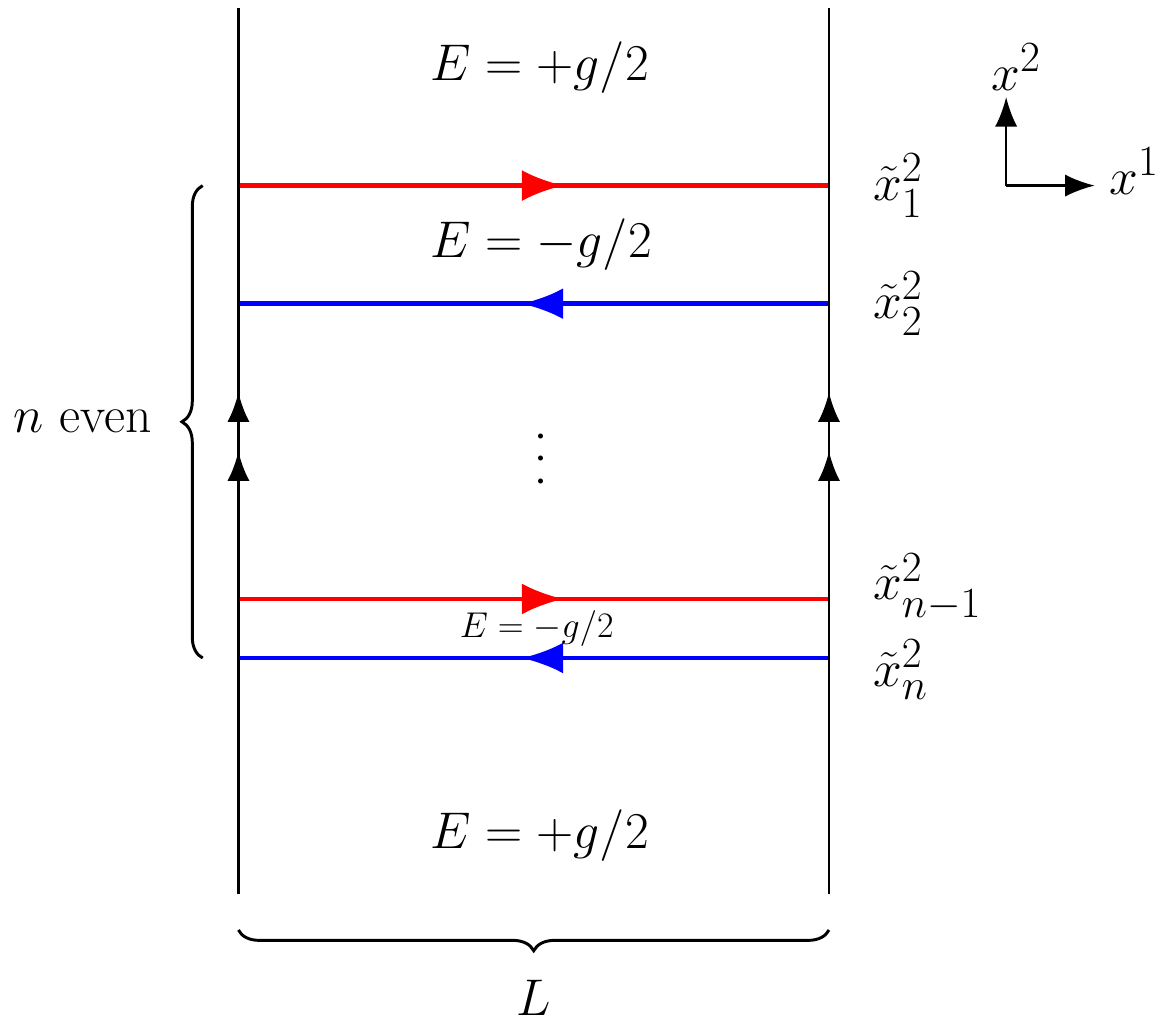}
\caption{$n$ even for $\langle + g/2|+g/2\rangle$.}\label{stlineeven}
\end{subfigure}%
\begin{subfigure}{.5\textwidth}
  \centering
  \hspace*{0.5cm}\includegraphics[width=1.\linewidth,keepaspectratio]{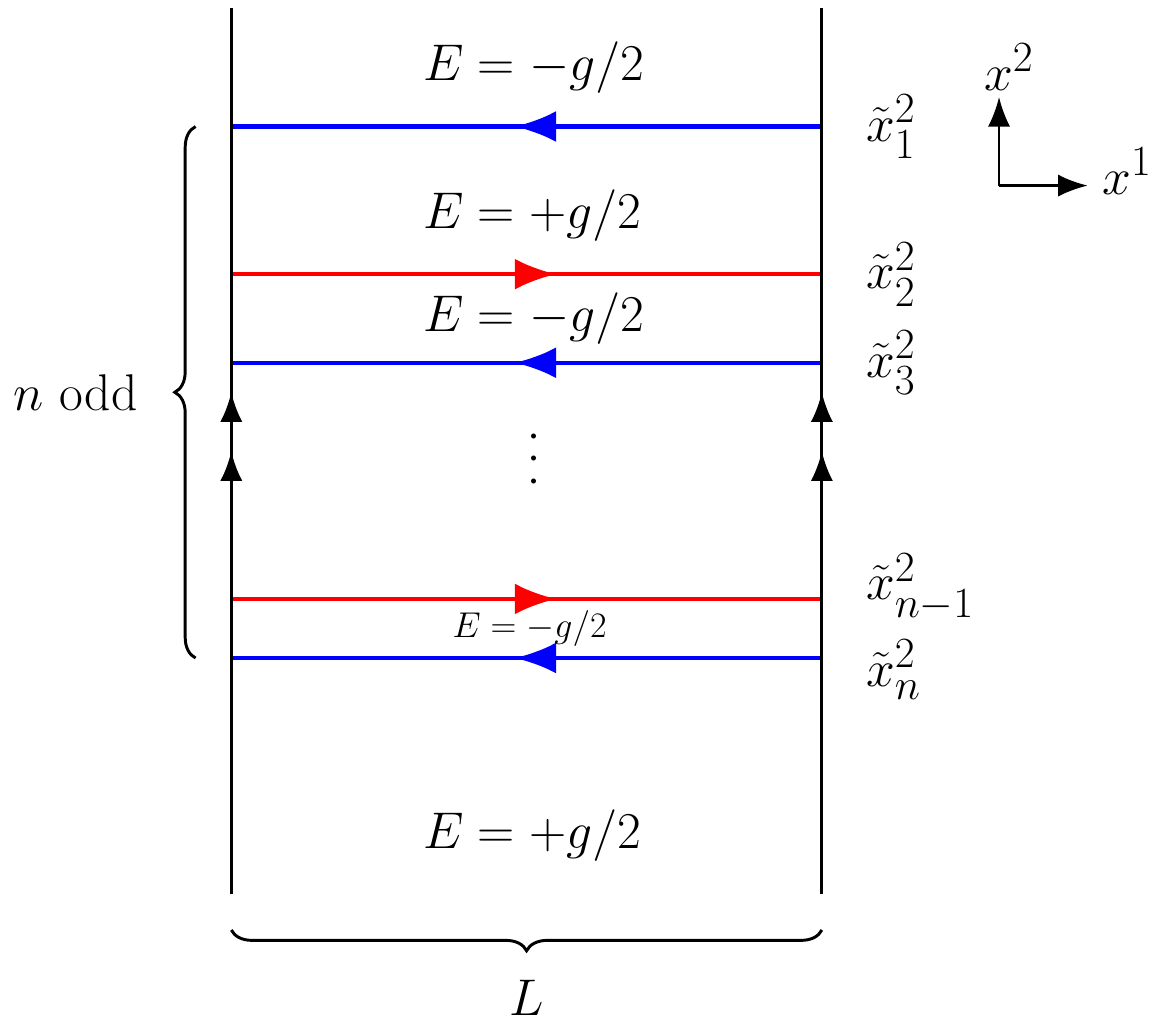}
  \caption{$n$ odd for $\langle - g/2|+g/2\rangle$.}
  \label{stlineodd}
\end{subfigure}
\caption{The multi-straight line instanton solutions contributing to $\langle \pm g/2|+g/2\rangle$.}
\label{stlineinst}
\end{figure}

The $n$-worldline instanton solution which leads to $(-\Gamma_E^{(n)})$ and satisfies the boundary condition $E(x^2\rightarrow \pm \infty)=+g/2$ only exists for $n$ even. It is given by $n$ straight lines running in the compact $x^1$-direction with alternating orientations, with $n/2$ of them in the $(+x^1)$ direction and $n/2$ in the $(-x^1)$ direction. This is illustrated in Figure \ref{stlineeven}. From \eqref{eucSeffspinor} and \eqref{Zs0}, we have
\begin{align}
\Gamma_E^{(n)}&=\prod_{i=1}^n \int_0^\infty\frac{ds_i}{s_i}e^{-m^2 s_i} \int_{-\infty}^{\tilde{x}^2_{i-1}} d\tilde{x}^2_i \int_0^L d\tilde{x}^1_i\ \tilde{Z}(s_i,\tilde{x}_i),\;\tilde{x}^2_0\equiv -\infty+t_E\label{GammaEngo2},\\
\tilde{Z}(s_i,\tilde{x}_i)&=\int_{\psi_i(s_i)=-\psi_i(0)} \DD \psi_i(u_i)\int_{x_i(s_i)=x_i(0)=\tilde{x}_i}\DD x_i(u_i) \notag \\
&\quad\quad \exp\left[-\int_0^{s_i}du_i\ \left(\frac{\dot{x}^2_i}{4}+igA\cdot\dot{x}_i +\frac{1}{2}\psi_i^\mu \dot{\psi}_{i\mu}-ig \psi_i^\mu\psi_i^\nu F_{\mu\nu}(x_i(u))\right)\right] .\label{Zsi}
\end{align}
We take the electric field $\bar{E}\equiv E(x_{i\text{cl}}(u))$ on the worldline to be the average (mean) of that at its two sides. On the trivial fermion solution $\psi_{\text{cl}}=0$, from Gauss' law \eqref{gauss} and $E(x^2\rightarrow \pm \infty)=+g/2$, we find that $\bar{E}\equiv E(x_{i\text{cl}}(u_i))=0$. The worldlines as solutions to (\ref{spinoreom}) are given by

\begin{equation}
\begin{aligned}
x_{i\text{cl}}^1(u_i)&=\pm L\frac{u_i}{s_i}+\tilde{x}_i^1 ,\; x_{i\text{cl}}^2(u_i)=\tilde{x}_i^2\in (-\infty,\tilde{x}_{i-1}^2) ,\\
 \bar{E}&\equiv E(x_{i\text{cl}}(u_i))=0,\; E(x^2\rightarrow \pm \infty)=+g/2.
\end{aligned}
\end{equation}
Since $E(x_{i\text{cl}}(u_i))=0$, the ratio of determinants due to the fluctuations (for both bosonic and fermionic) is exactly one, so we have
\begin{align}
\tilde{Z}(s_i,\tilde{x}_i)&=\frac{e^{-\frac{L^2}{4s_i}}}{4\pi s_i}.\label{Ztildecircle}
\end{align}
Hence, \eqref{GammaEngo2} becomes
\begin{align}
\Gamma_E^{(n)}&=\frac{(it L)^n}{n!} \prod_{i=1}^n \int_0^\infty\frac{ds_i}{4\pi s_i^2}e^{-m^2 s_i-\frac{L^2}{4s_i}}\\
&=\frac{(it L)^n}{n!}\left( \frac{m}{\pi L}K_1(mL) \right)^n\overset{mL\gg 1}{\approx}\frac{1}{n!}\left((it) \frac{1}{\sqrt{2\pi}}\sqrt{\frac{m}{L}} e^{-mL} \right)^n,
\end{align}
where $K_{\nu}(x)$ is the modified Bessel function of the second kind, $mL\gg 1$ is the semi-classical regime.
Moreover, from \eqref{Sm2d2}, the on-shell action of the gauge field is
\begin{align}
S_{A,n}[A_{{\text{cl}},n}]&=+\frac{1}{2}\int d^2x\  E_{\text{cl},n}^2=\frac{1}{2} \int d^2x \left(\frac{g}{2} \right)^2,
\label{S_A_onshell}
\end{align}
which can be absorbed in the normalization of the transition amplitude. Therefore, the survival amplitude \eqref{go2} is
\begin{align}
\langle +g/2 |+g/2\rangle &=\sum_{n\text{ even}}\frac{1}{n!}\left((it L) \frac{m}{\pi L}K_1(mL) \right)^n\\
&=\sum_{n=0}^\infty \frac{(-1)^n}{(2n)!}\left(t (m/\pi) K_1(mL) \right)^{2n}\\
&=\cos{\left( t (m/\pi) K_1(mL)\right)}\overset{mL\gg 1}{\approx}\cos{\left( t \frac{1}{\sqrt{2\pi}}\sqrt{\frac{m}{L}} e^{-mL}\right)}\label{ppgo2}
\end{align}
From this, it is apparent that we have normalized the amplitude correctly as we threw away the infinite gauge field action. The exponent of the time scale associated to this process is independent of the electric field value $E=\pm g/2$. Thus, this process cannot be mediated by the production of a real pair by taking energy out of the electric field. 

\begin{figure}[ht]
  \centering
    \includegraphics[width=.8\textwidth]{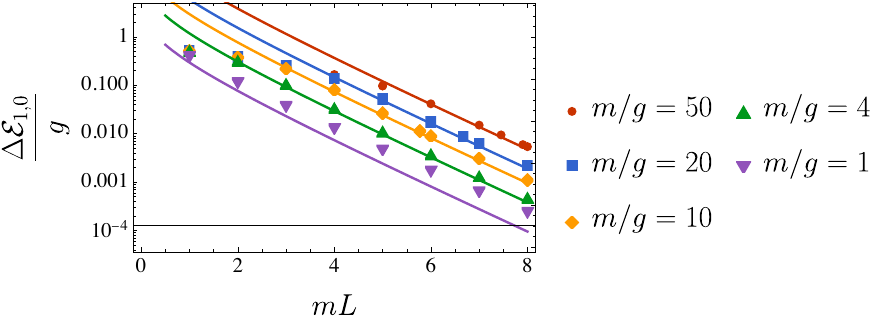}
    \includegraphics[width=.8\textwidth]{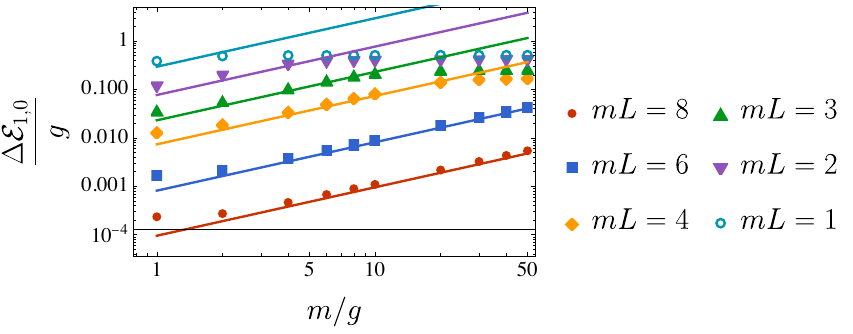}
  \caption{
  The continuum extrapolated ($N/\sqrt{x}$ fixed, $N\to \infty$) lattice results (dots) versus the straight line instanton estimate (\ref{eq: DE10_instanton}) (curves) of the
  energy difference $\Delta\EE_{1,0}/g$ between the ground and first excited states for background field $\alpha=0.5$. Top panel: each color represents a fixed value of $m/g$; bottom panel: each color represents a fixed value of $mL$.
  }
\label{fig: DE10og_instanton_vs_lattice}
\end{figure}

We can similarly compute $\langle -g/2 |+g/2\rangle$, in which case the multi-instanton solutions are $n$ straight lines, $n$ odd, along $x^1$, with $(n-1)/2$ running in the $(+x^1)$ direction and $(n+1)/2$ running in the $(-x^1)$ direction. This is illustrated in Figure \ref{stlineodd}. In the end, after an appropriate normalization, one gets 
\begin{align}
\langle -g/2 |+g/2\rangle =\sin{\left( t (m/\pi) K_1(mL)\right)}\overset{mL\gg 1}{\approx}\sin{\left( t \frac{1}{\sqrt{2\pi}}\sqrt{\frac{m}{L}}e^{-mL}\right)}.\label{mpg02}
\end{align}
The fact that $\langle +g/2 |+g/2\rangle$ and $\langle -g/2 |+g/2\rangle$ are given by cosine and sine and their square norms sum to one is in agreement with the bosonized description studied in \secref{sec4}, in which $E=\pm g/2$ are the two degenerate global minima of the sine-Gordon potential, and the transitions between them are mediated by the Coleman double-well instantons yielding cosine and sine. 
Comparing (\ref{mpg02}) with (\ref{eq: amplitude_DW_QM}), we extract the ground state energy splitting $\Delta \EE_{1,0}/g$ for background $E=g/2$ from the straight line instantons,
\begin{align}
  \left(\frac{\Delta \EE_{1,0}}{g}\right)_{\text{instanton}} = 2 \sqrt{\frac{(m/g)^2}{2\pi mL}} e^{-mL} \ .
  \label{eq: DE10_instanton}
\end{align}
In Figure \ref{fig: DE10og_instanton_vs_lattice}, we plot the instanton estimates (\ref{eq: DE10_instanton}) of $\EE_{1,0}/g$ for $\alpha=0.5$ in curves and compare them with the continuum extrapolated lattice results.
For relatively large $m/g$ and relatively large $mL$ where the instanton semiclassical computation applies, the instanton estimates agree remarkably well with the lattice results.

\subsubsection{\texorpdfstring{$|k|=2$}{|k|=2}: Lemon instantons}
Next, we wish to identify the worldline instantons that compute the amplitude $\langle +g | -g \rangle$. 
This corresponds to $|k|=2$ quenched tunneling when $mL\gg 1$ (cf. the top row of Figure \ref{fig: nf=2 quench_1}).
From \eqref{Sspinor} and \eqref{Eif}, we have the Euclidean action
\begin{align}
S&=\sum_{i=1}^n \int_{-s_i/2}^{s_i/2} du_i \left( \frac{\dot{x}_i^2}{4}+igA\cdot \dot{x}_i +\frac{1}{2}\psi_i^\mu \dot{\psi}_{i\mu}-ig \psi_i^\mu\psi_i^\nu F_{\mu\nu}(x_i(u_i)) \right)+S_{A}.\label{Sm2d2lemon}
\end{align}
Note that we have now taken $u_i\in[-s_i/2,s_i/2)$. For the trivial fermion background $\psi_{\text{cl}}=0$, the equations of motion \eqref{spinoreom} for each worldline (with the worldline index $i$ suppressed) become
\begin{subequations}
\begin{empheq}[left=\empheqlbrace]{align}
\ddot{x}^1_{\text{cl}}&=+2giF_{12}\dot{x}_{\text{cl}}^2\\
\ddot{x}^2_{\text{cl}}&=-2giF_{12}\dot{x}_{\text{cl}}^1.
\end{empheq}
\end{subequations}
As before, we take the electric field $E(x_{\text{cl}}(u))$ on the worldline to be the average (mean) of that at its two sides; Gauss' law \eqref{gauss} and $E(x^2\rightarrow \pm \infty)=\pm g$ imply that $E(x_i(u_i))=\pm g/2\equiv \pm \bar{E}$.

\begin{figure}[H]
\begin{center}
\includegraphics[width=.95\linewidth,keepaspectratio]{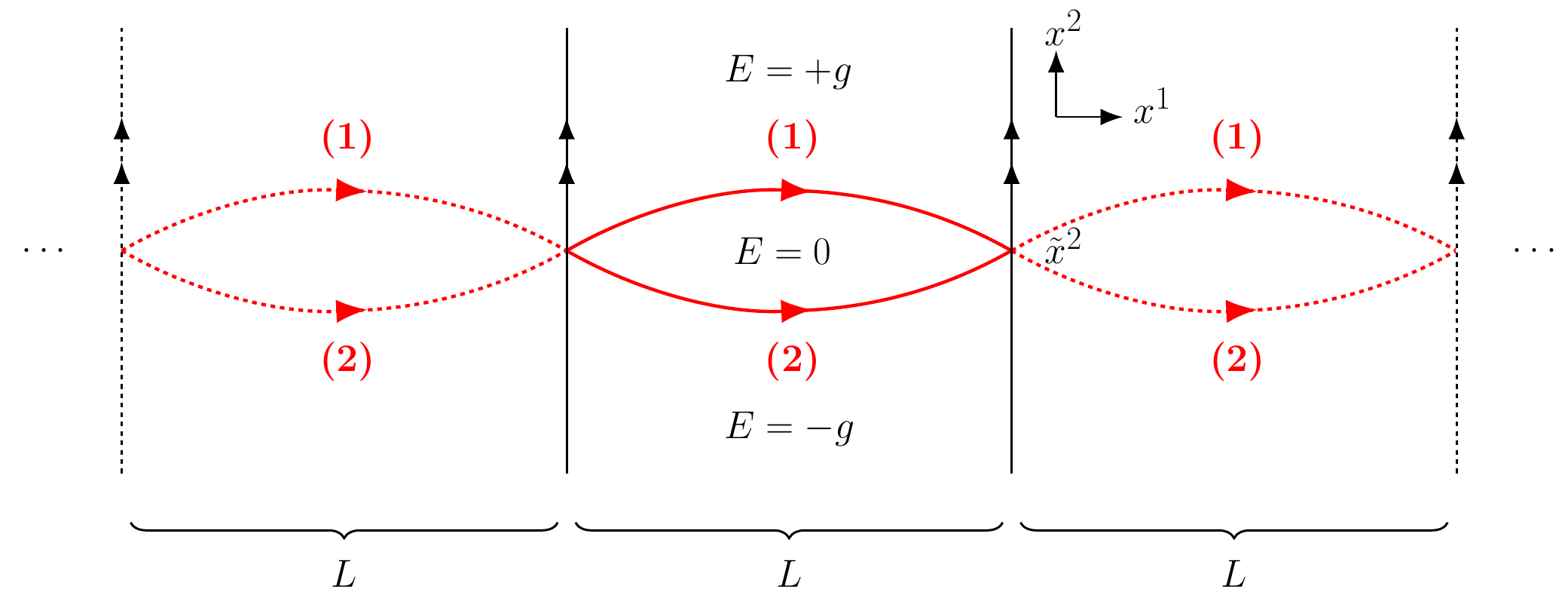}
\caption{The single-lemon instanton \eqref{lemon1}, \eqref{lemon2} contributing to $\langle +g|-g\rangle$.}\label{lemonsingle}
\end{center}
\end{figure}

We identify a single-instanton solution, illustrated in Figure \ref{lemonsingle}, which we dub the ``lemon instanton". It consists of two circle segments $(1)$ and $(2)$ running in the positive $x^1$-direction, patched together at $u=0$ and $u=\pm s/2$, where the worldline has a continuous zeroth and first derivatives, while the second derivative changes sign. This solution is periodic since $x^1\sim x^1+L$. The region enclosed by the worldline has $E=0$, while outside we have $E=\pm g$, and on the segments, $E(1)=-E(2)=\bar{E}\equiv g/2$. Explicitly, the solution is given by
\begin{align}
(1)\text{: }&\begin{dcases}
x^1_{\text{cl}}&=+a \sin{(b(u/s+1/4))}+\tilde{x}^1\\
x^2_{\text{cl}}&=+a(\cos{(b(u/s+1/4))}-\cos{(b/4)})+\tilde{x}^2
\end{dcases},\quad u\in [-s/2,0]\label{lemon1}\\
(2)\text{: }&\begin{dcases}
x^1_{\text{cl}}&=+a\sin{(b(u/s-1/4))}+\tilde{x}^1\\
x^2_{\text{cl}}&=-a(\cos{(b(u/s-1/4))}-\cos{(b/4)})+\tilde{x}^2
\end{dcases},\quad u\in (0,s/2).\label{lemon2}
\end{align}
Taking into account the periodicity in $x^1\sim x^1+L$, the parameters take the values
\begin{align}
b&=2g\bar{E}s,\quad a=\frac{L}{2\sin(\frac{g\bar{E}s}{2})}.\label{lemonab}
\end{align}

The on-shell action of \eqref{Sm2d2lemon} for this single-instanton solution takes the form
\begin{align}
S_{E1,0}&=\int_{-s/2}^{s/2} du\  \frac{\dot{x}_{\text{cl}}^2}{4}+\cancel{\frac{1}{2}\int d^2x\ g^2}-\frac{1}{2}g^2\times(\text{Lemon Area})\\
&=\frac{g\bar{E}}{2}L^2 \cot\left( \frac{g\bar{E}s}{2}\right).\label{lemononshell}
\end{align}

\subsubsection*{One-loop calculations}
Around the lemon instanton, the bosonic fluctuation operator in \eqref{Zs2} takes the form
\begin{align}
\Lambda_{\mu\nu}&\equiv  -\frac{1}{4}\delta_{\mu\nu}\frac{d^2}{du^2}+ \frac{1}{2} ig F_{\mu\nu}(x_{\text{cl}}(u)) \frac{d}{du}\nonumber \\
&= \frac{1}{4s^2} \begin{pmatrix}
-\frac{d^2}{dv^2} & -(2g\bar{E}s)\text{sign}(v)\frac{d}{dv}\\
+(2g\bar{E}s)\text{sign}(v)\frac{d}{dv}& -\frac{d^2}{dv^2}
\end{pmatrix}\\
&=U^{-1}\frac{1}{4s^2} \begin{pmatrix}
-\frac{d^2}{dv^2} -i(2g\bar{E}s)\text{sign}(v)\frac{d}{dv} &0\\
0&  -\frac{d^2}{dv^2}+i(2g\bar{E}s)\text{sign}(v)\frac{d}{dv}
\end{pmatrix}U,\;\\
U&\equiv \frac{1}{\sqrt{2}} \begin{pmatrix}
i&1\\-i&1
\end{pmatrix}.\label{Umatrix}
\end{align}
Here we have defined $u\equiv sv$, so that $v\in[-1/2,1/2)$. The eigenmodes of the fluctuation operator must have continuous zeroth and first derivatives. By diagonalizing the operator, the eigenmodes that satisfy Dirichlet boundary condition are easily found to be
\begin{subequations}
\begin{align}
y_{(1),n}&=\begin{cases}
(+\sqrt{2}\sin{(2\pi n v)}\sin{\left(\frac{bv}{2}\right)},+\sqrt{2}\sin{(2\pi n v)}\cos{\left(\frac{bv}{2}\right)})\qquad\qquad&,\; -1/2\leq  v \leq 0\\
(-\sqrt{2}\sin{(2\pi n v)}\sin{\left(\frac{bv}{2}\right)},+\sqrt{2}\sin{(2\pi nv )}\cos{\left(\frac{bv}{2}\right)})&,\; 0< v < 1/2
\end{cases},\nonumber\\
y_{(2),n}&=\begin{cases}
(-\sqrt{2}\sin{(2\pi n v)}\cos{\left(\frac{bv}{2}\right)},+\sqrt{2}\sin{(2\pi n v)}\sin{\left(\frac{bv}{2}\right)})\qquad\qquad&,\; -1/2\leq  v \leq 0\\
(-\sqrt{2}\sin{(2\pi n v)}\cos{\left(\frac{bv}{2}\right)},-\sqrt{2}\sin{(2\pi n v)}\sin{\left(\frac{bv}{2}\right)})&,\; 0< v < 1/2
\end{cases},\nonumber
\end{align}\label{y12}
\end{subequations}
where $b=2g\bar{E}s$ as in \eqref{lemonab}, and $n=1,2,\ldots$. The eigenvalues are
\begin{align}
\lambda_n&=\left[\left(\frac{2\pi}{2g\bar{E}s} \right)^2 n^2-\frac{1}{4}\right]\left(\frac{1}{4s^2}\right) (2g\bar{E}s)^2,\quad n =1,2,\ldots,\label{locallambdan}
\end{align}
with multiplicity two. The functional determinant can be computed by a Riemann zeta function regularization to become
\begin{align}
\det\nolimits_D \Lambda_{\mu\nu}&=\prod_{n=1}^\infty \left(\left[\left(\frac{2\pi}{2g\bar{E}s} \right)^2 n^2 -\frac{1}{4}\right] \left(\frac{1}{4s^2} \right)(2g\bar{E}s)^2 \right)^2=\frac{\sin^2{(g\bar{E}s/2)}}{(2g\bar{E}s/4)^2} (4s^2).
\end{align}
Furthermore, the determinant of the free operator $\Lambda_{\mu\nu}^{\text{free}}=-(1/4s^2)\delta_{\mu\nu}d^2/dv^2$ (with Dirichlet condition) is easily found to be $\det \Lambda_{\mu\nu}^{\text{free}}=4s^2$, so
\begin{align}
\sqrt{\frac{\det\nolimits_D \Lambda_{\mu\nu}^{\text{free}}}{\det\nolimits_D \Lambda_{\mu\nu}}}&=\frac{g\bar{E}s/2}{\sin{(g\bar{E}s/2)}}.\label{lemonlocalratio}
\end{align}
The fermionic fluctuation operator with anti-periodic boundary condition at $v=\pm 1/2$ reads
\begin{align}
&\quad \frac{d}{du}-2ig F_{\mu\nu}(x_{\text{cl}}(u))\nonumber\\
&=\frac{1}{s}\begin{pmatrix}
\frac{d}{dv} & +(g\bar{E}s)\text{sign}(v) \\
-(g\bar{E}s)\text{sign}(v) & \frac{d}{dv}
\end{pmatrix}\\
&=U^{-1} \frac{1}{s}\begin{pmatrix}
\frac{d}{dv}  +i(g\bar{E}s)\text{sign}(v) &0\\
0& \frac{d}{dv}-i(g\bar{E}s)\text{sign}(v)
\end{pmatrix} U ,
\end{align}
where $U$ is the same matrix as in \eqref{Umatrix}. Since the operator is first-order, the eigenmodes only need to be continuous at $v=0$ (in addition to being anti-periodic at $v=\pm 1/2$). They are easily found to be
\begin{align}
y_{\psi,n}&=\begin{cases}
U^{-1}(C_1 e^{(+ig\bar{E}s+i\pi(2n+1))v},C_2 e^{(-ig\bar{E}s+i\pi(2n+1))v})\qquad\qquad&,\; -1/2\leq  v \leq 0\\
U^{-1}(C_1 e^{(-ig\bar{E}s+i\pi(2n+1))v},C_2 e^{(+ig\bar{E}s+i\pi(2n+1))v})&,\; 0< v < 1/2
\end{cases},\label{spinoreigenlemon}
\end{align}
where $C_{1,2}$ are constants. They have eigenvalues $\lambda_{\psi,n}=i\pi(2n+1)/s$, where $n\in\mathbb{Z}$. Thus, the eigenvalues are independent of $g\bar{E}$, and so the determinant is canceled by that of the free operator.

Substituting \eqref{lemononshell}, \eqref{lemonlocalratio} and \eqref{Zs2} back to the effective action \eqref{eucSeffspinor}, we have 
\begin{align}
\Gamma_E&=\int d^2\tilde{x}\int_0^{\infty}\frac{ds}{s}e^{-m^2 s-\frac{g\bar{E}}{2}L^2 \cot\left( \frac{g\bar{E}s}{2}\right)}  \frac{1}{4\pi s}\frac{g\bar{E}s/2}{\sin{(g\bar{E}s/2)}}.
\end{align}
The integral over the proper time $s$ is very similar to that in \cite{Draper2018}. We evaluate it by the saddle point approximation. The saddle points are given by 
\begin{align}
s_p&= \frac{2}{g\bar{E}} \left[\sin^{-1}\left( L\left/ \frac{2m}{g\bar{E}} \right. \right) + \pi p\right],\;p=0,1,2,\ldots,
\end{align}
and have on-shell actions
\begin{align}
S_{p}&\equiv\left. m^2 s+\frac{g\bar{E}}{2}L^2 \cot\left( \frac{g\bar{E}s}{2}\right)\right|_{s=s_p}\\
&=mL\sqrt{1-\left( L\left/ \frac{2m}{g\bar{E}} \right.\right)^2}+m \left( \frac{2m}{g\bar{E}} \right)\left[\sin^{-1}\left( L\left/ \frac{2m}{g\bar{E}} \right. \right) + \pi p\right]\\
&\overset{L\ll \frac{2m}{g\bar{E}}}{\longrightarrow} 2mL+\pi p m \left( \frac{2m}{g\bar{E}} \right) .
\end{align}
Including the pre-factors, we finally get
\begin{align}
\Gamma_E&=\sum_{p=0}^\infty (it)\frac{(-1)^p}{4\sqrt{2\pi}}\sqrt{g\bar{E}}\sqrt{L\left/ \frac{2m}{g\bar{E}} \right.}\left( 1- \left( L\left/ \frac{2m}{g\bar{E}} \right.\right)^2\right)^{-\frac{1}{4}} \left[\sin^{-1}\left( L\left/ \frac{2m}{g\bar{E}} \right. \right) + \pi p\right]^{-1} e^{-S_p}\label{lemonfinalp}\\
&\overset{L\ll \frac{2m}{g\bar{E}}}{\longrightarrow} \frac{it}{4\sqrt{\pi}}\sqrt{\frac{m}{L}}e^{-2mL}.
\end{align}
The dominant saddle, $p=0$, has the same action as that of Brown's instanton in \cite{Brown2015}. It is instructive to consider different limits. When $L\rightarrow (2m/g\bar{E})$ where pair production becomes kinematically favored, $S_{p=0}\rightarrow \pi m^2/g\bar{E}$, which is the same as the on-shell action of Schwinger pair production. On the other hand, in the small circle limit $L\ll (2m/g\bar{E})$, the $p>0$ saddles are exponentially suppressed. The $p=0$ saddle dominates, with action approaching $S_{p=0}\rightarrow 2mL$. The asymptotic form of the action can be understood intuitively: in this limit, the area enclosed by the lemon vanishes, and the action is given by ($m$ times) the length of the worldline. As in the straight line case, the fact the on-shell action does not depend on the electric field value in the small circle limit suggests that the process does not involve the production of a real pair.

Moreover, in the present case, the gauge field fluctuation $\tilde{\delta}A$ not due to the variation of the worldline again decouples; its contribution can again be absorbed in the normalization of the amplitude. In the end, if the lemon instanton were the only solution contributing to the amplitude, then we would get, in the weak-field and small circle limits,
\begin{align}
\langle +g|-g\rangle= \sin\left(\frac{t}{4\sqrt{\pi}}\sqrt{\frac{m}{L}}e^{-2mL}\right).\label{lemonwrong}
\end{align}

Comparing (\ref{lemonwrong}) with (\ref{eq: amplitude_DW_QM}), the energy difference between the second and the first excited states $\Delta\EE_{2,1}/g$ for background $E=g$ can be obtained from the lemon instantons as
\begin{align}
  \left(\frac{\Delta \EE_{2,1}}{g}\right)_{\text{lemon}} = 2 \sqrt{\frac{(m/g)^2}{16 \pi mL}} e^{-2mL} \ .
  \label{eq: DE21_instanton}
\end{align}
However, by comparing the lemon instanton estimates (\ref{eq: DE21_instanton}) with the continuum extrapolated lattice results of $\frac{\Delta \EE_{2,1}}{g}$ for various $m/g$ and $mL$, we find that they do not agree.
Instead, if we modify the prefactor by multiplying it with an extra factor $\frac{R_c}{2L} = \frac{(m/g)^2}{mL}$, where $R_c \equiv \frac{m}{g \bar E} = \frac{2m}{g^2}$ is the critical radius of the lemon instanton, the modified lemon instanton estimate
\begin{align}
  \left(\frac{\Delta \EE_{2,1}}{g}\right)_{\text{modified lemon}}= 2\sqrt{\frac{1}{16\pi}} \frac{(m/g)^3}{(mL)^{3/2}} e^{-2mL}
  \label{eq: lemon_modified}
\end{align}
is then consistent with the lattice result for relatively large $m/g$ and relatively large $mL$ where the semiclassical calculation applies.

\begin{figure}[ht]
  \centering
    \includegraphics[width=.8\textwidth]{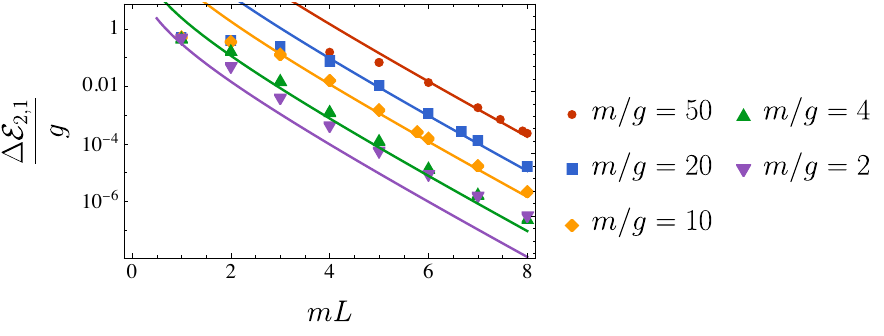}
    \includegraphics[width=.8\textwidth]{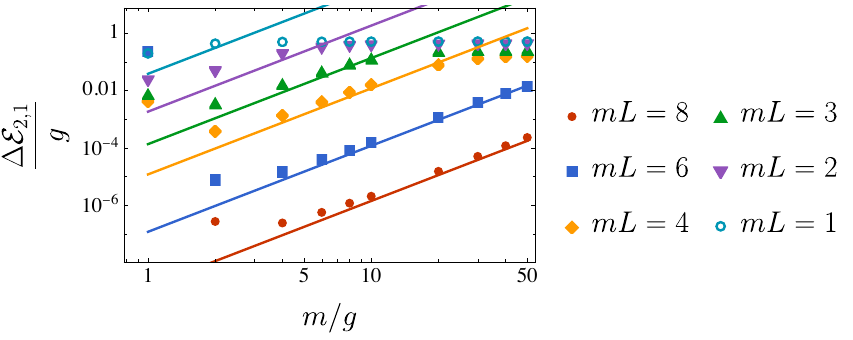}
  \caption{
  The continuum extrapolated ($N/\sqrt{x}$ fixed, $N\to \infty$) lattice results (dots) versus the modified lemon instanton estimate (\ref{eq: lemon_modified}) (curves) of the
  energy difference $\Delta\EE_{2,1}/g$ between the second and first excited states for background field $\alpha=0$. Top panel: each color represents a fixed value of $m/g$; bottom panel: each color represents a fixed value of $mL$.
  }
\label{fig: DE21og_instanton_vs_lattice}
\end{figure}

In Figure \ref{fig: DE21og_instanton_vs_lattice}, the modified lemon instanton estimates (\ref{eq: lemon_modified}) of $\EE_{2,1}/g$ for $\alpha=0$ (the spectrum of which is equivalent to that for $\alpha=1$) are plotted in curves and compared with the continuum extrapolated lattice results for various $m/g$ and $mL$.

\begin{figure}[H]
\begin{center}
\includegraphics[width=.75\linewidth,keepaspectratio]{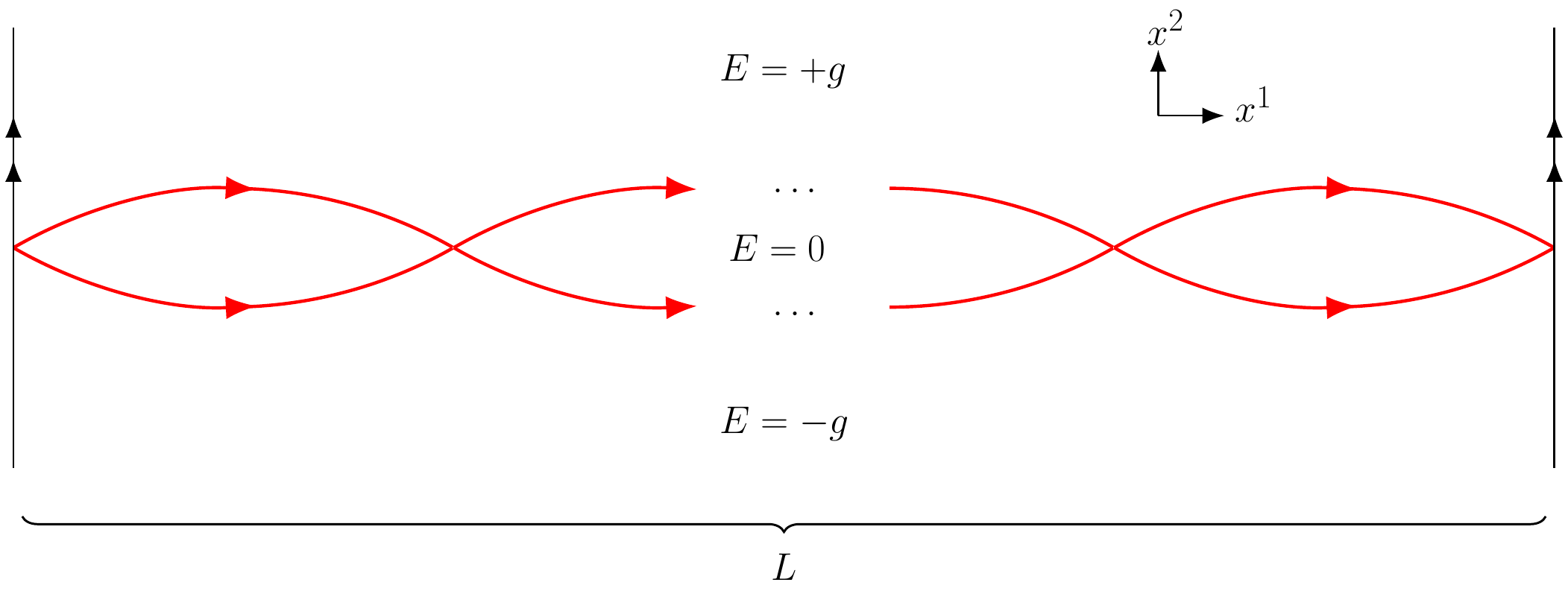}
\caption{A possible ``chain instanton" contributing to $\langle +g|-g\rangle$.}\label{lemonchain}
\end{center}
\end{figure}

\paragraph{Other instantons?\\}
The discrepancy in the pre-factor of the tunneling time scale between the numerical results and that given by the lemon instanton \eqref{lemon1} alone suggests that, among other possibilities, we may have missed some other contributing single-instanton solutions. One class of candidates is the ``chain instantons", illustrated in Figure \ref{lemonchain}. Each of them is a worldline intersecting itself some odd number $N_{\text{lemon}}$ times, generalizing the single-lemon. One can apply the one-loop calculations of the single-lemon to these ``$N_{\text{lemon}}$-chains". In the small circle limit, the area enclosed by each $N_{\text{lemon}}$-chain vanishes, and the action of each chain is given by $S_{\text{chain}}\rightarrow 2mL$, same as for the single-lemon. The one-loop pre-factor is found to be $N_{\text{lemon}}\sqrt{m/L}$, up to a numerical factor. Therefore, after including these chains, the effective action would still take the form $\Gamma_E\propto it\sqrt{m/L}e^{-2mL}$, not to mention that the sum would be divergent. In other words, this still does not agree with our numerical results. We will investigate it further in a future work.

\subsubsection{\texorpdfstring{$|k| \geq 3$}{|k|>=3}: Higher-order instantons}

\begin{figure}[H]
\begin{center}
\hspace*{2.cm}\includegraphics[width=.45\linewidth,keepaspectratio]{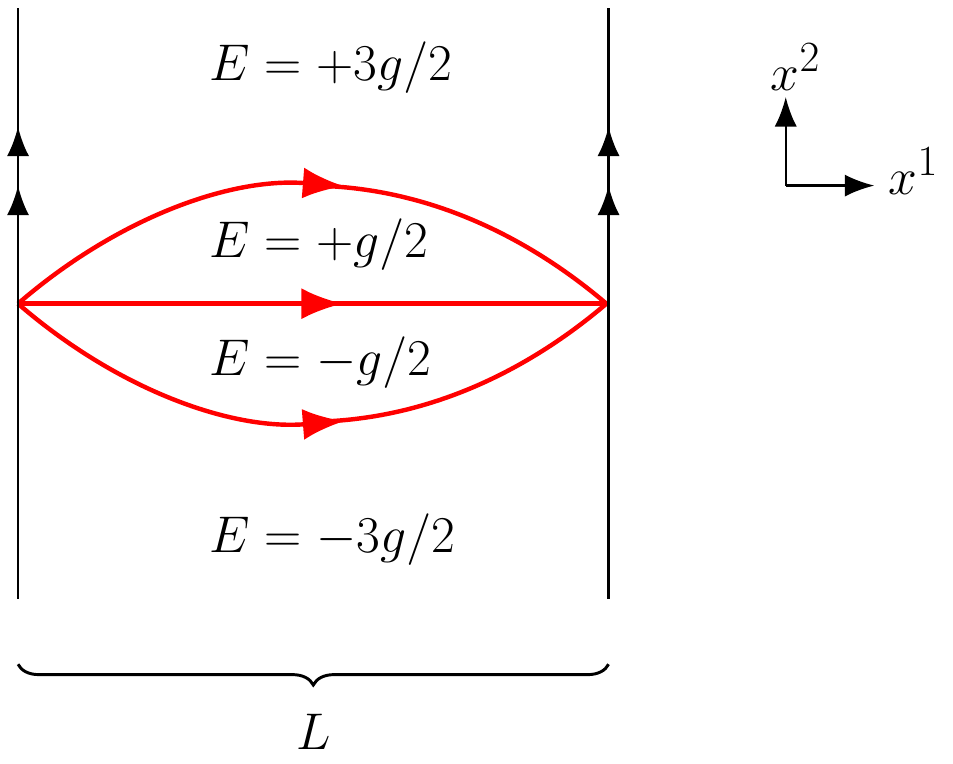}
\caption{A proposed ``lime wedge" instanton contributing to $\langle +3g/2|-3g/2\rangle$.}\label{lemonlinesingle}
\end{center}
\end{figure}

We have only discussed instantons mediating the transition amplitudes between eigenstates with $E=\pm g/2$ and $E=\pm g$ respectively. We propose that, for tunneling between higher electric field values (i.e. $|k|\geq 3$ quenched tunneling when $mL\gg 1$), the instantons can be constructed by combinations of the lemon and the straight line, in a way that obeys Gauss' law. For example, for $\langle +3g/2|-3g/2\rangle$ (cf. the top row of Figure \ref{fig: nf=3 quench_1}), we expect the ``lime wedge" instanton, illustrated in Figure \ref{lemonlinesingle}, to contribute. In the small circle limit, its action is approximately $3mL$. 

Following this pattern, we predict that, in the small circle limit, the time scale of the transition amplitude $\langle + k g/2|- k g/2\rangle$ for any positive integer $k$ has an exponential dependence given by $\exp(k mL)$.  This exponential dependence on $k$ appears to be consistent with the energy gaps computed via the lattice and numerical quantum mechanics at least up to $k=3$ (cf. Table \ref{table: c_nQM_vs_LAT_1_mog=50}).  When the circle is sufficiently small that Kaluza-Klein modes can be neglected, it  also follows from the bosonized quantum mechanical description in that there are $k$ barriers separating the  local minima corresponding to $E = \pm k g /2$ (cf. Figure \ref{fig: potential_and_wavefunction_Np=10_DW}).

\section{Summary and outlook}\label{sec6}

We have analyzed the massive Schwinger model with periodic boundary conditions using four distinct techniques, all of which agree quantitatively.  Our analysis reveals several novel features.  These results raise a number of questions for future investigations:

\begin{itemize}

\item Finding the correct method for computing the pre-factor for the novel instantons described in Section \ref{sec5} in the case $k > 1$ and including the chain instantons.

\item Exploring the relation between non-perturbative effects in the fermionic and bosonized descriptions.  Specifically, how are the (ferminonic) instantons described in Section \ref{sec5}  related to quantum mechanical tunneling in the bosonized description?

\item The quantization of the electric field in the Schwinger model  serves as a toy model for higher-dimensional theories with higher-form fluxes.  Such theories can be used to build  models of vacuum energy and inflation (for instance \cite{Bousso:2000xa, Douglas:2006es, DAmico:2012wal}).  It would be interesting to investigate the implications of our findings for these theories, and to explore the bosonized version of this quantization.

\end{itemize}


\acknowledgments 

We thank Tim Byrnes, Patrick Draper,  and Yucheng Zhang for very useful discussions, and especially Adam Brown and Lorenzo Sorbo for extensive comments on a draft.  We acknowledge the use of the NYU Prince and Greene clusters.  The work of M.K. and X.H. is partially supported by the NSF
through the grant PHY-1820814, and X.H. acknowledges support from the James Arthur Graduate Award and a Henry M. MacCracken Graduate Fellowship.

\appendix
\section{Lattice formulation of massive Schwinger model}
\label{sec: lattice}
In this appendix, we review the Kogut-Susskind approach \cite{KS1975,CKSS1975,Susskind1976} to the lattice Hamiltonian formulation of the massive Schwinger model. The lattice Hamiltonian (\ref{eq: lattice Schwinger Hamiltonian}), which plays a central role in our lattice simulation, will be derived. We also present here some facts about the lattice simulation that are omitted in the main text.

\subsection{Staggered fermions}
Naively putting spinors on a lattice results in the fermion doubling problem \cite{NIELSEN1981219}. In $1+1$ dimensions, this problem is completely resolved by the staggered fermion approach due to Kogut and Susskind \cite{KS1975}, wherein the spinor indices are identified with spatial ``indices''. That is, the two components of a local Dirac spinor are placed on adjacent spatial sites. We put the upper components on even $n$ sites while 
the lower components on odd $n$ sites, i.e.
\begin{align}
	\psi \equiv \begin{pmatrix}
		\psi_{\text{upper}} \\
		\psi_{\text{lower}}
	\end{pmatrix}
	\equiv 
	\begin{pmatrix}
		\psi_{e} \\
		\psi_{o}
	\end{pmatrix} \ ,   
\end{align}
\begin{align}
	\psi_e(n) = \chi(n) \quad \text{for } n \text{ even} \ , \quad 
	\psi_o(n) = \chi(n) \quad \text{for } n \text{ odd} \ ,
	\label{eq: staggered fermion}
\end{align}
where $\chi$ is a one-component spinor. The mass dimensions of the spinors defined above are 
$\left[\psi\right] = \left[\Dadj{\psi}\right] = \left[\psi_e\right] = \left[\psi_o\right] = \left[\chi\right] = \frac{1}{2}$. We introduce dimensionless spinor $\phi = a^{\frac{1}{2}} \chi$ for later use; it is not to be confused with the dual bosonic field in the main text.

\subsection{Discretization of the Hamiltonian}
Our goal is to obtain the discretized version of Hamiltonian (\ref{eq: hamiltonian_Schwinger}) in the temporal gauge $A_0 = 0$.

\paragraph{Fermion kinetic term\\}
The massless Dirac term in terms of staggered fermions reads
\begin{equation}
	H_D\equiv -\int \dd x\ \Dadj{\psi} i \gamma^1 \p_1\psi 
	= \int \dd x\ i \psi ^{\dagger} \gamma^0 \gamma_1 \p_1 \psi 
  = \int \dd x\ i \left(\psi^{\dagger}_{e} \p_1 \psi_{o} + \psi^{\dagger}_{o}
	\p_1 \psi_{e}\right)  \ .
  \label{eq: H_D}
\end{equation}
Here the gamma matrices in Dirac basis are given by
\begin{align}
	\gamma_0=\sigma_z=\begin{pmatrix}
		1 & 0 \\
		0 & -1
	\end{pmatrix} \ , \quad
	\gamma_1 = \begin{pmatrix}
		0 & 1 \\
		-1 & 0
	\end{pmatrix} \ , \quad
	\gamma_5=\gamma_0 \gamma_1 = \sigma_x =\begin{pmatrix}
		0 & 1 \\
		1& 0 
	\end{pmatrix} \ ,
\end{align}
with $\gamma^0=\gamma_0$ and $\gamma^1=-\gamma_1$.

\begin{figure}[th]
	\centering
    \includegraphics[width=0.8\textwidth]{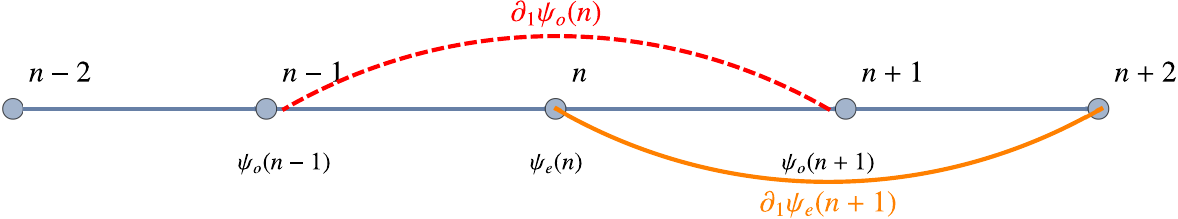}
    \includegraphics[width=0.8\textwidth]{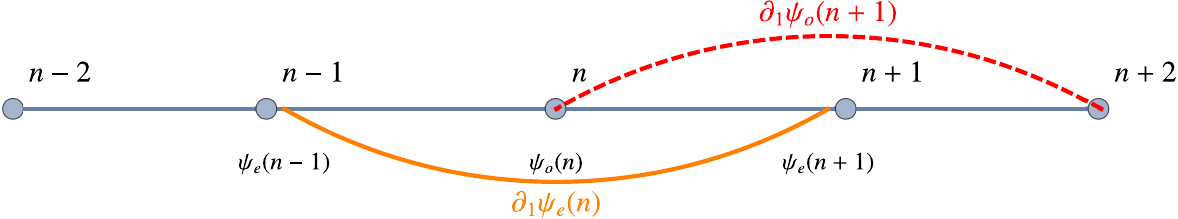}
	\caption{Spatial derivatives of one-component fermion fields on staggered lattice 
	for even $n$ (upper) and odd $n$ (lower) cases. The spatial derivatives in the figure are expressed 
	by $\p_1 \psi_o(n) \equiv \frac{\psi_o(n+1)-\psi_o(n-1)}{2a}$, 
	$\p_1 \psi_e(n+1) \equiv \frac{\psi_e(n+2)-\psi_e(n)}{2a}$, 
	$\p_1 \psi_o(n+1) \equiv \frac{\psi_o(n+2)-\psi_o(n)}{2a}$, and 
	$\p_1 \psi_e(n) \equiv \frac{\psi_e(n+1)-\psi_e(n-1)}{2a}$.}
\label{fig: spatial derivative on staggered lattice}
\end{figure}

A staggered fermion field is well-defined locally on a single lattice site whereas its spatial derivative contains 
information about the neighborhood of the site it sits on. Since the fermion kinetic term $H_D$ contains 
an ``interaction'' between the local field and its derivative $\psi^{\dagger} \p_1 \psi$, we need to specify where this ``interaction''
happens. 
It is convenient to make it happen on the same site $\psi^{\dagger}$ sits on. For instance,
$\psi_e^{\dagger}\p_1 \psi_o (n) \equiv \psi_e^{\dagger} (n) \p_1 \psi_o(n)$ is defined for even $n$,
with the spatial derivative of $\psi_o$ on even $n$ lattice given by \footnote{We stress that 
for even $n$, $\psi_e(n),\ \psi_o(n-1)$ and $\psi_o(n+1)$ are well-defined according to 
(\ref{eq: staggered fermion}). In contrast, $\psi_o(n)$, $\psi_e(n-1)$ and $\psi_e(n+1)$ are 
meaningless for even $n$.}
\begin{align}
	\p_1 \psi_o(n) \approx \frac{\psi_o (n+1)-\psi_o (n-1)}{2a}.
	\label{eq: ordinary derivative_disc}
\end{align}
For odd $n$, $\psi^{\dagger}_o \p_1 \psi_e(n) \equiv \psi^{\dagger}_o(n) \p_1 \psi_e(n)$ 
can be defined in the same manner. Diagrammatically illustrated in 
Figure \ref{fig: spatial derivative on staggered lattice} is how spatial derivatives on staggered lattice are defined.
The fermion field kinetic term (\ref{eq: H_D}) can then be written as ``hopping'' terms,
\begin{align}
	&i \left(\psi^{\dagger}_{e} \p_1 \psi_{o} + \psi^{\dagger}_{o} \p_1 \psi_{e}\right) \notag
	\\
	&\equiv
	\begin{cases}
		i \left[\psi^{\dagger}_{e} \p_1 \psi_{o} (n) + \psi^{\dagger}_{o} \p_1 \psi_{e}(n+1)\right]  & \mbox{if $n$ even} \\
		i \left[\psi^{\dagger}_{e} \p_1 \psi_{o} (n+1) + \psi^{\dagger}_{o} \p_1 \psi_{e}(n)\right] & \mbox{if $n$ odd}
	\end{cases}
	\notag
	\\
	&= 
	\begin{cases}
		\frac{i}{2a} \left[\chi^{\dagger}(n)\left(\chi(n+1)-\chi(n-1)\right) 
			+ \chi^{\dagger} (n+1) \left(\chi(n+2)-\chi(n)\right)\right] & \mbox{if $n$ even} \\
		\frac{i}{2a} \left[\chi^{\dagger}(n+1)\left(\chi(n+2)-\chi(n)\right) 
			+ \chi^{\dagger} (n) \left(\chi(n+1)-\chi(n-1)\right)\right] & \mbox{if $n$ odd}
	\end{cases}
	\label{eq: H_D_hopping}
\end{align}
On the lattice, the integral is translated into a summation of operators on all sites as \footnote{The summation runs over either all even $n$ sites or all odd $n$ sites, depending on which expression in (\ref{eq: H_D_hopping}) is used. It is easy to see that these two summations are equivalent by manipulating the dummy index.}
\begin{align}
	H_D \to&\ \frac{i}{2a} \cdot a \sum_{n\ \text {even}} \left[\chi^{\dagger}(n)\left(\chi(n+1)-\chi(n-1)\right) 
	+ \chi^{\dagger} (n+1) \left(\chi(n+2)-\chi(n)\right)\right] \notag \\
	=&\ \frac{i}{2} \left[\sum_{n\ \text{even}} \chi^{\dagger}(n) \chi(n+1) 
	-\sum_{n'\ \text{odd}} \chi^{\dagger}(n'+1)\chi(n') \right.\nonumber\\
	&\  \qquad\qquad \qquad\qquad  \left.
	+ \sum_{n''\ \text{odd}} \chi^{\dagger}(n'') \chi(n''+1) 
	-\sum_{n\ \text{even}} \chi^{\dagger}(n+1) \chi(n)\right] \notag \\
	=&\ \frac{i}{2} \sum_n \left[\chi^{\dagger}(n)\chi(n+1) - \chi^{\dagger}(n+1) \chi(n)\right] \notag \\
	=&\ \frac{i}{2a} \sum_n \left[\phi^{\dagger} (n) \phi(n+1) - \phi^{\dagger}(n+1)\phi(n)\right] \ .
	\label{eq: massless Dirac}
\end{align}
In the second line we redefine the dummy indices $n'\equiv n-1$ and $n''\equiv n+1$. In the last line the fermion field $\chi$ is transferred to the one-component dimensionless fermion field $\phi$.

\paragraph{Fermion mass term\\}
The discretization of the fermion mass term is
\begin{align}
	H_m\equiv &\int \dd x\ m \Dadj{\psi} \psi = \int \dd x\ m \psi^{\dagger} \gamma^0 \psi 
	= \int \dd x\ m\left[\psi^{\dagger}_{e} \psi_{e} 
	- \psi^{\dagger}_{o} \psi_{o} \right] \notag \\
	\to&\ a \sum_{n\ \text{even}} m \left[\chi^{\dagger}(n)\chi(n) 
	- \chi^{\dagger}(n+1) \chi(n+1)\right] \notag\\
	=&\ a\sum_n m \left[(-1)^n \chi^{\dagger}(n) \psi(n)\right] \notag \\
	=&\ \sum_n m \left[(-1)^n \phi^{\dagger}(n) \phi(n)\right] \ .
\end{align}

\paragraph{Gauge field kinetic term\\}
In temporal gauge $A^0 =0$, the field equation for gauge field is 
$-\dot{A^1}=E$, which indicates that the electric field $E$ is canonically conjugate to $A_1(n) \equiv A(n)$.
On the lattice this commutation relation is $\left[A(n), E(m)\right]=i \frac{1}{a} \delta_{n,m}$. In terms of the dimensionless link phase $\theta(n)\equiv a g A_1(n)$ and the dimensionless lattice electric field $L(n)$, the commutation relation reduces to $\left[\theta(n), L(m)\right] = i \delta_{n,m}$. The most general operator $L(n)$ takes the form of $L(n)=\frac{E(n)}{g}-\frac{F}{g} \equiv \frac{E(n)}{g}-\alpha$ 
where $F$ is a constant background field.

The canonical commutation relation immediately implies that $e^{\pm i \theta(n)}$ is a $L$-shift operator on the $n$-th link, 
\begin{align}
	e^{\pm i \theta(n)} \ket{l(n)} = \ket{l(n)\pm 1} \ ,
	\label{eq: L_ladder}
\end{align}
where $\ket{l(n)}$ is the eigenstate of $L(n)$ with eigenvalue $l(n)$. As seen from \eqref{eq: bosonized action},   physics is periodic in $\theta \equiv 2\pi \alpha$ with period $2\pi$. Without loss of generality, we take $\alpha \in [0,1)$ such that $L(n)$ is an integer on the $n$-th link for a certain $n$. By Gauss' law (\ref{eq: lattice Gauss' law}), $L(n)$ for all $n$ are automatically integers.

The gauge kinetic term is then discretized in terms of $L(n)$ and constant $\alpha$ as 
\begin{align}
	H_G \equiv \int \dd x\ \frac{1}{2} E^2 \to a \sum_n \frac{1}{2} E^2(n) = \frac{1}{2} a g^2 \sum_n \left[L(n)+ \alpha\right]^2 \ .
\end{align}

\begin{figure}[ht]
	\centering
    \includegraphics[width=0.8\textwidth]{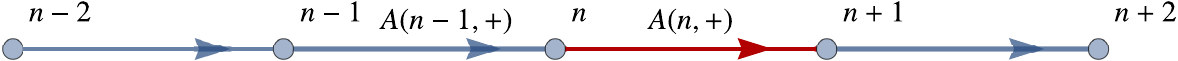}
    \includegraphics[width=0.8\textwidth]{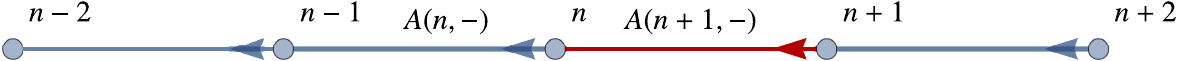}
	\caption{Gauge field $A(n,+)\equiv A(n)\equiv A_1(n)$ living between sites is specified by its starting site and its spatial direction.}
\label{fig: gauge field on staggered lattice}
\end{figure}

\paragraph{Gauge invariant interaction term\\}
The gauge field defined on a lattice lives on the links connecting adjacent sites and has certain direction on each link. The gauge field 
``starting'' from $n$-th site and pointing to $(n+1)$-th site is defined as 
\begin{align}
	A_1(n) \equiv A(n) \equiv A(n,+) \equiv A(n,n+1) \ .
\end{align}
A consistency condition for the gauge field living between $n$-th and $(n+1)$-th sites is 
$A(n,n+1)\equiv A(n,+)=-A(n+1,-) \equiv - A(n+1,n)$.
Figure \ref{fig: gauge field on staggered lattice} illustrates the definition of gauge field between sites.

One can further define the Wilson line (or the \emph{link field}) from $n$-th to $(n+1)$-th sites
on one dimensional spatial lattice is related to the gauge field $A_1$ as 
\begin{align}
	U(n,+)\equiv U(n,n+1)\equiv e^{i a g A(n,n+1)} = e^{i a g A(n,+)}  = e^{i a g A(n)} \equiv e^{i\theta(n)} \ ,
\end{align}
and the link from $n$-th to $(n-1)$-th sites is
\begin{align}
	U(n,-)&\equiv U(n,n-1)\equiv e^{i a g A(n,n-1)} = e^{i a g A(n,-)} \notag \\
	&= e^{-i a g A(n-1,+)} =e^{-i a g A(n-1)} = e^{-i\theta(n-1)} 
	=U^{\dagger}(n-1,+)\ .
\end{align}
To incorporate the gauge invariant 
interaction term $H_{\text{int}} = \int \dd x\ g \Dadj{\psi} \gamma^1 A_1 \psi$ into the full Hamiltonian, one can simply promote the ordinary derivative to the covariant derivative. This turns the fermion kinetic term $H_D$ to the covariant kinetic term $H_D+H_{\text{int}}$.

The covariant derivative of $\psi_o$ on even $n$ sites is defined by 
	\begin{align}
		D_1 \psi_o (n) = \frac{U(n,n+1)\psi_o(n+1)-U(n,n-1)\psi_o(n-1)}{2a} \ ,
		\label{eq: convariant derivative_disc}
	\end{align}
	which is generalized from (\ref{eq: ordinary derivative_disc}). The covariant 
	derivative acting on $\psi_e$ is obtained accordingly. 
	Substitution $\p_{\mu} \to D_{\mu}$ promotes (\ref{eq: H_D}) to 
	\begin{align}
		H_D+H_{\text{int}}
		&= \int \dd x\ i \left(\psi^{\dagger}_{e} D_1 \psi_{o} + \psi^{\dagger}_{o} D_1 \psi_{e}\right) \ .
		\label{eq: H_D+H_int_1}
	\end{align}

The covariant kinetic term becomes dressed ``hopping'' terms
\begin{align}
	i \left(\psi^{\dagger}_{e} D_1 \psi_{o} + \psi^{\dagger}_{o} D_1 \psi_{e}\right)
	\equiv \begin{cases}
		i \left[\psi^{\dagger}_{e} D_1 \psi_{o} (n) + \psi^{\dagger}_{o} D_1 \psi_{e}(n+1)\right] & \mbox{if $n$ even} \\
		i \left[\psi^{\dagger}_{e} D_1 \psi_{o} (n+1) + \psi^{\dagger}_{o} D_1 \psi_{e}(n)\right] & \mbox{if $n$ odd}
	\end{cases}  
\end{align}
where
\begin{itemize}
	\item for even $n$,
		\begin{align}
			&i \left[\psi^{\dagger}_{e} D_1 \psi_{o} (n) + \psi^{\dagger}_{o} D_1 \psi_{e}(n+1)\right]
			\notag \\
			=&\frac{i}{2a} \left[\chi^{\dagger}(n) e^{i\theta(n)} \chi(n+1)
			-\chi^{\dagger}(n) e^{-i \theta(n-1)} \chi(n-1)\right. \notag \\ 
			&\left.+\chi^{\dagger}(n+1) e^{i\theta(n+1)} \chi(n+2) 
			- \chi^{\dagger}(n+1) e^{-i \theta(n)} \chi(n) \right] \ ,
		\end{align}
	\item for odd $n$,
		\begin{align}
			&i \left[\psi^{\dagger}_{e} D_1 \psi_{o} (n+1) + \psi^{\dagger}_{o} D_1 \psi_{e}(n)\right]
			\notag \\
			=&\frac{i}{2a} \left[\chi^{\dagger}(n+1) e^{i\theta(n+1)} \chi(n+2) 
			- \chi^{\dagger}(n+1) e^{-i \theta(n)}\chi(n) \right. \notag \\
			& \left. \chi^{\dagger}(n) e^{i\theta(n)} \chi(n+1)
			- \chi^{\dagger}(n) e^{-i \theta(n-1)}\chi(n-1)\right] \ .
		\end{align}
\end{itemize}
The discretization of the covariant term is derived as
\begin{align}
	H_D+H_{\text{int}} \to& \frac{i}{2a} \cdot a  \sum_{n\ \text{even}}
	\left[\chi^{\dagger}(n) e^{i\theta(n)} \chi(n+1)
		-\chi^{\dagger}(n) e^{-i \theta(n-1)} \chi(n-1)\right. \notag \\ 
		&\left.+\chi^{\dagger}(n+1) e^{i\theta(n+1)} \chi(n+2) 
		- \chi^{\dagger}(n+1) e^{-i \theta(n)} \chi(n) \right] \notag \\
	=& \frac{i}{2} \sum_n \left[\chi^{\dagger}(n) e^{i\theta(n)} \chi(n+1)
	-\chi^{\dagger}(n+1)e^{-i \theta(n)} \chi(n)\right] \notag \\
	=& \frac{i}{2a} \sum_n \left[\phi^{\dagger}(n) e^{i\theta(n)} \phi(n+1) 
	-\phi^{\dagger}(n+1) e^{-i\theta(n)}\phi(n)\right]\ ,
\end{align}

\paragraph{Full lattice Hamiltonian\\}
Putting everything together, the full Schwinger Hamiltonian (\ref{eq: hamiltonian_Schwinger}) is discretized as 
\begin{align}
	H_{\text{lat}} =& H_D+H_{\text{int}} + H_m + H_G \notag \\
	=&  \frac{i}{2a} \sum_n \left[\phi^{\dagger}(n) e^{i\theta(n)} \phi(n+1) 
	-\phi^{\dagger}(n+1) e^{-i\theta(n)}\phi(n)\right] 	\notag \\
	&+ \sum_n m \left[(-1)^n \phi^{\dagger}(n) \phi(n)\right]
	+\frac{1}{2} a g^2 \sum_n \left[L(n)+ \alpha\right]^2 \ .
	\label{eq: lattice Schwinger model}
\end{align}

\subsection{Jordan-Wigner transformation}
In 1+1 dimensions, fermionic degrees of freedom can be mapped onto spin degrees of freedom via Jordan-Wigner transformation which is explicitly given by 
\begin{align}
	\phi(n)&=\prod_{l<n} \left[i \sigma_3(l)\right] \sigma^-(n)
	=\exp[i \frac{\pi}{2}\sum_{l<n} \sigma_3(l)] \sigma^-(n) \ ,
	\label{eq: J-W_1}
	\\
	\phi^{\dagger}(n)&= \prod_{l<n} \left[-i \sigma_3(l)\right] \sigma^+(n)
	=\exp[-i \frac{\pi}{2}\sum_{l<n} \sigma_3(l)] \sigma^+(n) \ ,
	\label{eq: J-W_2}
\end{align}
where $\sigma_3(n)$, 
$\sigma^{\pm}(n)\equiv \frac{1}{2}\left(\sigma_1(n)\pm i \sigma_2(n)\right)$ are Pauli operators acting on $n$-th site. This map faithfully captures the internal dynamical structure of a spinor in the sense that it preserves the canonical anti-commutation relations $\{\phi^{\dagger}(m),\phi(n)\}=\delta_{m,n}$, 
$\{\phi(m),\phi(n)\}=0$ and $\{\phi^{\dagger}(m),\phi^{\dagger}(n)\}=0$.

Applying the Jordan-Wigner transformation (\ref{eq: J-W_1}) and (\ref{eq: J-W_2}) to the lattice Schwinger model, we obtain
\begin{align}
	\phi^{\dagger}(n) \phi(n+1) 
	&= -i \sigma^+(n) \sigma^-(n+1) \ , \\
	\phi^{\dagger}(n+1) \phi(n)
	&= i \sigma^+(n+1) \sigma^-(n) \ , \\
	(-1)^n \phi^{\dagger}(n) \phi(n)
	&= (-1)^n \sigma^+(n) \sigma^-(n) =(-1)^n\frac{1}{2} \left[1+\sigma_3(n)\right] \ .
\end{align}
In terms of Pauli operators and other bosonic operators, the lattice Schwinger Hamiltonian (\ref{eq: lattice Schwinger model}) is given by
\begin{align}
	H_{\text{lat}}=&\frac{1}{2a} \sum_n \left[\sigma^+(n) e^{i\theta(n)} \sigma^-(n+1) 
	+ \sigma^+(n+1) e^{-i \theta(n)} \sigma^-(n)\right] \notag \\
	&+\sum_n (-1)^n \frac{m}{2} \left[1+\sigma_3(n)\right]
	+\frac{1}{2} a g^2 \sum_n \left[L(n)+ \alpha\right]^2 \ ,
\end{align}
which is eq.(\ref{eq: lattice Schwinger Hamiltonian}) in the main text.

\section{The cutoff \texorpdfstring{$l_{\text{max}}$}{lmax} and the continuum extrapolation of energy levels}
\label{sec: lmax and extrapolation}
In our lattice simulation implementation, the dimension of the lattice Hilbert space is determined by the number of lattice sites $N$ and the electric field cutoff $l_{\text{max}}$. An excessive $l_{\text{max}}$ can be computationally expensive, while one that is too small could lead to physical effects being missed out from simulation results. Throughout this work, we use ``sufficiently large'' values of $l_{\text{max}}$ that achieve error $\lesssim \Op(10^{-10})$ for the lowest four energy levels $\EE/g$. More explicitly, our choices of $l_{\text{max}}$ for the entire parameter regime of $1\leq m/g\leq 50$ and $1\leq mL\leq 8$ are
\begin{align}
  l_{\text{max}}=
  \begin{cases}
    30, & \text{if }30 < m/g \leq 50,\ 1\leq mL<2 \\
    25, & \text{if }10 < m/g \leq 30,\ 1\leq mL<2 \\
    20, & \text{if }30 < m/g \leq 50,\ 2\leq mL<4 \\
    15, & \text{if }10 < m/g \leq 30,\ 2\leq mL<4\ \text{or}\ 4 < m/g \leq 10,\ 1\leq mL<2\\
    10, & \text{otherwise}
  \end{cases} 
\end{align}
As briefly mentioned in Section \ref{sec2}, we perform the finite volume continuum extrapolation with $\frac{N}{\sqrt{x}}$ fixed and $N\to \infty$ to extract physical quantities which are independent of lattice parameters. Among the important quantities is the energy gap $\Delta \EE_{i,i-1}/g$. We fit the lattice results of the energy difference $\Delta \EE_{i,i-1}/g(m/g,mL;N)$ to 
\begin{align}
  \frac{\Delta \EE_{i,i-1}}{g}(m/g,mL;N) = \frac{\Delta \EE_{i,i-1}^{[N\to \infty]}}{g} +\beta N^{-\xi}
\end{align}
using the functions \texttt{FindFit} and \texttt{NonlinearModelFit} in \textit{Mathematica}. The fitting parameters are $\beta$ and $\xi$. And $\Delta \EE_{i,i-1}^{[N\to \infty]}/g$ is the extrapolated value of the energy difference between the $i$-th and $(i-1)$-th energy eigenstates.

For $10\leq m/g\leq 50$ and $4\leq mL\leq 8$, we use $N=8,10,12,14,16$ data to carry out the extrapolation fits while for other parameter regimes, we use $N=8,10,12,14$ data.

\section{More about the prefactor \texorpdfstring{$c$}{c}}\label{sec: prefactor}
In this appendix, we present more details about the prefactor $c$ shown up in the bosonized action (\ref{eq: bosonized action}). We determine the physical values of $c$ via the continuum extrapolation of the lattice data of $c$. Its dependence on the mass-coupling ratio $m/g$ and the size of the spatial circle $mL$ is then investigated.

As proposed in Section \ref{sec4}, we find the optimal $c$ up to $\Op(0.01)$ such that the numerical bosonized quantum mechanics generates the closest value of the ground energy difference $\Delta \EE_{1,0}/g$ for $\alpha=0.5$ produced by lattice simulation for given $m/g$, $mL$ and $N$. The prefactor $c$ determined this way are denoted by $c(m/g,mL;N)$.
In Tables \ref{table: summary_c_part1} -- \ref{table: summary_c_part4}, we summarize values of $c(m/g,mL;N)$ for $1\leq m/g\leq 50$, $1\leq mL\leq 8$ and various lattice sites $N$.

In order to get the physical values of $c$ which do not depend on $N$, we perform the finite volume continuum extrapolation by fitting $c(m/g,mL;N)$ to
\begin{align}
	c(m/g,mL;N) = c^{[N\to\infty]}(m/g,mL) - v N^{-u}
	\label{eq: cc_largeN_extrapolation}
\end{align}
for fixed $m/g$ and $mL$ and get the fitting parameters $c^{[N\to\infty]}(m/g,mL)$, $v$ and $u$ by using the functions \texttt{FindFit} and \texttt{NonlinearModelFit} in \textit{Mathematica}.
The data of $c(m/g,mL;N)$ fit the extrapolation model (\ref{eq: cc_largeN_extrapolation}) quite well, indicating that $c(m/g,mL;N)$ converges to $c^{[N\to\infty]}(m/g,mL)$ as $N\to \infty$. The values of $c^{[N\to\infty]}(m/g,mL)$ are summarized in the last columns of Tables \ref{table: summary_c_part1} -- \ref{table: summary_c_part4}.

\begin{figure}[ht]
	\centering
		\includegraphics[width=1\textwidth]{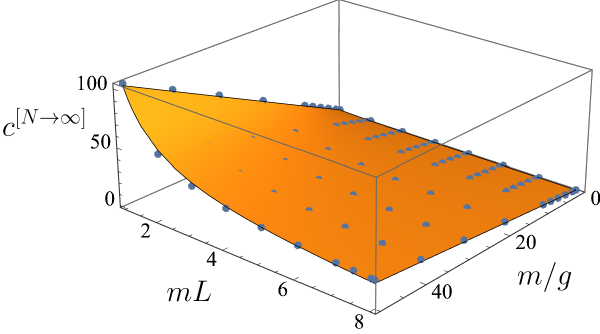}
		\includegraphics[width=1\textwidth]{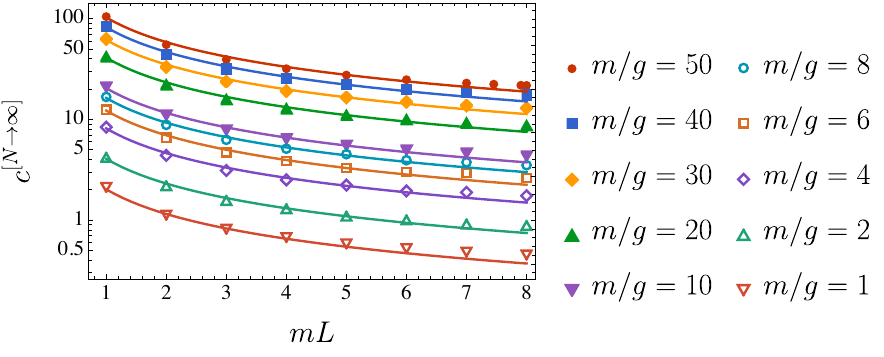}
		\includegraphics[width=1\textwidth]{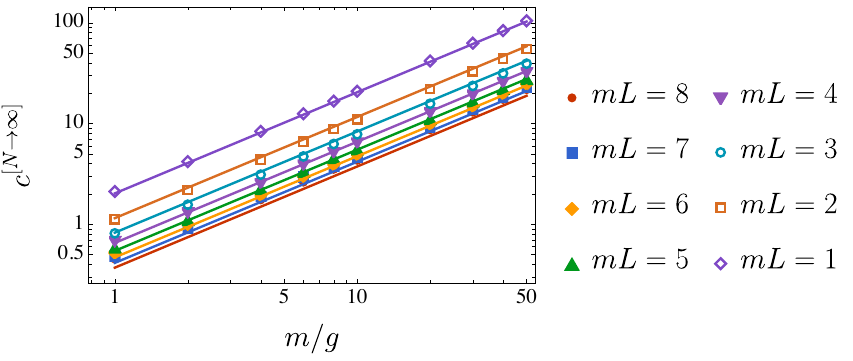}
	\caption{The continuum extrapolated prefactor $c^{[N\to \infty]}$ as a function of $m/g$ and $mL$. Top panel: the blue dots are the continuum extrapolated prefactors $c^{[N\to \infty]}$ summarized in Tables \ref{table: summary_c_part1} -- \ref{table: summary_c_part4} while the orange surface is the function (\ref{eq: c_dependence_on_m/g_mL}) with the best fit (\ref{eq: best fit}). Middle panel: the dependence of $c^{[N\to \infty]}$ on $mL$ for various fixed $m/g$ values.
	Bottom panel: the dependence of $c^{[N\to \infty]}$ on $m/g$ for various fixed $mL$ values.
	}
\label{fig: prefactor_c_mog_mL}
\end{figure}

To study the dependence of $c$ on the mass-coupling ratio $m/g$ and the size of the spatial circle $mL$, we further fit $c^{[N\to\infty]}(m/g,mL)$ to the model
\begin{align}
		c^{[N\to\infty]}(m/g,mL) = \eta \cdot \left(\frac{m}{g}\right)^{\gamma} \cdot (mL)^{\delta}  \ ,
		\label{eq: c_dependence_on_m/g_mL}
\end{align}
using the functions \texttt{FindFit} and \texttt{NonlinearModelFit} in \textit{Mathematica}. The best fit is found to be
\begin{align}
	\eta \approx 2.0187\ , \quad
	\gamma \approx 1.00447 \ , \quad
	\delta \approx -0.816478 \ .
	\label{eq: best fit}
\end{align}
In Figure \ref{fig: prefactor_c_mog_mL}, we plot the extrapolated prefactor $c^{[N\to \infty]}$ as a function of $m/g$ and $mL$.

\begin{landscape}
\begin{table}[ht]
	\centering
	\begin{tabular}{cc|ccccc|cccccccc}
	\hline
		$\frac{m}{g}$ & $mL$ & $c(m/g,mL;N=8)$ &  
		$c(m/g,mL;N=10)$ & $c(m/g,mL;N=12)$ & $c(m/g,mL;N=14)$ & $c(m/g,mL;N=16)$ & $c^{[N\to \infty]}$
	\\ 
	\hline
		50 & 8 & 16.20 & 17.51 & 18.36 & 18.93 & 19.34 & 21.65
	\\
		50 & 7.90569 & 16.37 & 17.67 & 18.51 & 19.08 & 19.48 & 21.75
	\\
		50 & 7.45356 & 17.21 & 18.49 & 19.30 & 19.85 & 20.23 & 22.27
	\\
		50 & 7 & 18.15 & 19.39 & 20.18 & 20.70 & 21.05 & 22.87
	\\
		50 & 6 & 20.64 & 21.80 & 22.50 & 22.96 & 23.27 & 24.68
	\\
		50 & 5 & 23.98 & 25.03 & 25.66 & 26.06 & 26.32 & 27.44
	\\
		50 & 4 & 28.79 & 29.75 & 30.30 & 30.64 & 30.87 & 31.73
	\\
		50 & 3 & 36.59 & 37.50 & 38.00 & 38.31 & -- & 39.21
	\\
		50 & 2 & 51.96 & 53.03 & 53.61 & 53.97 & -- & 54.98
	\\
		50 & 1 & 95.18 & 98.61 & 100.45 & 101.55 & -- & 104.49
	\\
	\hline
	\hline
		40 & 8 & 12.96 & 14.01 & 14.69 & 15.15 & 15.47 & 17.28
	\\
		40 & 7 & 14.52 & 15.52 & 16.14 & 16.56 & 16.84 & 18.26
	\\
		40 & 6 & 16.52 & 17.44 & 18.00 & 18.37 & 18.62 & 19.79
	\\
		40 & 5 & 19.18 & 20.03 & 20.53 & 20.85 & 21.06 & 21.93
	\\
		40 & 4 & 23.03 & 23.80 & 24.24 & 24.51 & 24.70 & 25.39
	\\
		40 & 3 & 29.28 & 30.00 & 30.40 & 30.65 & -- & 31.40
	\\
		40 & 2 & 41.57 & 42.43 & 42.89 & 43.17 & -- & 43.92
	\\
		40 & 1 & 76.16 & 78.90 & 80.36 & 81.23 & -- & 83.52
	\\
	\hline
	\hline
		30 & 8 & 9.72 & 10.51 & 11.02 & 11.36 & 11.60 & 12.92
	\\
		30 & 7 & 10.89 & 11.64 & 12.11 & 12.42 & 12.63 & 13.68
	\\
		30 & 6 & 12.39 & 13.08 & 13.50 & 13.78 & 13.97 & 14.87
	\\
		30 & 5 & 14.39 & 15.02 & 15.40 & 15.63 & 15.80 & 16.48
	\\
		30 & 4 & 17.28 & 17.85 & 18.18 & 18.38 & 18.52 & 19.04
	\\
		30 & 3 & 21.96 & 22.50 & 22.80 & 22.99 & -- & 23.56
	\\
		30 & 2 & 31.19 & 31.82 & 32.17 & 32.38 & -- & 32.99
	\\
		30 & 1 & 57.14 & 59.19 & 60.28 & 60.93 & -- & 62.63
	\\
	\hline
	\end{tabular}
	\caption{The prefactors $c(m/g,mL;N)$ determined based on the numerical bosonized quantum mechanics and the $N$-site lattice calculation for $m/g=30,40,50$ and $1\leq mL\leq 8$, and the continuum extrapolated values $c^{[N\to \infty]}$ obtained from fitting data to (\ref{eq: cc_largeN_extrapolation}).}
	\label{table: summary_c_part1}
\end{table}

\begin{table}[ht]
	\centering
	\begin{tabular}{cc|ccccc|cccccccc}
	\hline
		$\frac{m}{g}$ & $mL$ & $c(m/g,mL;N=8)$ &  
		$c(m/g,mL;N=10)$ & $c(m/g,mL;N=12)$ & $c(m/g,mL;N=14)$ & $c(m/g,mL;N=16)$ & $c^{[N\to \infty]}$
	\\ 
	\hline
		20 & 8 & 6.48 & 7.01 & 7.35 & 7.58 & 7.74 & 8.63
	\\
		20 & 7 & 7.27 & 7.76 & 8.07 & 8.28 & 8.42 & 9.17
	\\
		20 & 6 & 8.26 & 8.72 & 9.00 & 9.19 & 9.31 & 9.90
	\\
		20 & 5 & 9.60 & 10.02 & 10.27 & 10.42 & 10.53 & 10.95
	\\
		20 & 4 & 11.52 & 11.90 & 12.12 & 12.26 & 12.35 & 12.71
	\\
		20 & 3 & 14.65 & 15.00 & 15.20 & 15.32 & -- & 15.69
	\\
		20 & 2 & 20.80 & 21.22 & 21.45 & 21.59 & -- & 21.99
	\\
		20 & 1 & 38.13 & 39.48 & 40.20 & 40.62 & -- & 41.72
	\\
	\hline
	\hline
		10 & 8 & 3.25 & 3.51 & 3.68 & 3.79 & 3.87 & 4.30
	\\
		10 & 7 & 3.64 & 3.89 & 4.04 & 4.15 & 4.22 & 4.60
	\\
		10 & 6 & 4.14 & 4.37 & 4.51 & 4.60 & 4.66 & 4.93
	\\
		10 & 5 & 4.80 & 5.01 & 5.14 & 5.22 & 5.27 & 5.49
	\\
		10 & 4 & 5.77 & 5.96 & 6.06 & 6.13 & 6.18 & 6.36
	\\
		10 & 3 & 7.33 & 7.51 & 7.60 & 7.66 & -- & 7.80
	\\
		10 & 2 & 10.41 & 10.62 & 10.73 & 10.79 & -- & 10.94
	\\
		10 & 1 & 19.12 & 19.77 & 20.11 & 20.32 & -- & 20.86
	\\
	\hline
	\hline
		8 & 8 & 2.60 & 2.81 & 2.95 & 3.04 & -- & 3.50
	\\
		8 & 7 & 2.92 & 3.11 & 3.24 & 3.32 & -- & 3.74
	\\
		8 & 6 & 3.31 & 3.50 & 3.61 & 3.68 & -- & 3.92
	\\
		8 & 5 & 3.85 & 4.01 & 4.11 & 4.18 & -- & 4.50
	\\
		8 & 4 & 4.62 & 4.77 & 4.85 & 4.91 & -- & 5.10
	\\
		8 & 3 & 5.87 & 6.01 & 6.09 & 6.13 & -- & 6.25
	\\
		8 & 2 & 8.34 & 8.50 & 8.58 & 8.64 & -- & 8.80
	\\
		8 & 1 & 15.31 & 15.83 & 16.10 & 16.26 & -- & 16.66
	\\
	\hline
	\end{tabular}
	\caption{The prefactors $c(m/g,mL;N)$ determined based on the numerical bosonized quantum mechanics and the $N$-site lattice calculation for $m/g=8,10,20$ and $1\leq mL\leq 8$, and the continuum extrapolated values $c^{[N\to \infty]}$ obtained from fitting data to (\ref{eq: cc_largeN_extrapolation}).}
	\label{table: summary_c_part2}
\end{table}

\begin{table}[ht]
	\centering
	\begin{tabular}{cc|ccccc|cccccccc}
	\hline
		$\frac{m}{g}$ & $mL$ & $c(m/g,mL;N=8)$ &  
		$c(m/g,mL;N=10)$ & $c(m/g,mL;N=12)$ & $c(m/g,mL;N=14)$ & $c(m/g,mL;N=16)$ & $c^{[N\to \infty]}$
	\\ 
	\hline
		6 & 8 & 1.96 & 2.12 & 2.22 & 2.29 & -- & 2.61
	\\
		6 & 7 & 2.20 & 2.34 & 2.43 & 2.50 & -- & 2.93
	\\
		6 & 6 & 2.49 & 2.63 & 2.71 & 2.77 & -- & 3.00
	\\
		6 & 5 & 2.89 & 3.02 & 3.09 & 3.14 & -- & 3.29
	\\
		6 & 4 & 3.47 & 3.58 & 3.65 & 3.69 & -- & 3.85
	\\
		6 & 3 & 4.41 & 4.51 & 4.57 & 4.60 & -- & 4.70
	\\
		6 & 2 & 6.26 & 6.38 & 6.44 & 6.48 & -- & 6.58
	\\
		6 & 1 & 11.51 & 11.89 & 12.08 & 12.20 & -- & 12.48
	\\
	\hline
	\hline 
		4 & 8 & 1.32 & 1.42 & 1.49 & 1.53 & -- & 1.74 
	\\
		4 & 7 & 1.48 & 1.57 & 1.63 & 1.67 & -- & 1.88
	\\
		4 & 6 & 1.68 & 1.77 & 1.82 & 1.85 & -- & 1.94
	\\
		4 & 5 & 1.94 & 2.02 & 2.07 & 2.10 & -- & 2.22 
	\\
		4 & 4 & 2.32 & 2.40 & 2.44 & 2.46 & -- & 2.50
	\\
		4 & 3 & 2.95 & 3.02 & 3.05 & 3.07 & -- & 3.10
	\\
		4 & 2 & 4.19 & 4.26 & 4.30 & 4.32 & -- & 4.38
	\\
		4 & 1 & 7.71 & 7.94 & 8.07 & 8.14 & -- & 8.34
	\\
	\hline
	\hline
		2 & 8 & 0.70 & 0.75 & 0.78 & 0.80 & -- & 0.88
	\\
		2 & 7 & 0.78 & 0.82 & 0.85 & 0.86 & -- & 0.91
	\\
		2 & 6 & 0.87 & 0.91 & 0.94 & 0.95 & -- & 1.00
	\\
		2 & 5 & 1.00 & 1.04 & 1.06 & 1.07 & -- & 1.09
	\\
		2 & 4 & 1.19 & 1.22 & 1.24 & 1.25 & -- & 1.29
	\\
		2 & 3 & 1.50 & 1.53 & 1.55 & 1.55 & -- & 1.56
	\\
		2 & 2 & 2.12 & 2.15 & 2.17 & 2.17 & -- & 2.18
	\\
		2 & 1 & 3.92 & 4.01 & 4.06 & 4.09 & -- & 4.18
	\\
	\hline
	\end{tabular}
	\caption{The prefactors $c(m/g,mL;N)$ determined based on the numerical bosonized quantum mechanics and the $N$-site lattice calculation for $m/g=2,4,6$ and $1\leq mL\leq 8$, and the continuum extrapolated values $c^{[N\to \infty]}$ obtained from fitting data to (\ref{eq: cc_largeN_extrapolation}).}
	\label{table: summary_c_part3}
\end{table}

\begin{table}[ht]
	\centering
	\begin{tabular}{cc|ccccc|cccccccc}
	\hline
		$\frac{m}{g}$ & $mL$ & $c(m/g,mL;N=8)$ &  
		$c(m/g,mL;N=10)$ & $c(m/g,mL;N=12)$ & $c(m/g,mL;N=14)$ & $c(m/g,mL;N=16)$ & $c^{[N\to \infty]}$
	\\ 
	\hline
		1 & 8 & 0.42 & 0.44 & 0.45 & 0.45 & -- & 0.45
	\\
		1 & 7 & 0.46 & 0.47 & 0.48 & 0.48 & -- & 0.48
	\\
		1 & 6 & 0.50 & 0.52 & 0.52 & 0.52 & -- & 0.52
	\\
		1 & 5 & 0.56 & 0.58 & 0.58 & 0.58 & -- & 0.58
	\\
		1 & 4 & 0.66 & 0.66 & 0.67 & 0.67 & -- & 0.67
	\\
		1 & 3 & 0.80 & 0.81 & 0.81 & 0.81 & -- & 0.81
	\\
		1 & 2 & 1.11 & 1.11 & 1.11 & 1.12 & -- & 1.12
	\\
		1 & 1 & 2.03 & 2.05 & 2.06 & 2.07 & -- & 2.11
	\\
	\hline
	\end{tabular}
	\caption{The prefactors $c(m/g,mL;N)$ determined based on the numerical bosonized quantum mechanics and the $N$-site lattice calculation for $m/g=1$ and $1\leq mL\leq 8$, and the continuum extrapolated values $c^{[N\to \infty]}$ obtained from fitting data to (\ref{eq: cc_largeN_extrapolation}).}
	\label{table: summary_c_part4}
\end{table}
\end{landscape}

\section{Alternative approach to computing worldline instantons}\label{appworldlinealt}
In this appendix, we provide alternative calculations for the worldline instantons studied in \secref{sec5}, but for scalar quantum electrodynamics. Specifically, we first integrate over the proper time, and then perform the path integral, as in \cite{Affleck:1981bma,MO2015}. 
The calculations in \secref{sec5} suggest that the spin factor does not make any contributions to the effective actions for the class of instantons we investigated (see \eqref{Ztildecircle} and \eqref{spinoreigenlemon}). Thus, the scalar results here can provide a direct check against the spinor results in \secref{sec5}.

Recall from \eqref{Zs} that 
\begin{align}
-\Gamma_{E}[A]&=\int_0^\infty \frac{ds}{s} e^{-m^2s}\int_{x(s)=x(0)} \DD x(u)\ \exp\left(-\int_0^s du \left( \frac{\dot{x}^2}{4}+igA\cdot \dot{x}\right)\right) \label{eucSeffb}.
\end{align}
Integrating out the proper time $s$, in the weak-field regime $m\sqrt{\int_0^1 du \dot{x}^2}\gg 1$, this becomes approximately
\begin{align}
-\Gamma_E[A]&\approx \sqrt{\frac{2\pi}{m}}\int_{x(1)=x(0)}\DD x\ \frac{1}{(\textstyle{\int_0^1 du \dot{x}^2})^{\frac{1}{4}}}\exp\left( -m \sqrt{\int_0^1 du \dot{x}^2}-ig\int_0^1 du A\cdot\dot{x}  \right).\label{Seffnonlocal}
\end{align}
We now have a non-local worldline action,
\begin{align}
S&=m\sqrt{\int_0^1 du\ \dot{x}^2}+ig\int_0^1 du\ A\cdot\dot{x},\label{Snonlocal}
\end{align}
from which the equations of motion satisfied by the worldline instantons are
\begin{align}
\frac{m\ddot{x}_{\text{cl}\mu}}{\textstyle{\sqrt{\int_0^1 du\ \dot{x}^2_{\text{cl}}}}}&=igF_{\mu\nu}(x_{\text{cl}}(u)) \dot{x}_{\text{cl}}^\nu,\quad\dot{x}^2_{\text{cl}}=\text{constant}\equiv a^2,\;a>0.\label{eomnonlocal}
\end{align}
Around a single-instanton $x_{\text{cl}}$, the quadratic action is
\begin{align}
\delta^2 S&=\int_0^1 du \int_0^1 du'\ \delta x^\mu(u)M_{\mu\nu}(u,u')\delta x^\nu(u'),\text{ where}\\
M_{\mu\nu}(u,u')&=\left[\left(\frac{m}{(\textstyle{\int \dot{x}^2_{\text{cl}}})^{\frac{1}{2}}} \right)\delta_{\mu\nu}\left(-\frac{d^2}{du^2} \right)+igF_{\mu\nu}(x_{\text{cl}}(u))\frac{d}{du} \right]\delta(u-u')\label{Mmunu}\\
&\qquad\qquad -\frac{m}{(\textstyle{\int \dot{x}^2_{\text{cl}}})^{\frac{3}{2}}}\ddot{x}_{\text{cl}\;\mu}(u)\ddot{x}_{\text{cl}\;\nu}(u')\nonumber\\
&\equiv L_{\mu\nu}(u)\delta(u-u')-\frac{m}{(\textstyle{\int \dot{x}^2_{\text{cl}}})^{\frac{3}{2}}}\ddot{x}_{\text{cl}\;\mu}(u)\ddot{x}_{\text{cl}\;\nu}(u').
\end{align}
In the last line, we separated the operator $M_{\mu\nu}(u,u')$ into the local part with $L_{\mu\nu}(u)$ and the non-local part.\footnote{Studies of eigenvalue problems of non-local Sturm-Liouville operators such as \eqref{Mmunu} exist in the literature; see e.g. \cite{freitas_1994} and \cite{doi:10.1080/00036810600555171}.} Since we have not factored the endpoint $x(0)=x(1)$ out of the path integral the fluctuation $\delta x^\mu(u)$ satisfies the periodic boundary condition,
\begin{align}
\text{Periodic Condition: }\delta x^\mu (0)=\delta x^\mu (1).\label{PBC}
\end{align}
Under this, we take $M_{\mu\nu}$ as \textit{self-adjoint}, so that the spectral theorem applies. The quadratic operator $M_{\mu\nu}(u,u')$ around a saddle point $x_{\text{cl}}$ typically has zero modes, which contribute to the path integral by some factor $N_0$ \cite{Vandoren:2008xg}. The non-zero eigenmodes contribute to the path integral \eqref{Seffnonlocal} by the functional determinant with zero modes discarded, $\left(\det\nolimits'{M_{\mu\nu}^{(n)}}\right)^{-1/2}$.

Therefore, the path integral in \eqref{Seffnonlocal} is, at the quadratic level,
\begin{align}
-\Gamma_E[A]&\approx \sqrt{\frac{2\pi}{m}}\sum_{n} \frac{1}{(\textstyle{\int_0^1 du\ \dot{x}^2_{\text{cl,n}}})^{\frac{1}{4}}} e^{-S_{n,0}} N_{n,0}\left(\det\nolimits'{M_{\mu\nu}^{(n)}}\right)^{-\frac{1}{2}}.\label{Seffnonlocaln}
\end{align} 
Here for each inequivalent single-instantons $x_{\text{cl},n}$, we have denoted the on-shell actions by $S_{n,0}$ and the zero mode factors by $N_{n,0}$. We have approximated $(\int_0^1 du\ \dot{x}^2)^{-\frac{1}{4}}$ by its on-shell value at each instanton $x_{\text{cl,n}}$.

\subsection{The straight line instantons}
The effective action \eqref{GammaEngo2} for the amplitude $\langle +g/2|+g/2\rangle$ is
\begin{align}
-\Gamma_E^{(n)}&=\prod_{i=1}^n \int_0^\infty\frac{ds_i}{s_i}e^{-m^2 s_i} \int_{-\infty}^{\tilde{x}^2_{i-1}} d\tilde{x}^2_i \int_0^L d\tilde{x}^1_i\ \tilde{Z}(s_i,\tilde{x}_i),\;\tilde{x}^2_0\equiv -\infty+t_E\label{GammaEngo2b},\\
\tilde{Z}(s_i,\tilde{x}_i)&=\int_{x_i(s)=x_i(0)=\tilde{x}_i}\DD x_i\ \exp\left(-\int_0^{s_i}du_i\left(\frac{\dot{x}^2_i}{4}+igA\cdot\dot{x}_i \right)\right) .\label{Zsib}
\end{align}
We have adopted the formula for scalar quantum electrodynamics. From \eqref{Seffnonlocaln}, for $m\sqrt{\int_0^1 du\ \dot{x}^2}\gg 1$, we have
\begin{align}
-\Gamma^{(n)}_E&\approx\left(\frac{2\pi}{m}\right)^{\frac{n}{2}} \left(\prod_{i=1}^n\frac{1}{(\textstyle{\int_0^1 du_i\ \dot{x}^2_{\text{cl,i}}})^{\frac{1}{4}}}\right) e^{-S_{n,0}} N_{n,0}\left(\det\nolimits'{M_{\mu\nu}^{(n)}}\right)^{-\frac{1}{2}}.\label{Seffnonlocalstlinen}
\end{align}
Since the electric field on the instanton worldlines is zero, $\bar{E}\equiv E(x_{i\text{cl}}(u_i))=0$, the equation of motion \eqref{eomnonlocal} for $n$ instantons ($n$ even) is again solved by straight lines,
\begin{equation}
\begin{aligned}
x_{i\text{cl}}^1(u_i)&=\pm L u_i+\tilde{x}_i^1 ,\; x_{i\text{cl}}^2(u_i)=\tilde{x}_i^2\in (-\infty,\tilde{x}_{i-1}^2) ,\; \\
\bar{E}&\equiv E(x_{i\text{cl}}(u_i))=0,\; E(x^2\rightarrow \pm \infty)=+g/2.
\end{aligned}
\end{equation}
From this, we have the on-shell action
\begin{align}
m\sqrt{\int_0^1 du_i\ \dot{x}^2_{i\text{cl}}}=mL \quad\Rightarrow \quad S_{n,0}=nmL.
\end{align}
This also means that the regime $m\sqrt{\int_0^1 du\ \dot{x}^2}\gg 1$ corresponds to the semi-classical regime $mL\gg 1$, i.e. the on-shell action $S_{n,0}=nmL\gg \hbar$. To find the zero mode factor $N_{n,0}$, note that for our straight line instantons, the reparametrization mode redundancy $u_i\longrightarrow u_i+\delta u_i$ coincides with the translation of $x_{i\text{cl}}^1$, thus we only have a volume factor,
\begin{align}
N_{n,0}&=\frac{1}{n!}\left(\frac{it L}{2\pi}\right)^n.
\end{align}
Next, since $\bar{E}\equiv E(x_{i\text{cl}}(u_i))=0$ and $x_{i\text{cl}}(u_i)$ are linear, the quadratic operator $M^{(n)}_{\mu\nu}$ is simply
\begin{align}
M^{(n)}_{\mu\nu}&=\mathop{\otimes}_{i=1}^n \left(-\delta_{\mu\nu}\frac{m}{L}\frac{d^2}{du^2_i} \right)\quad \Rightarrow \quad \left(\det\nolimits'{M_{\mu\nu}^{(n)}}\right)^{-\frac{1}{2}}=\left(\frac{m}{L} \right)^n.
\end{align}
Therefore, \eqref{Seffnonlocalstlinen} (in the regime $mL\gg 1$) becomes
\begin{align}
-\Gamma^{(n)}_E[A]&=\left(\frac{2\pi}{m}\right)^{\frac{n}{2}} \left(\frac{1}{L}\right)^{\frac{n}{2}} e^{-nmL} \left(\frac{1}{n!}\left(\frac{it L}{2\pi}\right)^n \right)\left(\frac{m}{L} \right)^n=\frac{1}{n!}\left((it) \frac{1}{\sqrt{2\pi}}\sqrt{\frac{m}{L}} e^{-mL} \right)^n,
\end{align}
from which \eqref{ppgo2} follows.

\subsection{The lemon instantons}

\subsubsection*{Classical Solutions}
Focusing on the single-instanton solution, the equations of motion \eqref{eomnonlocal} explicitly read
\begin{subequations}
\begin{empheq}[left=\empheqlbrace]{align}
\ddot{x}^1_{\text{cl}}&=+\frac{iga}{m}F_{12}\dot{x}_{\text{cl}}^2\\
\ddot{x}^2_{\text{cl}}&=-\frac{iga}{m}F_{12}\dot{x}_{\text{cl}}^1\\
a^2&=(\dot{x}^1_{\text{cl}})^2+(\dot{x}^2_{\text{cl}})^2,\;a>0.
\end{empheq}
\end{subequations}
The solution is given by (see Figure \ref{lemonsingle})
\begin{align}
(1)\text{: }&\begin{dcases}
x^1_{\text{cl}}&=(a/b)\sin{(b(u+1/4))}+\tilde{x}^1\\
x^2_{\text{cl}}&=+(a/b)(\cos{(b(u+1/4))}-\cos{(b/4)})+\tilde{x}^2
\end{dcases},\quad u\in [-1/2,0]\label{lemon1b}\\
(2)\text{: }&\begin{dcases}
x^1_{\text{cl}}&=(a/b)\sin{(b(u-1/4))}+\tilde{x}^1\\
x^2_{\text{cl}}&=-(a/b)(\cos{(b(u-1/4))}-\cos{(b/4)})+\tilde{x}^2
\end{dcases},\quad u\in (0,1/2).\label{lemon2b}
\end{align}
We have taken $u\in[-1/2,1/2)$. Periodicity in $x^1\sim x^1+L$, i.e. $x^1_{\text{cl}}(0)-x^1_{\text{cl}}(-1/2)=L=x^1_{\text{cl}}(1/2)-x^1_{\text{cl}}(0)$, implies
\begin{align}
b&=4\sin^{-1}{\left(L\left/\left( \frac{2m}{g\bar{E}}\right)\right. \right)}\quad,\quad a=\sqrt{\int \dot{x}^2_{\text{cl}}}=\left(\frac{m}{g\bar{E}}\right) 4\sin^{-1}{\left(L\left/\left( \frac{2m}{g\bar{E}}\right)\right.\right)}.\label{lemonab2}
\end{align}
The weak-field limit becomes
\begin{align}
m\sqrt{\int \dot{x}^2}\gg 1 \quad &\Rightarrow \quad \left(\frac{m^2}{g\bar{E}}\right) 4\sin^{-1}{\left(L\left/\left( \frac{2m}{g\bar{E}}\right)\right.\right)}\gg 1 \quad \overset{L\ll \frac{2m}{g\bar{E}}}{\Rightarrow} \quad 2mL\gg 1.\label{lemonweakfield}
\end{align}
Note that 
\begin{align}
L\longrightarrow \frac{2m}{g\bar{E}}\quad\Rightarrow\quad b\longrightarrow 2\pi.
\end{align}
The on-shell action for this single-instanton solution takes the form
\begin{align}
S_{E1,0}&=m\sqrt{\int_{-\frac{1}{2}}^{\frac{1}{2}} du\ \dot{x}_{\text{cl}}^2}+\cancel{\frac{1}{2}\int d^2x\ g^2}-\frac{1}{2}g^2\times(\text{Lemon Area})\\
&=mL\sqrt{1-\left( L\left/\left(\frac{2m}{g\bar{E}}\right) \right.\right)^2}+m \left(\frac{2m}{g\bar{E}} \right)\sin^{-1}{\left(L\left/\left( \frac{2m}{g\bar{E}}\right)\right. \right)}\label{lemonSonshell}.
\end{align}

\subsubsection*{One-Loop Corrections}

In the quadratic action \eqref{Mmunu}, the fluctuations must have continuous zeroth and first derivatives. The path integral \eqref{Seffnonlocal} is defined with periodic boundary condition $x(+1/2)=x(-1/2)$. Alternatively, we can pull the endpoint $x(+1/2)=x(-1/2)=\tilde{x}$ out from the path integral, perform the path integral with Dirichlet condition on $\delta x^\mu(u)$, and then integrate over $\tilde{x}$. This was done when we integrated over the proper time later, in \eqref{DBC}, and in \cite{MO2015}. In this choice, $M_{\mu\nu}(u,u')$ has no zero modes and the path integral is contributed by $\det M_{\mu\nu}$. Following \cite{MO2015}, we make use of the matrix determinant lemma to write the determinant $\det M_{\mu\nu}$ (for either choice of boundary condition) as
\begin{align}
\det M_{\mu\nu}&=\det L_{\mu\nu}\times\left(1-\frac{m}{(\textstyle{\int \dot{x}^2_{\text{cl}}})^{\frac{3}{2}}}\int_{-\frac{1}{2}}^{\frac{1}{2}}du\int_{-\frac{1}{2}}^{\frac{1}{2}}du'\ \ddot{x}_{\text{cl}}^{\mu}(u) G_{\mu\nu}(u,u') \ddot{x}_{\text{cl}}^{\nu}(u') \right).\label{localnonlocal}
\end{align}
Here $G_{\mu\nu}(u,u')$ is the Green's function of $L_{\mu\nu}$ under the specified boundary condition. 

\subsubsection*{The Local Part}
We wish to compute the determinant $\det\nolimits{L_{\mu\nu}}$ for either periodic or Dirichlet boundary condition, where
\begin{align}
L_{\mu\nu}&=\begin{pmatrix}
-\frac{m}{\textstyle{(\int \dot{x}^2_{\text{cl}}})^{\frac{1}{2}}}\frac{d^2}{du^2} & ig F_{12}(x_{\text{cl}}(u))\frac{d}{du}\\
-ig F_{12}(x_{\text{cl}}(u))\frac{d}{du}  &  -\frac{m}{\textstyle{(\int \dot{x}^2_{\text{cl}}})^{\frac{1}{2}}}\frac{d^2}{du^2}
\end{pmatrix}=\frac{m}{a}\begin{pmatrix}
-\frac{d^2}{du^2}  &  -\text{sign}(u)b \frac{d}{du}\\
+\text{sign}(u)b \frac{d}{du}  &  -\frac{d^2}{du^2}
\end{pmatrix}\label{Llemon},
\end{align}
$u\in[-1/2,1/2)$, and $a$ and $b$ were defined in \eqref{lemonab2}. We define the periodic function $\text{sign}(u)$ as
\begin{align}
\text{sign}(u)&\equiv \begin{cases}
+1&,\quad -1<u<-1/2,0<u<1/2,\ldots\\
-1&,\quad -1/2<u<0,1/2<u<1,\ldots\\
c\in[-1,1]&,\quad u=\ldots-1/2,0,1/2,\ldots
\end{cases}.
\end{align}
Crucially, in order that the determinant $\det L_{\mu\nu}$ be well-defined, $L_{\mu\nu}$ must be self-adjoint acting on the space $\Omega$ of fluctuations with inner product
\begin{align}
(\vec{x},\vec{y})\equiv \int^{1/2}_{-1/2} du\; \vec{x}\cdot \vec{y}\quad,\quad(\vec{x} , L \vec{y})\overset{!}{=}(L \vec{x} ,\vec{y}),\quad\forall \vec{x},\vec{y}\in\Omega,\label{Linnerwrong}
\end{align}
so that the spectral theorem applies. Let us denote $L_{\mu\nu}$ in \eqref{Llemon} as
\begin{align}
L_{\mu\nu}&=\frac{m}{a}\begin{pmatrix}
-\frac{d^2}{du^2}  &  +b(u) \frac{d}{du}\\
-b(u) \frac{d}{du}  &  -\frac{d^2}{du^2}
\end{pmatrix},\quad \text{where }b(u)\equiv -\text{sign}(u)b.
\end{align}
For any $\vec{x},\vec{y}\in\Omega$, (we shifted the range of $u$ by $\varepsilon$ to avoid the ambiguity in $b(u)$ at $u=\pm 1/2$)
\begin{align}
&\quad \frac{a}{m}(\vec{x} , L \vec{y})=\frac{a}{m}\int_{-\frac{1}{2}-\varepsilon}^{\frac{1}{2}-\varepsilon}du\; \vec{x} \cdot L \vec{y}\\
&=\frac{a}{m}(L\vec{x} , \vec{y})+\int_{-\frac{1}{2}-\varepsilon}^{\frac{1}{2}-\varepsilon}du\;\partial_u \left( -x^1\partial_u y^1+\partial_u x^1 y^1-x^2\partial_u y^2+\partial_u x^2 y^2 +b(u)(x^1 y^2-x^2 y^1)\right)\nonumber\\
&\qquad\qquad-\int_{-\frac{1}{2}-\varepsilon}^{\frac{1}{2}-\varepsilon}du\; \partial_u b(u)(x^1 y^2-x^2 y^1).
\end{align}
Imposing that the fluctuations $\delta x^\mu(u)$ have continuous zeroth and first derivatives at $u=0$ and $u=\pm 1/2$, the boundary terms all cancel, and so
\begin{align}
\frac{a}{m}(\vec{x} , L \vec{y})-\frac{a}{m}(L\vec{x} , \vec{y})&=-\int_{-\frac{1}{2}-\varepsilon}^{\frac{1}{2}-\varepsilon}du\; \partial_u b(u)(x^1 y^2-x^2 y^1)=2b\left.  \left(x^1 y^2-x^2 y^1\right)\right|^{u=0}_{u=-\frac{1}{2}}.
\end{align}
Therefore, we conclude that for $L_{\mu\nu}$ to be self-adjoint, the space $\Omega$ of fluctuations on which it acts consists of those with continuous zeroth and first derivatives at $u=0$ and $u=\pm 1/2$, and must satisfy
\begin{align}
\text{self-adjoint $L_{\mu\nu}$: }(\vec{x} , L \vec{y})-(L\vec{x} , \vec{y})=\left(\frac{m}{a}\right)2b\left.\left(x^1 y^2-x^2 y^1\right)\right|^{u=0}_{u=-\frac{1}{2}}\overset{!}{=}0,\quad\forall \vec{x},\vec{y}\in\Omega.\label{selfadL}
\end{align}

With the self-adjointness condition of $L_{\mu\nu}$ sorted out, we now find its eigenvalues and eigenmodes explicitly, for either choice of boundary condition. This is conveniently achieved by first diagonalizing $L_{\mu\nu}$ in \eqref{Llemon}:
\begin{align}
L_{\mu\nu}&=\frac{m}{a}\begin{pmatrix}
-\frac{d^2}{du^2}  &  -\text{sign}(u)b \frac{d}{du}\\
+\text{sign}(u)b \frac{d}{du}  &  -\frac{d^2}{du^2}
\end{pmatrix}\\
&=U^{-1} \frac{m}{a}\begin{pmatrix}
-\frac{d^2}{du^2}-\text{sign}(u) ib \frac{d}{du} &0 \\
0 & -\frac{d^2}{du^2}+\text{sign}(u) ib \frac{d}{du}
\end{pmatrix} U= U^{-1}\mathcal{D}U,\\
U&\equiv \frac{1}{\sqrt{2}}\begin{pmatrix}
i & 1\\
-i & 1
\end{pmatrix}.
\end{align}
Continuity of zeroth and first derivatives at $u=0$ and $u=\pm 1/2$ imposes strong constraints on the eigenmodes. The only zero modes allowed are $(1,0)$ and $(0,1)$ which generate the global translations. They are zero modes with periodic boundary condition; if we impose Dirichlet condition, there are no zero modes. 

The eigenmodes are given by
\begin{subequations}
\begin{align}
y_{(1),n}&= \begin{cases}
(+\sqrt{2}\sin{(2\pi n u)}\sin{\left(\frac{bu}{2}\right)},+\sqrt{2}\sin{(2\pi n u)}\cos{\left(\frac{bu}{2}\right)})\qquad\qquad&,\; -1/2\leq  u \leq 0\\
(-\sqrt{2}\sin{(2\pi n u)}\sin{\left(\frac{bu}{2}\right)},+\sqrt{2}\sin{(2\pi n u)}\cos{\left(\frac{bu}{2}\right)})&,\; 0< u < 1/2
\end{cases},\nonumber\\
y_{(2),n}&=\begin{cases}
(-\sqrt{2}\sin{(2\pi n u)}\cos{\left(\frac{bu}{2}\right)},+\sqrt{2}\sin{(2\pi n u)}\sin{\left(\frac{bu}{2}\right)})\qquad\qquad&,\; -1/2\leq  u \leq 0\\
(-\sqrt{2}\sin{(2\pi n u)}\cos{\left(\frac{bu}{2}\right)},-\sqrt{2}\sin{(2\pi n u)}\sin{\left(\frac{bu}{2}\right)})&,\; 0< u < 1/2
\end{cases},\nonumber\\
y_{(3),n}&=\begin{cases}
\frac{1}{\sqrt{2}n}(+2n \cos(2\pi n u) \sin{\left(\frac{bu}{2} \right)}-\frac{b}{2\pi}\sin(2\pi n u)\cos{\left(\frac{bu}{2} \right)},\\
\qquad\qquad\qquad +2n \cos(2\pi n u) \cos{\left(\frac{bu}{2} \right)}+\frac{b}{2\pi}\sin(2\pi n u)\sin{\left(\frac{bu}{2} \right)} )&,\; -1/2\leq  u \leq 0\\
\frac{1}{\sqrt{2}n}(-2n \cos(2\pi n u) \sin{\left(\frac{bu}{2} \right)}+\frac{b}{2\pi}\sin(2\pi n u)\cos{\left(\frac{bu}{2} \right)},\\
\qquad\qquad\qquad +2n \cos(2\pi n u) \cos{\left(\frac{bu}{2} \right)}+\frac{b}{2\pi}\sin(2\pi n u)\sin{\left(\frac{bu}{2} \right)} )&,\; 0< u < 1/2
\end{cases},\nonumber\\
y_{(4),n}&=\begin{cases}
\frac{1}{\sqrt{2}n}(-2n \cos(2\pi n u) \cos{\left(\frac{bu}{2} \right)}-\frac{b}{2\pi}\sin(2\pi n u)\sin{\left(\frac{bu}{2} \right)},\\
\qquad\qquad\qquad +2n \cos(2\pi n u) \sin{\left(\frac{bu}{2} \right)}-\frac{b}{2\pi}\sin(2\pi n u)\cos{\left(\frac{bu}{2} \right)} ) &,\; -1/2\leq  u \leq 0\\
\frac{1}{\sqrt{2}n}(-2n \cos(2\pi n u) \cos{\left(\frac{bu}{2} \right)}-\frac{b}{2\pi}\sin(2\pi n u)\sin{\left(\frac{bu}{2} \right)},\\
\qquad\qquad\qquad -2n \cos(2\pi n u) \sin{\left(\frac{bu}{2} \right)}+\frac{b}{2\pi}\sin(2\pi n u)\cos{\left(\frac{bu}{2} \right)} )&,\; 0< u < 1/2
\end{cases},\nonumber
\end{align}\label{y1234}
\end{subequations}
where $n=1,2,\ldots$. They all have eigenvalues
\begin{align}
\lambda_n&=\left[\left(\frac{2\pi}{b} \right)^2 n^2-\frac{1}{4}\right]\left(\frac{m}{a}\right) b^2,\quad n =1,2,\ldots.
\end{align}
Of these, $y_{(1),n}$ and $y_{(2),n}$ satisfy the Dirichlet condition vanishing at $u=\pm 1/2$, and moreover vanish at $u=0$. 

We need to check if $y_{(i),n}$ are orthogonal under the inner product $(y_{(i),n},y_{(j),n'})\equiv \int_{-1/2}^{1/2}y_{(i),n}\cdot y_{(j),n'}$. It can be easily shown that the inner products are orthogonal--- all save one: the inner product
\begin{align}
(y_{(3),n},y_{(4),n'})&=\frac{b}{(n^2-{n'}^2)\pi^2}((-1)^{n+n'}-1),\quad n\neq n',\label{normnonorth}
\end{align}
cannot be made vanishing. This means that the modes $y_{(3),n}$ and $y_{(4),n'}$ are not orthogonal when $n$ and $n'$ are not both even or both odd ($n\neq n'$ in particular). By the spectral theorem, this implies that $L_{\mu\nu}$ acting on the space spanned by $y_{(3),n}$ and $y_{(4),n'}$ is not self-adjoint. Indeed, we can compute the obstruction \eqref{selfadL} to the self-adjointness of $L_{\mu\nu}$,
\begin{align}
(y_{(3),n} , L y_{(4),n'})-(Ly_{(3),n} , y_{(4),n'})=\left(\frac{m}{a}\right)2b\left.\left(y_{(1),n}\times y_{(2),n'}\right)\right|^{u=0}_{u=-\frac{1}{2}}=\left(\frac{m}{a}\right)4b(1-(-1)^{n+n'}),
\end{align}
which is non-vanishing precisely for the cases where their inner products \eqref{normnonorth} do not vanish. In fact, $y_{(3),n}$ and $y_{(4),n}$ (which have positive eigenvalues) are not orthogonal to the constant zero modes $(1,0)$ and $(0,1)$ either.\\

We thus reach the conclusion that, $L_{\mu\nu}$ is not self-adjoint on the space $\Omega$ of \textit{periodic}, first-order differentiable vectors with the inner product \eqref{Linnerwrong}. The corresponding functional determinant is then ill-defined. To avoid this problem, we therefore choose to impose Dirichlet condition on the fluctuations by pulling the endpoint $x(+1/2)=x(-1/2)=\tilde{x}$ out from the path integral and integrate over it in the end. This space is spanned by $y_{(1),n}$ and $y_{(2),n}$ in \eqref{y1234}. Since they vanish at $u=\pm 1/2$, and also at $u=0$, they satisfy the self-adjoint condition \eqref{selfadL}. Thus, in the Dirichlet problem, $L_{\mu\nu}$ has no zero modes and has eigenvalues 
\begin{align}
\lambda_n&=\left[\left(\frac{2\pi}{b} \right)^2 n^2-\frac{1}{4}\right]\left(\frac{m}{a}\right) b^2,\quad n =1,2,\ldots.\label{locallambdanb}
\end{align}
with multiplicity two. The functional determinant is then
\begin{align}
\det L_{\mu\nu}&=\prod_{n=1}^\infty \left(\left[\left(\frac{2\pi}{b} \right)^2 n^2 -\frac{1}{4}\right] \left(\frac{m}{a} \right)b^2 \right)^2=\frac{\sin^2{(b/4)}}{(b/4)^2} \left( \frac{m}{a} \right)^{-1}.
\end{align}
Furthermore, the determinant of the free operator $L_{\mu\nu}^{\text{free}}=-(m/a)\delta_{\mu\nu}d^2/du^2$ (with Dirichlet condition) is easily found to be $\det L_{\mu\nu}^{\text{free}}=(m/a)^{-1}$, so
\begin{align}
\sqrt{\frac{\det L_{\mu\nu}^{\text{free}}}{\det L_{\mu\nu}}}&=\frac{b/4}{\sin{(b/4)}}.\label{lemonlocalratio2}
\end{align}
We can also reproduce \eqref{lemonlocalratio2} by the Gel'fand-Yaglom  method \cite{GY1960}, similar to what was done in \cite{MO2015}. 

\subsubsection*{The Non-Local Part}
Next, we compute the non-local part of the function determinant $\det M_{\mu\nu}$ \eqref{localnonlocal} with Dirichlet condition,
\begin{align}
1-\frac{m}{(\textstyle{\int \dot{x}^2_{\text{cl}}})^{\frac{3}{2}}}\int_{-\frac{1}{2}}^{\frac{1}{2}}du\int_{-\frac{1}{2}}^{\frac{1}{2}}du'\ \ddot{x}_{\text{cl}}^{\mu}(u) G_{\mu\nu}(u,u') \ddot{x}_{\text{cl}}^{\nu}(u') .
\end{align}
Recall that the Dirichlet eigenmodes are $y_{(1),n}$ and $y_{(2),n}$ in \eqref{y1234}, which are real and \textit{orthonormal}. The Green's function $G_{\mu\nu}(u,u')$ of $L_{\mu\nu}$ (for the Dirichlet problem) is then
\begin{align}
G^{\mu\nu}(u,u')&=\sum_{\substack{i=1,2\\n\in\mathbb{N}}}\frac{1}{\lambda_n} y_{(i),n}^\mu(u)y_{(i),n}^\nu(u'),
\end{align}
which results in
\begin{align}
&\quad 1-\frac{m}{(\textstyle{\int \dot{x}^2_{\text{cl}}})^{\frac{3}{2}}}\int_{-\frac{1}{2}}^{\frac{1}{2}}du\int_{-\frac{1}{2}}^{\frac{1}{2}}du'\ \ddot{x}_{\text{cl}}^{\mu}(u) G_{\mu\nu}(u,u') \ddot{x}_{\text{cl}}^{\nu}(u')\\
&=1-\frac{m}{a^3}  \sum_{\substack{i=1,2\\n\in\mathbb{N}}}\frac{1}{\lambda_n} \int_{-\frac{1}{2}}^{\frac{1}{2}}du\int_{-\frac{1}{2}}^{\frac{1}{2}}du'\ \ddot{x}_{\text{cl}\mu}(u) y_{(i),n}^\mu(u)y_{(i),n}(u')^\nu \ddot{x}_{\text{cl}\nu}(u')=\frac{b}{4}\cot{(b/4)}.\label{nonlocalans}
\end{align}
Therefore, substituting \eqref{lemonlocalratio2} and \eqref{nonlocalans}, the (Dirichlet) determinant of the full operator $M_{\mu\nu}$ is
\begin{align}
\sqrt{\frac{\det M^{\text{free}}_{\mu\nu}}{\det M_{\mu\nu}}}&=\sqrt{\frac{\det L^{\text{free}}_{\mu\nu}}{\det L_{\mu\nu}}}\times \left(1-\frac{m}{(\textstyle{\int \dot{x}^2_{\text{cl}}})^{\frac{3}{2}}}\int_{-\frac{1}{2}}^{\frac{1}{2}}du\int_{-\frac{1}{2}}^{\frac{1}{2}}du'\ \ddot{x}_{\text{cl}}^{\mu}(u) G_{\mu\nu}(u,u') \ddot{x}_{\text{cl}}^{\nu}(u') \right)^{-\frac{1}{2}}\\
&= \frac{b/4}{\sin{(b/4)}} \left(\frac{b}{4}\cot{(b/4)} \right)^{-\frac{1}{2}}.\label{lemonM}
\end{align}

Finally, plugging in \eqref{lemonSonshell}, \eqref{lemonab2} and \eqref{lemonM}, the effective action \eqref{Seffnonlocal} around the lemon instanton \eqref{lemon1b} in the approximation $2mL\gg 1$ is
\begin{align}
-\Gamma_E[A]&\approx \sqrt{\frac{2\pi}{m}}\int_{x(1)=x(0)}\DD x\ \frac{1}{(\textstyle{\int_0^1 du\ \dot{x}^2})^{\frac{1}{4}}}\exp\left( -m \sqrt{\int_0^1 du\ \dot{x}^2}-ig\int_0^1 du\ A\cdot\dot{x}  \right)\\
&=\sqrt{\frac{2\pi}{m}}\frac{1}{(\textstyle{\int_0^1 du\ \dot{x}_{\text{cl}}^2})^{\frac{1}{4}}}e^{-S_{E1,0}}(itL)\left[\left(\frac{m}{2\pi  a} \right)\sqrt{\frac{\det M^{\text{free}}_{\mu\nu}}{\det M_{\mu\nu}}}\right]\\
&=(it) \frac{\sqrt{g\bar{E}}}{4\sqrt{2\pi}} \sqrt{L\left/\frac{2m}{g\bar{E}}\right.}   \left(1-\left(L \left/\frac{2m}{g\bar{E}} \right.\right)^2 \right)^{-\frac{1}{4}}\left[\sin^{-1}{\left(L \left/\frac{2m}{g\bar{E}} \right.\right)} \right]^{-1} e^{-S_{E1,0}}
\end{align}
which is the same as the contribution from the dominant $p=0$ saddle in \eqref{lemonfinalp}. In the second equality, the factor $m/2\pi a $ gives the correct normalization to the ratio of functional determinants, cf. \eqref{scalardetnorm}.

\bibliography{Schwinger.bib}

\end{document}